\documentclass[twoside,10pt]{article}
\usepackage{amsmath}
\usepackage{amssymb}
\usepackage{graphicx}

\renewcommand{\theequation}{\thesection.\arabic{equation}}

\newtheorem{satz}{Theorem}[section]
\newtheorem{definition}[satz]{Definition}
\newtheorem{lemma}[satz]{Lemma}

\newtheorem{proposition}[satz]{Proposition}

\newtheorem{notation}{Notation}[section]
\newtheorem{assumption}{Assumption}
\renewcommand{\section}{\secdef\sct\sect}
\newcommand{\sct}[2][default]{\refstepcounter{section}
\vspace{0.5cm} \setcounter{equation}{0}
\centerline{ \scshape \arabic{section}.\ #1} \vspace{0.3cm}}
\newcommand{\sect}[1]{
\vspace{0.5cm} \centerline{\large\scshape #1} \vspace{0.3cm}}
\renewcommand{\subsection}{\secdef \subsct\sbsect}
\newcommand{\subsct}[2][default]{\refstepcounter{subsection}
\nopagebreak \vspace{0.5\baselineskip} {\flushleft\bf
\arabic{section}.\arabic{subsection}~\bf #1  } \nopagebreak}
\newcommand{\sbsect}[1]{\vspace{0.1cm}\noindent
{\bf #1}\vspace{0.1cm}}
\renewcommand{\subsubsection}{\secdef \subsubsect\sbsbsect}
\newcommand{\subsubsect}[2][default]{\refstepcounter{subsubsection} \nopagebreak
\vspace{0.1\baselineskip} \nopagebreak {\flushleft
\sffamily\slshape
\arabic{section}.\arabic{subsection}.\arabic{subsubsection}
\ \sffamily #1\/.}\ }
\newcommand{\sbsbsect}[1]{\vspace{0.1cm}\noindent
{\bf #1}\ }
\input{tcilatex}
\begin{document}

\title{Characterization of the Quasi-Stationary State of an Impurity Driven
by Monochromatic Light {II} - {M}icroscopic Foundations}
\author{J.-B. Bru and W. de Siqueira Pedra}
\date{\today }
\maketitle

\begin{abstract}
From quantum mechanical first principles only, we rigorously study the
time--evolution of a $N$--level atom (impurity) interacting with an external
monochromatic light source within an infinite system of free electrons at
thermal equilibrium (reservoir). In particular, we establish the relation
between the full dynamics of the compound system and the effective dynamics
for the $N$--level atom, which is studied in detail in \cite%
{BruPedraWestrich2011a}. Together with \cite{BruPedraWestrich2011a} the
present paper yields a purely microscopic theory of optical pumping in laser
physics. The model we consider is general enough to describe gauge invariant
atom--reservoir interactions.
\end{abstract}

\section{Introduction\label{Section intro}}

Optical pumping is an important method in laser technology to produce the
so--called \textquotedblleft inversion of population\textquotedblright\ of
some optically active (quantum) system, as for instance impurities in
crystals \cite{Laser1, Laser2}. Such an inversion is then used to obtain
optical amplification through stimulated emission of photons. We aim to
derive the phenomenon of optical pumping by (i) being coherent with the
phenomenological description of physics textbooks and experimental facts,
(ii) starting from first principles of quantum mechanics only, and applying
mathematically rigorous methods to study microscopic models.

We analyzed in \cite{BruPedraWestrich2011a} the \emph{effective}
time--evolution of a $N$--level atom (an impurity) interacting with an
external monochromatic light source within a host environment (reservoir),
which is represented by an infinite system of free fermions at thermal
equilibrium. The external monochromatic light source is a time--periodic
classical field stimulating transitions between two energy levels of the
impurity. We showed \cite{BruPedraWestrich2011a} from this effective
dynamics how an inversion of population of energy levels of the impurity can
appear and derive a dynamical law for the evolution of populations under the
influence of the external oscillating field (optical pumping). We proved
\cite{BruPedraWestrich2011a} that a generalization of the celebrated \emph{%
Pauli master equation,} used in all standard textbooks on laser physics,
correctly describes the time--evolution of populations. In contrast to the
usual Pauli equation, this generalization takes memory effects into account.
This proof uses \cite[Theorem 3.3]{BruPedraWestrich2011a}, which states that
the restriction of the full unitary dynamics (of the
impurity--reservoir--pump system) to the $N$--level atom is properly
described -- up to small corrections for moderate pump strengths -- by an
effective non--autonomous time--evolution involving atomic degrees of
freedom only. The detailed proof of this assertion, which is not performed
in \cite{BruPedraWestrich2011a}, is the main issue of the present paper.

Thus, together with \cite{BruPedraWestrich2011a}, the results presented here
give a complete microscopic derivation of optical pumping and the induced
inversion of population from quantum mechanical first principles only.
Indeed, to our knowledge there is only one framework in which aspects of
laser phenomenology has been mathematically rigorously analyzed from first
principles, namely for some versions of the Dicke model \cite{Dic}, see \cite%
{HeppLieb1973,HeppLieb1973-2,HeppLieb1973-3,AlliSewell1995}. Nevertheless,
Dicke--type models are based on two--level atoms whereas the phenomenology
of lasers as described in physics textbooks is based on three-- or
four--level atoms \cite{Levine1968}. Moreover, they cannot explain the
inversion of population at finite number of impurities. For more details,
see \cite[Section 1]{BruPedraWestrich2011a} as well as \cite[Chap. 11]%
{Sewell}. Note that so--called \textquotedblleft one--atom
lasers\textquotedblright\ are object of recent research, both experimentally
and theoretically. See for instance \cite{mcgyver}. In a future work, we aim
to couple the impurity--reservoir--pump system considered here to a cavity
(few--mode bosonic field) in order to study light amplification in such
devices, directly from the microscopic quantum dynamics.

The derivation of the effective atomic dynamics is conceptually similar to
what is done in \cite[Section 4]{BruPedraWestrich2011a}, but technically
more involved: We represent the non--autonomous evolution as an autonomous
dynamics on some enlarged Hilbert space of periodic functions
(Floquet--Howland method). Next, we perform an analytic translation $%
\mathcal{G}(\theta )$ of the generator $\mathcal{G}$ of the autonomous
dynamics and prove that the dynamics driven by both operators coincide with
each other when restricted to the atomic subspace. We then study the
discrete spectrum and eigenspaces of $\mathcal{G}(\theta )$ through Kato's
perturbation theory \cite{Kato}. Finally, by using the inverse Laplace
transform for strongly continuous semigroups together with Riesz
projections, we analyze the action of the semigroup $\{\mathrm{e}^{\alpha
\mathcal{G}(\theta )}\}_{\alpha \geq 0}$ on vectors of the atomic subspace.
This analysis leads to the main result of the paper, that is, Theorem \ref%
{ninja thm cool}. Notice that, as compared to the model used in \cite%
{BruPedraWestrich2011a} to illustrate the microscopic origin of the
effective dissipative dynamics of the impurity, we consider here a more
general atom--reservoir coupling in order to include gauge invariant
interactions. The results of \cite{BruPedraWestrich2011a} only concern the
effective dynamics of the impurity and are very general. They also hold for
such more physical microscopic interactions.

To finish, we would like to drive the attention of the reader to \cite{BMPS}
which complements the present work in the following sense:\ The model
considered in \cite{BMPS} is exactly the same one treated here, up to the
fact that the time--dependent perturbation $H_{c}(t)$ of the atomic part
commutes with the unperturbed atomic Hamiltonian $H_{c}$ and is very strong.
See \cite[Section \textquotedblleft model and main results\textquotedblright
]{BMPS}. By contrast, in the situation considered here the time--dependent
perturbation $\eta H_{\mathrm{p}}(t)$ does not commute with the atomic
Hamiltonian $H_{\mathrm{at}}$ and is very weak. See Section \ref{section
math framework} below. Moreover, here we obtain an effective purely atomic
dynamics which well approximates the full evolution of populations uniformly
for all times, whereas in \cite{BMPS} only the time scale $t\sim \left(
\lambda T\right) ^{-1}$ is considered. In \cite{BMPS}, as well as in the
present work, $\lambda $ denotes the reservoir--atom coupling and is small,
but not vanishing. $T$ stands in \cite{BMPS} for the time--period of the
\textquotedblleft control term\textquotedblright\ $H_{c}(t)$ and the regime
of interest is the limit $T\downarrow 0$ with $H_{c}(t)=\mathcal{O}(T^{-1})$%
. By contrast, here the time--period of $H_{\mathrm{p}}(t)$ is fixed once
for all (the pump frequency is set to be equal to the largest Bohr frequency
of the atom) and the pump--atom coupling $\eta $ is of order $\mathcal{O}%
(\lambda ^{2})$. Mathematical methods like time--dependent $C$%
--Liouvilleans, evolution groups and Howland operators, complex spectral
deformation, Riesz projections, and others used in \cite{BMPS} will be again
employed here. Because of the big difference between the analyzed regimes
explained above, note however that both studies differ from each other in
which concerns technical aspects.

The paper is organized as follows. In Section \ref{section math framework}
we introduce the microscopic model. Then, in Section \ref{Section Atomic
Microscopic Dynamics} we define the dynamics of the
impurity--reservoir--pump system and state our main result (Theorem \ref%
{ninja thm cool}), which is proven in Section \ref{Section technical proofs}%
. Finally, Section \ref{sectino appendix} is an appendix\ briefly reviewing,
for the reader's convenience, some useful mathematical objects.

\begin{notation}
\label{remark constant}\mbox{
}\newline
To simplify notation, we denote by $C,D\in \mathbb{R}^{+}$ generic positive
and finite constants. Note that these constants do not need to be the same
from one statement to another. A norm on space $\mathcal{X}$ is denoted by $%
\Vert \cdot \Vert _{\mathcal{X}}$. A similar notation is used for scalar
products in Hilbert spaces. $\mathbf{1}_{\mathcal{X}}$ denotes either the
identity of a $C^{\ast }$--algebra $\mathcal{X}$ or the identity operator
acting on $\mathcal{X}$.
\end{notation}

\section{The host environment--impurity--light source microscopic model\label%
{section math framework}}

For completeness and to fix notations, we recall below the setting of \cite%
{BruPedraWestrich2011a}. We keep the discussions as short as possible and
refer to \cite[Sections 2--3]{BruPedraWestrich2011a} for details.

\subsection{The host environment as a thermal reservoir of free fermions}

Let $\mathfrak{h}_{1}:=L^{2}(\mathbb{R}^{3})$ and $\mathbf{E}:\mathbb{R}%
^{3}\rightarrow \mathbb{R}$ be any measurable rotationally invariant
function that, up to some diffeomorphism, behaves like $|p|$. In the sequel,
to simplify discussions, we set $\mathbf{E}(p)=|p|$. Define the
(multiplication) operator $h_{1}=h_{1}(\mathbf{E})$ on $\mathfrak{h}_{1}$ by
$f(p)\mapsto \mathbf{E}(p)f(p)$. Let ${\mathcal{V}}_{\mathcal{R}}$ be the CAR%
$\ C^{\ast }$--algebra generated by the annihilation and creation operators $%
a(f),$ $a^{+}(f):=a(f)^{\ast }$, $f\in \mathfrak{h}_{1}$, acting on the
antisymmetric Fock space ${\mathcal{F}}_{-}(\mathfrak{h}_{1})$ (with
one--particle Hilbert space $\mathfrak{h}_{1}$) and fulfilling canonical
anti--commutation relations. The dynamics of the reservoir is given by the
strongly continuous group $\{\tau _{t}^{{\mathcal{R}}}\}_{t\in \mathbb{R}}$
of Bogoliubov automorphisms on ${\mathcal{V}}_{{\mathcal{R}}}$ uniquely
defined by the condition:
\begin{equation}
\tau _{t}^{{\mathcal{R}}}\left( a(f)\right) =a(\mathrm{e}^{ith_{1}}f),\quad
f\in \mathfrak{h}_{1},\quad t\in \mathbb{R}\ .  \label{eq:Bogoliubov}
\end{equation}%
$\delta _{{\mathcal{R}}}$ denotes the symmetric derivation generating $\tau
_{t}^{{\mathcal{R}}}$. The initial state of the reservoir $\omega _{\mathcal{%
R}}$ at inverse temperature $\beta \in \mathbb{R}^{+}$ is the unique $\left(
\tau ^{{\mathcal{R}}},\beta \right) $--KMS\ state (thermal equilibrium
state).

\subsection{The impurity as a $N$--level atom\label{part.syst}}

Let $d\in \mathbb{N}$ and $\mathcal{B}(\mathbb{C}^{d})$ be the finite
dimensional $C^{\ast }$--algebra of all linear operators on $\mathbb{C}^{d}$
and take any self--adjoint element $H_{\mathrm{at}}=H_{\mathrm{at}}^{\ast
}\in \mathcal{B}(\mathbb{C}^{d})$ with $H_{\mathrm{at}}\notin \mathbb{R}%
\mathbf{1}_{\mathcal{B}(\mathbb{C}^{d})}\subset \mathcal{B}(\mathbb{C}^{d})$%
. Eigenvalues and eigenspaces of $H_{\mathrm{at}}$ are denoted by $E_{k}\in
\mathbb{R}$ and ${\mathcal{H}}_{k}\subset \mathbb{C}^{d}$, $k\in \{1,\ldots
,N\}$ ($N\geq 2$), respectively. $E_{k}$ is chosen such that $E_{j}<E_{k}$
whenever $j<k$. The dimension $n_{k}$ of the eigenspace ${\mathcal{H}}_{k}$
is the degeneracy of the $k$th atomic level. Then, the free atomic dynamics
is given by the group $\tau ^{\mathrm{at}}:=\{\tau _{t}^{\mathrm{at}%
}\}_{t\in \mathbb{R}}$ of automorphisms of the $C^{\ast }$--algebra $%
\mathcal{B}(\mathbb{C}^{d})$ defined by
\begin{equation}
\tau _{t}^{\mathrm{at}}(A):=\mathrm{e}^{itH_{\mathrm{at}}}A\mathrm{e}^{-itH_{%
\mathrm{at}}}\ ,\quad A\in \mathcal{B}(\mathbb{C}^{d})\ ,
\label{atom atomorph}
\end{equation}%
for all $t\in \mathbb{R}$.

Let $\omega _{\mathrm{at}}$ be any faithful state on $\mathcal{B}(\mathbb{C}%
^{d})$ and denote by $\rho _{\mathrm{at}}\in \mathcal{B}(\mathbb{C}^{d})$
its unique density matrix, i.e.,
\begin{equation*}
\omega _{\mathrm{at}}(A)=\mathrm{Tr}_{\mathbb{C}^{d}}\left( \rho _{\mathrm{at%
}}\ A\right) \ ,\qquad A\in \mathcal{B}(\mathbb{C}^{d})\ .
\end{equation*}%
For any inverse temperature $\beta \in \mathbb{R}^{+}$, the thermal
equilibrium state of the (free) atom is the (Gibbs) state $\mathfrak{g}_{%
\mathrm{at}}^{(\beta )}\equiv \mathfrak{g}_{\mathrm{at}}$ associated to the
density matrix
\begin{equation}
\rho _{\mathfrak{g}}:=\frac{\mathrm{e}^{-\beta H_{\mathrm{at}}}}{\mathrm{Tr}%
_{\mathbb{C}^{d}}\left( \mathrm{e}^{-\beta H_{\mathrm{at}}}\right) }\ .
\label{Gibbs.init}
\end{equation}

The triplet $(\mathfrak{H}_{\mathrm{at}},\pi _{\mathrm{at}},\Omega _{\mathrm{%
at}})$ stands for the standard GNS representation of $\omega _{\mathrm{at}}$%
:
\begin{equation}
\mathfrak{H}_{\mathrm{at}}:=\left( \mathcal{B}(\mathbb{C}^{d}),\left\langle
\cdot ,\cdot \right\rangle _{\mathfrak{H}_{\mathrm{at}}}\right)  \label{H at}
\end{equation}%
with
\begin{equation}
\left\langle A,B\right\rangle _{\mathfrak{H}_{\mathrm{at}}}:=\mathrm{Tr}_{%
\mathbb{C}^{d}}(A^{\ast }B)\ ,\qquad A,B\in \mathcal{B}(\mathbb{C}^{d})\ ,
\label{trace h at}
\end{equation}%
while
\begin{equation}
\Omega _{\mathrm{at}}:=\rho _{\mathrm{at}}^{1/2}\in \mathfrak{H}_{\mathrm{at}%
}\qquad \text{and}\qquad \pi _{\mathrm{at}}\left( A\right) =\underrightarrow{%
A}\ ,\quad A\in \mathcal{B}(\mathbb{C}^{d})\ .  \label{defnition vaccum at}
\end{equation}%
Here, for any $A\in \mathcal{B}(\mathbb{C}^{d})$, the left and right
multiplication operators $\underrightarrow{A}$ and $\underleftarrow{A}$ are
respectively defined on $\mathcal{B}(\mathbb{C}^{d})$ by
\begin{equation}
B\mapsto \underrightarrow{A}B:=AB\quad \text{and}\quad B\mapsto
\underleftarrow{A}B:=BA\ .  \label{left multi}
\end{equation}%
Note that the dynamics of the atom defined by (\ref{atom atomorph}) can be
represented in the Schr{\"{o}}dinger picture of Quantum Mechanics through
the Lindbladian
\begin{equation}
\mathfrak{L}_{\mathrm{at}}:=-i\left( \underrightarrow{H_{\mathrm{at}}}-\,%
\underleftarrow{H_{\mathrm{at}}}\right) =-i[H_{\mathrm{at}},\ \cdot \ ]=-%
\mathfrak{L}_{\mathrm{at}}^{\ast }  \label{Lat}
\end{equation}%
acting on the Hilbert space $\mathfrak{H}_{\mathrm{at}}$ since, for all $%
t\in \mathbb{R}$,
\begin{equation*}
\omega _{\mathrm{at}}\left( \tau _{t}^{\mathrm{at}}(A)\right) =\langle
\mathrm{e}^{t\mathfrak{L}_{\mathrm{at}}}\Omega _{\mathrm{at}},\,\pi _{%
\mathrm{at}}(A)\,\mathrm{e}^{t\mathfrak{L}_{\mathrm{at}}}\Omega _{\mathrm{at}%
}\rangle _{\mathfrak{H}_{\mathrm{at}}}\ ,\qquad A\in \mathcal{B}(\mathbb{C}%
^{d})\ .
\end{equation*}

Define the sets%
\begin{equation}
\mathfrak{t}_{\epsilon }:=\{(j,k):E_{j}-E_{k}=\epsilon \}\subset
\{1,2,\ldots ,N\}\times \{1,2,\ldots ,N\}  \label{t eps}
\end{equation}%
for each eigenvalue%
\begin{equation}
\epsilon \in \sigma (i\mathfrak{L}_{\mathrm{at}})=\left\{ E_{j}-E_{k}:j,k\in
\{1,2,\ldots ,N\}\right\} \ .  \label{definition spectre atomique}
\end{equation}%
Here, $\sigma \left( A\right) $ denotes the spectrum of the operator $A$.
Furthermore, for all $\epsilon \in \sigma \left( i\mathfrak{L}_{\mathrm{at}%
}\right) $, denote by $P_{\mathrm{at}}^{(\epsilon )}\in \mathcal{B}(%
\mathfrak{H}_{\mathrm{at}})$ the orthogonal projection on $\mathfrak{H}_{%
\mathrm{at}}$ associated to the eigenspace of $i\mathfrak{L}_{\mathrm{at}}$
with eigenvalue $\epsilon $. $\mathcal{B}(\mathfrak{H}_{\mathrm{at}})$
stands for the set of all linear operators on $\mathfrak{H}_{\mathrm{at}}$.

\subsection{The uncoupled reservoir--atom system}

Define the $C^{\ast }$--algebra ${\mathcal{V}}:=\mathcal{B}(\mathbb{C}%
^{d})\otimes {\mathcal{V}}_{\mathcal{R}}$. Then the free dynamics of the
atom--reservoir compound system is given by the strongly continuous group $%
\tau :=\{\tau _{t}\}_{t\in \mathbb{R}}$ of automorphisms of ${\mathcal{V}}$
defined by
\begin{equation}
\tau _{t}:=\tau _{t}^{\mathrm{at}}\otimes \tau _{t}^{{\mathcal{R}}}\ ,\qquad
t\in \mathbb{R}\ .  \label{free dynamics}
\end{equation}%
The generator of this dynamics is the symmetric derivation denoted by $%
\delta $ and acts on a dense sub--$\ast $--algebra $\mathrm{Dom}(\delta )$
of ${\mathcal{V}}$. The initial state of the atom--reservoir system is
\begin{equation}
\omega _{0}:=\omega _{\mathrm{at}}\otimes \omega _{{\mathcal{R}}}\ .
\label{initial state}
\end{equation}

As the state $\omega _{0}$ is faithful, the map ${\mathcal{V}}\times {%
\mathcal{V}}\rightarrow \mathbb{C}$,%
\begin{equation*}
(A,B)\mapsto \omega _{0}(A^{\ast }B)\ ,
\end{equation*}%
defines a scalar product on ${\mathcal{V}}$. For any fixed inverse
temperature $\beta \in \mathbb{R}^{+}$, set $\omega _{\mathrm{at}}:=%
\mathfrak{g}_{\mathrm{at}}^{(\beta )}$ and let the Hilbert space $\mathfrak{H%
}^{(\beta )}$ be the completion of ${\mathcal{V}}$ with respect to (w.r.t.)
the above scalar product. $\mathfrak{\tilde{P}}_{\mathrm{at}}^{(\beta )}$
stands for the orthogonal projection on $\mathfrak{H}^{(\beta )}$ with
(finite dimensional) range
\begin{equation*}
\mathrm{ran}(\mathfrak{\tilde{P}}_{\mathrm{at}}^{(\beta )})=\mathcal{B}(%
\mathbb{C}^{d})\otimes \mathbf{1}_{{\mathcal{V}}_{\mathcal{R}}}\subset {%
\mathcal{V}}\subset \mathfrak{H}^{(\beta )}\ .
\end{equation*}%
As $\mathrm{ran}(\mathfrak{\tilde{P}}_{\mathrm{at}}^{(\beta )})\subset {%
\mathcal{V}}\subset \mathfrak{H}^{(\beta )}$, the restriction $\mathfrak{%
\tilde{P}}_{\mathrm{at}}^{(\beta )}$ to ${\mathcal{V}}$ defines a projection
$\mathfrak{P}_{\mathrm{at}}^{(\beta )}\equiv \mathfrak{P}_{\mathrm{at}}$ on $%
{\mathcal{V}}$. Notice that in the sequel we identify $\mathrm{ran}(%
\mathfrak{P}_{\mathrm{at}}^{(\beta )})=\mathcal{B}(\mathbb{C}^{d})\otimes
\mathbf{1}_{{\mathcal{V}}_{\mathcal{R}}}\subset {\mathcal{V}}$ and $\mathcal{%
B}(\mathbb{C}^{d})$.

\subsection{Classical optical pump}

Define
\begin{equation*}
H_{\mathrm{p}}:=h_{\mathrm{p}}+h_{\mathrm{p}}^{\ast }\in \mathcal{B}(\mathbb{%
C}^{d})
\end{equation*}%
for some $h_{\mathrm{p}}\in \mathcal{B}(\mathbb{C}^{d})$ satisfying
\begin{eqnarray*}
\ker \left( h_{\mathrm{p}}\right) ^{\perp } &\subseteq &{\mathcal{H}}_{1}:=%
\mathrm{ran}\left( \mathbf{1}\left[ H_{\mathrm{at}}=E_{1}\right] \right) \ ,
\\
\mathrm{ran}\left( h_{\mathrm{p}}\right) &\subseteq &{\mathcal{H}}_{N}:=%
\mathrm{ran}\left( \mathbf{1}\left[ H_{\mathrm{at}}=E_{N}\right] \right) \ .
\end{eqnarray*}%
Here, $\mathbf{1}\left[ H_{\mathrm{at}}=E\right] $ stands for the orthogonal
projection on the eigenspace of $H_{\mathrm{at}}$ with corresponding
eigenvalue $E$. The coupling of the optical pump to the atom is represented
by a time--dependent perturbation of the form
\begin{equation*}
\eta \cos (\varpi t)\mathfrak{L}_{\mathrm{p}}\ ,\qquad \varpi
:=E_{N}-E_{1}>0\ ,\qquad t\in \mathbb{R}\ ,
\end{equation*}%
to the atomic Lindbladian $\mathfrak{L}_{\mathrm{at}}$. Here,
\begin{equation}
\mathfrak{L}_{\mathrm{p}}:=-i\left( \underrightarrow{H_{\mathrm{p}}}-\,%
\underleftarrow{H_{\mathrm{p}}}\right) =-i[H_{\mathrm{p}},\ \cdot \ ]=-%
\mathfrak{L}_{\mathrm{p}}^{\ast }  \label{L pump}
\end{equation}%
and $\eta \in \mathbb{R}$ is a coupling constant.

\subsection{Field form factors of the atom--reservoir interaction}

Let $\mathrm{K},m\in \mathbb{N}$ and, for any $\kappa =\{1,\ldots ,\mathrm{K}%
\}$, let $\{f_{\ell }^{(\kappa )}\}_{\ell =1}^{m}\subset \mathfrak{h}_{1}$
be a family of rotationally invariant functions, i.e., $f_{\ell }^{(\kappa
)}(p)=\mathrm{f}_{\ell }^{(\kappa )}(|p|)$ for all $p\in \mathbb{R}^{3}$ and
$\ell \in \{1,\ldots ,m\}$ with $\mathrm{f}_{\ell }^{(\kappa )}:\mathbb{R}%
_{0}^{+}\rightarrow \mathbb{C}$. For all $\kappa =\{1,\ldots ,\mathrm{K}\}$
and $\ell \in \{1,\ldots ,m\}$, the complex--valued functions $\mathrm{g}%
_{\ell }^{(\kappa )}$ and ${\mathrm{g}_{\ell }^{(\kappa )\#}}$ on $\mathbb{R}
$ are respectively defined by%
\begin{equation}
\forall x\in \mathbb{R}:\quad \mathrm{g}_{\ell }^{(\kappa )}(x):=|x|\left( 1+%
\mathrm{e}^{-\beta x}\right) ^{-1/2}\left\{
\begin{array}{lcr}
\mathrm{f}_{\ell }^{(\kappa )}(x) & , & x\geq 0\ , \\
&  &  \\
\overline{\mathrm{f}_{\ell }^{(\kappa )}(-x)} & , & x<0\ ,%
\end{array}%
\right.  \label{function gl}
\end{equation}%
and%
\begin{equation}
\mathrm{g}_{\ell }^{(\kappa )\#}(x):=i\overline{\mathrm{g}_{\ell }^{(\kappa
)}(-x)}\ ,\qquad x\in \mathbb{R}\ .  \label{g diese}
\end{equation}%
We assume the following:

\begin{assumption}
\label{d Hardy}\mbox{
}\newline
There is $\mathrm{r}_{\max }>0$ such that $\mathrm{g}_{\ell }^{(\kappa )}$
and ${\mathrm{g}_{\ell }^{(\kappa )\#}}$ have an analytic continuation to
the strip $\mathbb{R+}i(-\mathrm{r}_{\max },\mathrm{r}_{\max })$ and satisfy%
\begin{equation*}
\underset{y\in (-\mathrm{r}_{\max },\mathrm{r}_{\max })}{\sup }\left\{ \int_{%
\mathbb{R}}(|\mathrm{g}_{\ell }^{(\kappa )}(x+iy)|+|\mathrm{e}^{-\frac{\beta
}{2}(x+iy)}\mathrm{g}_{\ell }^{(\kappa )\#}(x+iy)|)^{2}\mathrm{d}x\right\}
<\infty \text{ .}
\end{equation*}
\end{assumption}

\noindent \lbrack The factor $\mathrm{e}^{-\frac{\beta }{2}(x+iy)}$ in this
assumption is, by mistake, missing in the condition below \cite[Eq. (2.16)]%
{BruPedraWestrich2011a}.]

To satisfy this condition one may, for instance, choose the functions $%
\mathrm{f}_{\ell }^{(\kappa )}(x)$ as linear combinations of terms of the
form $|x|^{2p-1}\exp (-Cx^{2})$ with $p\in \mathbb{N}_{0}$. Finally, at any
fixed inverse temperature $\beta \in \mathbb{R}^{+}$ of the fermionic
reservoir and for any $\kappa =\{1,\ldots ,\mathrm{K}\}$, let $\{f_{\ell
}^{(\kappa ,\beta )}\}_{\ell =1}^{m}$ be the family of functions $\mathbb{R}%
\rightarrow \mathbb{R}_{0}^{+}$ defined by%
\begin{equation}
f_{\ell }^{(\kappa ,\beta )}(x):=4\pi \frac{\left\vert x\text{ }\mathrm{f}%
_{\ell }^{(\kappa )}\left( \left\vert x\right\vert \right) \right\vert ^{2}}{%
1+\mathrm{e}^{-\beta x}}=4\pi \left\vert \mathrm{g}_{\ell }^{(\kappa
)}(x)\right\vert ^{2}\ .  \label{definition V1bis}
\end{equation}

\subsection{Atom--reservoir interaction}

For all $\kappa \in \{1,\ldots ,\mathrm{K}\}$ choose a finite collection $%
\{Q_{_{\ell _{1},\ldots ,\ell _{k}}}^{(\kappa )}\}_{\ell _{1},\ldots ,\ell
_{k}=1}^{m}\subset \mathcal{B}(\mathbb{C}^{d})$ satisfying%
\begin{equation*}
\left( Q_{_{\ell _{1},\ldots ,\ell _{\kappa }}}^{(\kappa )}\right) ^{\ast
}=Q_{_{\ell _{\kappa },\ldots ,\ell _{1}}}^{(\kappa )}\ .
\end{equation*}%
Then, the atom--reservoir interaction is implemented by the bounded
symmetric derivation $\lambda \delta _{\mathrm{at},\mathcal{R}}$ with
coupling constant $\lambda \in \mathbb{R}$ and%
\begin{equation}
\delta _{\mathrm{at},\mathcal{R}}:=i\sum_{\kappa =1}^{\mathrm{K}}\sum_{\ell
_{1},\ldots ,\ell _{\kappa }=1}^{m}\left[ Q_{_{\ell _{1},\ldots ,\ell
_{\kappa }}}^{(\kappa )}\otimes \Phi (f_{_{\ell _{1}}}^{(\kappa )})\cdots
\Phi (f_{\ell _{\kappa }}^{(\kappa )}),\;\cdot \;\right] \ .
\label{delta ar R}
\end{equation}%
Here, for all $f\in \mathfrak{h}_{1}$,
\begin{equation}
\Phi (f):=\frac{1}{\sqrt{2}}(a^{+}(f)+a(f))=\Phi (f)^{\ast }\in \mathcal{B}({%
\mathcal{F}}_{-}(\mathfrak{h}_{1}))\ .  \label{field operator}
\end{equation}

\noindent For any normed space $\mathcal{X}$, $\mathcal{B}(\mathcal{X})$
stands for the set of all linear bounded operators on $\mathcal{X}$.

To simplify discussions and proofs, as also done by other authors in similar
situations (see, for instance, \cite[Section 2.2, Assumption 2.4]%
{DerezinskiFruboes2006}), without loss of generality (w.l.o.g.), we assume:

\begin{assumption}
\label{assumption important 0}\quad $\mathfrak{P}_{\mathrm{at}}^{(\beta
)}\delta _{\mathrm{at},\mathcal{R}}\mathfrak{P}_{\mathrm{at}}^{(\beta )}=0.$
\end{assumption}

\noindent This technical condition is not essential for the results below
and does not exclude most physically relevant atom--reservoir couplings.
Note that this condition is automatically fulfilled if $\delta _{\mathrm{at},%
\mathcal{R}}$ is a linear combination of odd monomials in the fermionic
fields $\Phi (f)$, $f\in \mathfrak{h}_{1}$. This is the case for the model
considered in \cite{BruPedraWestrich2011a} and hence, this technical
condition is not explicitly imposed there. Moreover, Assumption \ref%
{assumption important 0} does not force the component of $\delta _{\mathrm{at%
},\mathcal{R}}$ which is even w.r.t. to the fermionic fields to vanish.

For all $\varepsilon \in \mathbb{R}^{+}$, define the linear operator $%
\mathcal{L}_{\mathcal{R}}^{(\varepsilon )}\in \mathcal{B}(\mathfrak{H}_{%
\mathrm{at}})$ by%
\begin{equation}
\mathfrak{L}_{\mathcal{R}}^{(\varepsilon )}:=\sum_{\epsilon \in \sigma (i%
\mathfrak{L}_{\mathrm{at}})}\left( P_{\mathrm{at}}^{(\epsilon )}\ \mathfrak{P%
}_{\mathrm{at}}^{(\beta )}\delta _{\mathrm{at},\mathcal{R}}(1-\mathfrak{P}_{%
\mathrm{at}}^{(\beta )})(\varepsilon +i\epsilon -\delta )^{-1}(1-\mathfrak{P}%
_{\mathrm{at}}^{(\beta )})\delta _{\mathrm{at},\mathcal{R}}\mathfrak{P}_{%
\mathrm{at}}^{(\beta )}P_{\mathrm{at}}^{(\epsilon )}\right) ^{\ast }\ .
\label{L(r)}
\end{equation}%
Recall that we identify the spaces $\mathrm{ran}(\mathfrak{P}_{\mathrm{at}%
}^{(\beta )})=\mathcal{B}(\mathbb{C}^{d})\otimes \mathbf{1}_{{\mathcal{V}}_{%
\mathcal{R}}}\subset {\mathcal{V}}$ and $\mathcal{B}(\mathbb{C}^{d})$. Note
further that $\varepsilon ,\epsilon \in \mathbb{R}$ and $\sigma (\delta
)\subset i\mathbb{R}$. [$\delta $ generates a group of contractions, i.e., $%
\pm \delta $ are generators of semigroups of contractions.] From the
analyticity assumptions on the field form factors of the atom--reservoir
interactions, one shows that the following limit exists:%
\begin{equation}
\mathfrak{L}_{\mathcal{R}}:=\lim_{\varepsilon \searrow 0}\mathfrak{L}_{%
\mathcal{R}}^{(\varepsilon )}\in \mathcal{B}(\mathfrak{H}_{\mathrm{at}%
})\equiv \mathcal{B}(\mathcal{B}(\mathbb{C}^{d}))\ .  \label{Lindblad gol}
\end{equation}%
See for instance Lemma \ref{pert G copy(1)}. It is known that, in this case,
the limit $\mathcal{L}_{\mathcal{R}}$ is the generator of a completely
positive group, see for instance \cite[Section 6.1]{DerezinskiFruboes2006}.
By the Stinespring theorem,
\begin{equation}
\mathfrak{L}_{\mathcal{R}}=-i\left( \underrightarrow{H_{\mathrm{Lamb}}}-\,%
\underleftarrow{H_{\mathrm{Lamb}}}\right) +\mathfrak{L}_{d}\ ,  \label{L R}
\end{equation}%
where $H_{\mathrm{Lamb}}=H_{\mathrm{Lamb}}^{\ast }\in \mathcal{B}(\mathbb{C}%
^{d})$ and $\mathfrak{L}_{d}$ has the form%
\begin{equation*}
\mathfrak{L}_{d}(\rho )=\sum_{\alpha \in \mathrm{A}}(2V_{\alpha }\rho
V_{\alpha }^{\ast }-V_{\alpha }^{\ast }V_{\alpha }\rho -\rho V_{\alpha
}^{\ast }V_{\alpha })\ ,\text{\qquad }\rho \in \mathfrak{H}_{\mathrm{at}}%
\text{ },
\end{equation*}%
for some fixed $V_{\alpha }\in \mathcal{B}(\mathbb{C}^{d})$, $\alpha \in
\mathrm{A}$, $|\mathrm{A}|<\infty $. Here, $H_{\mathrm{Lamb}}$ is the
so--called \emph{atomic Lamb shift} and $\mathfrak{L}_{d}\in \mathcal{B}(%
\mathfrak{H}_{\mathrm{at}})$ is the \emph{effective atomic dissipation}. The
atom--reservoir interaction yields an effective atomic dynamics generated by
$\mathfrak{L}_{\mathrm{at}}+\lambda ^{2}\mathfrak{L}_{\mathcal{R}}$ (without
pump). Notice furthermore that $H_{\mathrm{Lamb}}$ commutes with the atomic
Hamiltonian $H_{\mathrm{at}}$.

Explicit expressions for $H_{\mathrm{Lamb}}$ and $\mathfrak{L}_{d}$ in terms
of the microscopic couplings $Q_{_{\ell _{1},\ldots ,\ell _{\kappa
}}}^{(\kappa )}$ and $f_{\ell }^{(\kappa )}$ are quite cumbersome in the
general case but can straightforwardly be obtained. In the simplest
non--trivial case, i.e., if $Q_{_{\ell _{1},\ldots ,\ell _{\kappa
}}}^{(\kappa )}\equiv 0$ for $\kappa \geq 2$ and the family of functions
(form factors) $\left\{ f_{\ell }\right\} _{\ell =1}^{m}\equiv \{f_{_{\ell
}}^{(1)}\}_{\ell =1}^{m}\subset \mathfrak{h}_{1}$ is orthogonal, we obtain
the following: Let $\{\tilde{V}_{j,k}^{(\ell )}\}_{j,k,\ell }\subset
\mathcal{B}(\mathbb{C}^{d})$ be the family of operators defined by
\begin{equation}
\tilde{V}_{j,k}^{(\ell )}:=\mathbf{1}\left[ H_{\mathrm{at}}=E_{j}\right] \
Q_{\ell }\ \mathbf{1}\left[ H_{\mathrm{at}}=E_{k}\right]
\label{definition V1}
\end{equation}%
for $j,k\in \{1,2,\ldots ,N\}$ and $\ell \in \{1,2,\ldots ,m\}$. Then, the
atomic Lamb shift $H_{\mathrm{Lamb}}\in \mathcal{B}(\mathbb{C}^{d})$ is the
self--adjoint operator defined by
\begin{equation}
H_{\mathrm{Lamb}}=-\frac{1}{2}\sum_{\epsilon \in \sigma (i\mathfrak{L}_{%
\mathrm{at}})\backslash \{0\}}\ \sum_{(j,k)\in \mathfrak{t}_{\epsilon
}}\sum_{\ell =1}^{m}d_{j,k}^{(\ell )}\tilde{V}_{j,k}^{(\ell )\ast }\tilde{V}%
_{j,k}^{(\ell )}\ .  \label{Atomic Lamb shift}
\end{equation}%
The real coefficient
\begin{equation*}
d_{j,k}^{(\ell )}:=\mathcal{PP}\left[ f_{\ell }^{(1,\beta )}\left( \cdot
+(E_{k}-E_{j})\right) \right]
\end{equation*}%
is the principal part $\mathcal{PP}[f]$ of the function $f\equiv f_{\ell
}^{(1,\beta )}\left( \cdot +(E_{k}-E_{j})\right) $. See (\ref{definition
V1bis}) for the definition of $f_{\ell }^{(1,\beta )}$. Meanwhile,
\begin{equation}
\mathfrak{L}_{d}:=\frac{1}{2}\sum_{\epsilon \in \sigma (i\mathfrak{L}_{%
\mathrm{at}})}\ \sum_{(j,k)\in \mathfrak{t}_{\epsilon }}\sum_{\ell
=1}^{m}c_{j,k}^{(\ell )}\mathfrak{L}_{j,k}^{\left( \ell \right) }\ ,
\label{Effective atomic dissipation}
\end{equation}%
where $c_{j,k}^{(\ell )}:=\pi f_{\ell }^{(1,\beta )}(E_{k}-E_{j})\geq 0$ and%
\begin{equation}
\mathfrak{L}_{j,k}^{\left( \ell \right) }\left( \rho \right) :=2\tilde{V}%
_{j,k}^{(\ell )}\rho \tilde{V}_{j,k}^{(\ell )\ast }-\tilde{V}_{j,k}^{(\ell
)\ast }\tilde{V}_{j,k}^{(\ell )}\rho -\rho \tilde{V}_{j,k}^{(\ell )\ast }%
\tilde{V}_{j,k}^{(\ell )}\ ,\qquad \rho \in \mathfrak{H}_{\mathrm{at}}\ .
\label{eq:Effective atomic dissipation 2}
\end{equation}%
Note that these expressions for $H_{\mathrm{Lamb}}$ and $\mathfrak{L}_{d}$
appear in the heuristic derivation of the time--dependent Lindbladian given
in \cite[Section 6.1]{BruPedraWestrich2011a}. See also \cite[1.3. Remarks.]%
{Pillet-Jacblabla}.

To control Rabi oscillations (caused by the optical pump, whose strength is
of order $\mathcal{O}(\eta )$) via the effective atom--reservoir interaction
$\lambda ^{2}\mathfrak{L}_{\mathcal{R}}$ we assume:

\begin{assumption}[Moderate optical pump]
\label{assumption important}\mbox{
}\newline
For a fixed (but arbitrary) constant $C\in \mathbb{R}^{+}$ the couplings $%
\eta ,\lambda \in \mathbb{R}$ are chosen such that $|\eta |\leq C\lambda
^{2} $.
\end{assumption}

\noindent Moreover, we impose $0$ to be a non--degenerated eigenvalue of $%
\frac{\eta }{2}\mathfrak{L}_{\mathrm{p}}+\lambda ^{2}\mathfrak{L}_{\mathcal{R%
}}$ with some non--trivial \emph{real} spectral gap, that is, more
precisely,
\begin{equation*}
\max \left\{ \mathrm{\mathop{\rm Re}}\left\{ w\right\} \,|\,w\in \sigma
\left( \frac{\eta }{2}\mathfrak{L}_{\mathrm{p}}+\lambda ^{2}\mathfrak{L}_{%
\mathcal{R}}\right) \backslash \{0\}\right\} \leq -\lambda ^{2}C<0
\end{equation*}%
with $C\in \mathbb{R}^{+}$ being some fixed constant not depending on $%
\lambda $ and $\eta $. This allows the study of the dynamics of the atom at
large times. By the results of \cite{Spohn1977}, the following assumption on
the dissipative part $\mathfrak{L}_{d}$ of $\mathfrak{L}_{\mathcal{R}}$
suffices to ensure the above spectral property:

\begin{assumption}[Irreducibility of quantum Markov chains]
\label{assumption3}\mbox{
}\newline
The family $\{V_{\alpha }\}_{\alpha \in \mathrm{A}}\subset \mathcal{B}(%
\mathbb{C}^{d})$ satisfies
\begin{equation*}
\Big(\bigcup\limits_{\alpha \in \mathrm{A}}\left\{ V_{\alpha }\right\} \Big)%
^{\prime \prime }=\mathcal{B}(\mathbb{C}^{d})
\end{equation*}%
with $M^{\prime \prime }$ being the bicommutant of $M\subset \mathcal{B}(%
\mathbb{C}^{d})$.
\end{assumption}

\noindent See the proof of \cite[Lemma 6.3]{BruPedraWestrich2011a} for the
detailed arguments. This last assumption highlights the role played by
dissipative effects of the reservoir on the atom in order to get an
appropriate asymptotic evolution of populations of atomic levels. For
further discussions, see \cite[Section 3.2]{BruPedraWestrich2011a}.

From now on, we assume Assumptions \ref{d Hardy}--\ref{assumption3} to be
satisfied.

\section{Microscopic Dynamics\label{Section Atomic Microscopic Dynamics}}

Let
\begin{equation}
\delta _{t}^{(\lambda ,\eta )}:=\delta +\eta \cos (\varpi t)\delta _{\mathrm{%
at},\mathrm{p}}+\lambda \delta _{\mathrm{at},\mathcal{R}}\ ,\qquad t\in
\mathbb{R}\ ,  \label{derivation full nija}
\end{equation}%
with $\lambda ,\eta \in \mathbb{R}$ and%
\begin{equation*}
\delta _{\mathrm{at},\mathrm{p}}:=i[H_{\mathrm{p}}\otimes \mathbf{1}_{{%
\mathcal{V}}_{\mathcal{R}}},\;\cdot \;]\ .
\end{equation*}%
Recall that the generator of the group $\tau :=\{\tau _{t}\}_{t\in \mathbb{R}%
}$ (\ref{free dynamics}) is the symmetric derivation $\delta $, while $%
\delta _{\mathrm{at},\mathcal{R}}$ is defined by (\ref{delta ar R}).
Therefore, the time--dependent symmetric derivation $\delta _{t}^{(\lambda
,\eta )}$ corresponds to a bounded and smooth (w.r.t. $t\in \mathbb{R}$)
perturbation of $\delta $. For all $\lambda ,\eta \in \mathbb{R}$, the
(non--autonomous full) microscopic dynamics is defined by the unique
strongly continuous two--parameter family $\{\tau _{t,s}^{(\lambda ,\eta
)}\}_{t\geq s}$ of $\ast $--automorphisms of $\mathcal{V}$ satisfying the\
evolution equation%
\begin{equation}
\forall s,t\in {\mathbb{R}},\ t\geq s:\quad \partial _{t}\tau
_{t,s}^{(\lambda ,\eta )}=\tau _{t,s}^{(\lambda ,\eta )}\circ \delta
_{t}^{(\lambda ,\eta )}\ ,\qquad \tau _{s,s}^{(\lambda ,\eta )}:=\mathbf{1}_{%
\mathcal{V}}\ ,  \label{Cauchy problem 1}
\end{equation}%
on the (time--constant) domain of $\delta _{t}^{(\lambda ,\eta )}$. The
existence, uniqueness and strong continuity of the two--parameter family $%
\{\tau _{t,s}^{(\lambda ,\eta )}\}_{t\geq s}$ is shown by standard
arguments, see Proposition \ref{time.dep.trotter} and Definition \ref%
{Time--dependent C--Liouvillean copy(1)}.

The time--evolving state of the compound system is then given by
\begin{equation*}
\omega _{t}:=\omega _{0}\circ \tau _{t,0}^{(\lambda ,\eta )}=\left( \omega _{%
\mathrm{at}}\otimes \omega _{{\mathcal{R}}}\right) \circ \tau
_{t,0}^{(\lambda ,\eta )}\ ,\qquad t\in \mathbb{R}_{0}^{+}\ .
\end{equation*}%
The restriction of this state to the atomic degrees of freedom yields a
time--dependent atomic state defined by%
\begin{equation}
\omega _{\mathrm{at}}\left( t\right) \left( A\right) :=\omega _{t}(A\otimes
\mathbf{1}_{{\mathcal{V}}_{\mathcal{R}}}),\qquad A\in \mathcal{B}(\mathbb{C}%
^{d})\ ,  \label{ninja1}
\end{equation}%
for all $t\in \mathbb{R}_{0}^{+}$. The corresponding family of density
matrices of $\{\omega _{\mathrm{at}}\left( t\right) \}_{t\in \mathbb{R}%
_{0}^{+}}$ is denoted by $\{\rho _{\mathrm{at}}\left( t\right) \}_{t\in
\mathbb{R}_{0}^{+}}$.

The aim of this paper is to prove \cite[Theorem 3.3]{BruPedraWestrich2011a}.
This amounts to study the orthogonal projection $P_{\mathfrak{D}}\left( \rho
_{\mathrm{at}}\left( t\right) \right) $ of the atomic density matrix $\rho _{%
\mathrm{at}}\left( t\right) $ on the subspace
\begin{equation}
\mathfrak{D}\equiv \mathfrak{D}(H_{\mathrm{at}}):=\mathcal{B}({\mathcal{H}}_{%
\mathrm{1}})\oplus \cdots \oplus \mathcal{B}({\mathcal{H}}_{\mathrm{N}%
})\subset \mathfrak{H}_{\mathrm{at}}  \label{ninja0}
\end{equation}%
of block--diagonal matrices. In other words, we analyze the density matrix
\begin{equation}
P_{\mathfrak{D}}\left( \rho _{\mathrm{at}}\left( t\right) \right) =\overset{N%
}{\underset{k=1}{\sum }}\ \mathbf{1}\left[ H_{\mathrm{at}}=E_{k}\right] \
\rho _{\mathrm{at}}\left( t\right) \ \mathbf{1}\left[ H_{\mathrm{at}}=E_{k}%
\right]  \label{ninja0bis}
\end{equation}%
for any $t\in \mathbb{R}_{0}^{+}$. In fact, we compare the above
time--evolving density matrix with the unique solution of the following
(effective) atomic master equation on $\mathfrak{H}_{\mathrm{at}}$:
\begin{equation}
\forall t\in \mathbb{R}_{0}^{+}:\qquad \dfrac{\mathrm{d}}{\mathrm{d}t}\rho
(t)=\mathfrak{L}_{t}^{\left( \lambda ,\eta \right) }(\rho (t))\ ,\qquad \rho
(0)=\rho _{\mathrm{at}}\left( 0\right) \equiv \rho _{\mathrm{at}}\in
\mathfrak{H}_{\mathrm{at}}\ .  \label{effective atomic master equation}
\end{equation}%
Here, for any $\lambda ,\eta ,t\in \mathbb{R}$, $\mathfrak{L}_{t}^{\left(
\lambda ,\eta \right) }$ is a time--dependent Lindbladian defined on $%
\mathfrak{H}_{\mathrm{at}}$ by
\begin{equation}
\mathfrak{L}_{t}^{\left( \lambda ,\eta \right) }:=\mathfrak{L}_{\mathrm{at}%
}+\eta \cos (\varpi t)\mathfrak{L}_{\mathrm{p}}+\lambda ^{2}\mathfrak{L}_{%
\mathcal{R}}\ ,  \label{lindblad ninja1}
\end{equation}%
see (\ref{Lat}), (\ref{L pump}) and (\ref{Lindblad gol})--(\ref{L R}).

Since the Lindbladian $\mathfrak{L}_{t}^{\left( \lambda ,\eta \right) }$ is
continuous in time and generates a Markov completely positive (CP) semigroup
at any fixed $t\in \mathbb{R}$, there is a continuous two--parameter family
denoted by $\{\hat{\tau}_{t,s}^{(\lambda ,\eta )}\}_{t\geq s}\subset
\mathcal{B}(\mathfrak{H}_{\mathrm{at}})$ of CP trace--preserving maps
satisfying the evolution equation:%
\begin{equation*}
\forall s,t\in {\mathbb{R}},\ t\geq s:\quad \partial _{t}\hat{\tau}%
_{t,s}^{(\lambda ,\eta )}=\mathfrak{L}_{t}^{\left( \lambda ,\eta \right)
}\circ \hat{\tau}_{t,s}^{(\lambda ,\eta )}\ ,\qquad \hat{\tau}%
_{s,s}^{(\lambda ,\eta )}:=\mathbf{1}_{\mathfrak{H}_{\mathrm{at}}}\ .
\end{equation*}%
For any initial condition $\rho _{\mathrm{at}}\in \mathfrak{H}_{\mathrm{at}}$
on has:%
\begin{equation*}
\rho (t)=\hat{\tau}_{t,0}^{(\lambda ,\eta )}(\rho _{\mathrm{at}})\ ,\qquad
t\in \mathbb{R}_{0}^{+}\ .
\end{equation*}

The evolution of the true density matrix $\rho _{\mathrm{at}}\left( t\right)
$ of the atom at time $t$ depends on the (infinite) degrees of freedom of
the reservoir. By contrast, the time evolution of $\rho (t)$ only involves
atomic degrees of freedom and hence the effective density matrix $\rho (t)$
evolves in a space of finite dimension. The main interest of the initial
value problem (\ref{effective atomic master equation}) is that -- at small
couplings -- its solution $\rho (t)$ accurately approximates, for all $t\in
\mathbb{R}_{0}^{+}$, the density matrix $\rho _{\mathrm{at}}\left( t\right) $
of the time--dependent state $\omega _{\mathrm{at}}\left( t\right) $ on the
subspace $\mathfrak{D}$ ((\ref{ninja0}), space of populations of the atomic
energy levels):

\begin{satz}[Validity of the effective atomic master equation]
\label{ninja thm cool}\mbox{ }\newline
Assume that $\rho _{\mathrm{at}}\in \mathfrak{D}$ and let $\varepsilon \in
(0,1)$. The unique solution $\{\rho (t)\}_{t\geq 0}$ of the effective atomic
master equation (\ref{effective atomic master equation}) and the atomic
density matrix $\{\rho _{\mathrm{at}}\left( t\right) \}_{t\geq 0}$ satisfy
the bound
\begin{equation*}
\left\Vert P_{\mathfrak{D}}\left( \rho _{\mathrm{at}}(t)-\rho (t)\right)
\right\Vert _{\mathfrak{H}_{\mathrm{at}}}\leq C_{\varpi ,\varepsilon
}\left\vert \lambda \right\vert ^{1-\varepsilon }
\end{equation*}%
for some constant $C_{\varpi ,\varepsilon }\in \mathbb{R}^{+}$ depending on $%
\varpi ,\varepsilon $, but not on the initial state $\omega _{\mathrm{at}}$
of the atom and the parameters $t$, $\lambda $, and $\eta $, provided $%
|\lambda |$ is sufficiently small.
\end{satz}

Note that \cite[Theorem 3.3]{BruPedraWestrich2011a} asserts the above bound
with $\varepsilon =0$ but for the special case $\mathrm{K}=1$. We prove here
the slightly weaker bound with $\varepsilon \in (0,1)$, for technical
simplicity. [To get the bound for $\varepsilon =0$ one may improve the
estimate (\ref{estimate to improve}).]

\section{Technical Proofs\label{Section technical proofs}}

The remaining part of the paper\ is devoted to the proof of Theorem \ref%
{ninja thm cool}, which is concluded in Section \ref{thm final proof}. We
essentially follow \cite{Pillet-Jacblabla}, where the notion of $C$%
--Liouvilleans has been first introduced. Like in \cite{BMPS, S, FS}, the
setting of \cite{Pillet-Jacblabla} is extended here to allow time--dependent
$C$--Liouvilleans.

\subsection{Existence of the non--autonomous dynamics}

Recall that the generator of the group $\tau :=\{\tau _{t}\}_{t\in \mathbb{R}%
}$ (\ref{free dynamics}) is the symmetric derivation $\delta $. Then, any
time--dependent and self--adjoint family $\{W_{t}\}_{t\in \mathbb{R}}\subset
\mathcal{V}$ defines a family of symmetric derivations
\begin{equation}
\delta _{W_{t}}:=\delta +i[W_{t},\ \cdot \ ]\ ,\qquad t\in {\mathbb{R}}\ .
\label{derivatiion time dependent}
\end{equation}%
If $\{W_{t}\}_{t\in \mathbb{R}}\in C^{1}(\mathbb{R},\mathcal{V})$, then such
time--dependent derivations generate a unique fundamental solution $\{%
\mathfrak{W}_{s,t}\}_{s,t\in \mathbb{R}}$, which is in our case a family of $%
\ast $--automorphisms of $\mathcal{V}$.\ By fundamental solution, we mean
here that the family $\{\mathfrak{W}_{s,t}\}_{s,t\in \mathbb{R}}$ of bounded
operators acting on $\mathcal{V}$ is strongly continuous, preserves the
domain
\begin{equation*}
\mathrm{Dom}(\delta _{W_{t}})=\mathrm{Dom}(\delta )\ ,
\end{equation*}%
satisfies%
\begin{eqnarray*}
\mathfrak{W}_{\cdot ,t}(A) &\in &C^{1}(\mathbb{R};(\mathrm{Dom}(\delta
),\left\Vert \cdot \right\Vert _{\mathcal{V}}))\ , \\
\mathfrak{W}_{s,\cdot }(A) &\in &C^{1}(\mathbb{R};(\mathrm{Dom}(\delta
),\left\Vert \cdot \right\Vert _{\mathcal{V}}))\ ,
\end{eqnarray*}%
for all $A\in \mathrm{Dom}(\delta )$, and solves the corresponding Cauchy
initial value\ problem%
\begin{equation}
\forall s,t\in {\mathbb{R}}:\quad \partial _{s}\mathfrak{W}_{s,t}=-\delta
_{W_{s}}\circ \mathfrak{W}_{s,t}\ ,\quad \mathfrak{W}_{t,t}=\mathbf{1}_{%
\mathcal{V}}\mathbf{\ },  \label{sol fund}
\end{equation}%
in the strong sense on $(\mathrm{Dom}(\delta ),\left\Vert \cdot \right\Vert
_{\mathcal{V}})$:

\begin{proposition}[Non--autonomous $C^{\ast }$--dynamics--I]
\label{time.dep.trotter}\mbox{ }\newline
If $\{W_{t}\}_{t\in \mathbb{R}}\in C^{1}(\mathbb{R},\mathcal{V})$, then
there is a unique evolution family $\{\mathfrak{W}_{s,t}\}_{s,t\in \mathbb{R}%
}$ of $\ast $--automorphisms$\ $with the following properties: \newline
\emph{(i)}\ It satisfies the property%
\begin{equation*}
\forall t,r,s\in \mathbb{R}:\quad \mathfrak{W}_{s,t}=\mathfrak{W}_{s,r}%
\mathfrak{W}_{r,t}\ .
\end{equation*}%
\emph{(ii)}\ It is the fundamental solution of (\ref{sol fund}).\newline
\emph{(iii)}\ It solves in the strong sense on $(\mathrm{Dom}(\delta
),\left\Vert \cdot \right\Vert _{\mathcal{V}})$ the abstract Cauchy initial
value\ problem
\begin{equation*}
\forall s,t\in {\mathbb{R}}:\quad \partial _{t}\mathfrak{W}_{s,t}=\mathfrak{W%
}_{s,t}\circ \delta _{W_{t}}\ ,\quad \mathfrak{W}_{s,s}=\mathbf{1}_{\mathcal{%
V}}\ .
\end{equation*}%
\emph{(iv)}\ If $\{W_{t}\}_{t\in \mathbb{R}}$ is periodic with period $T>0$,
then%
\begin{equation*}
\forall s,t\in {\mathbb{R}},\ p\in \mathbb{Z}:\quad \mathfrak{W}_{s,t}=%
\mathfrak{W}_{s+pT,t+pT}\ .
\end{equation*}
\end{proposition}

\noindent \textit{Proof.} The proof of Assertions (i)--(iii) can be done by
standard arguments as $\{W_{t}\}_{t\in \mathbb{R}}\subset \mathcal{V}$. We
recommend the proof of \cite[Proposition 5.4]{OhmI} where we additionally
show \cite[Eq. (5.24)]{OhmI}, that is,
\begin{equation}
\mathfrak{W}_{s,t}\left( A\right) :=\tau _{-s}\left( \mathfrak{V}_{t,s}\tau
_{t}(A)\mathfrak{V}_{t,s}^{\ast }\right)  \label{w fract}
\end{equation}%
for any $A\in \mathcal{V}$ and $s,t\in \mathbb{R}$. Here, $\mathfrak{V}%
_{t,s}\in \mathcal{V}$ is given by the absolutely convergent series%
\begin{equation}
\mathfrak{V}_{t,s}:=\mathbf{1}_{\mathcal{V}}\mathbf{+}\sum\limits_{k\in {%
\mathbb{N}}}i^{k}\int_{s}^{t}\mathrm{d}s_{1}\cdots \int_{s}^{s_{k-1}}\mathrm{%
d}s_{k}\tau _{s_{k}}\left( W_{s_{k}}\right) \cdots \tau _{s_{1}}\left(
W_{s_{1}}\right)  \label{w fractbis}
\end{equation}%
in the Banach space $(\mathrm{Dom}(\delta ),\left\Vert \cdot \right\Vert _{%
\mathrm{Dom}(\delta )})$, where $\left\Vert \cdot \right\Vert _{\mathrm{Dom}%
(\delta )}$ stands for the graph norm of the closed operator $\delta $. In
particular, if $\{W_{t}\}_{t\in \mathbb{R}}$ is also periodic with period $%
T>0$, $\tau _{-pT}\left( \mathfrak{V}_{t+pT,s+pT}\right) =\mathfrak{V}_{t,s}$
and we obtain Assertion (iv) by using Equation (\ref{w fract}). \hfill $\Box
$

Then, the microscopic dynamics (\ref{Cauchy problem 1}) corresponds to the
following definition:

\begin{definition}[Non--autonomous $C^{\ast }$--dynamics--II]
\label{Time--dependent C--Liouvillean copy(1)}\mbox{ }\newline
The non--autonomous dynamics $\{\tau _{t,s}\}_{t\geq s}$ is defined by $\tau
_{t,s}:=\mathfrak{W}_{s,t}$ for $t\geq s$,\ provided $\{W_{t}\}_{t\in
\mathbb{R}}\in C^{1}(\mathbb{R},\mathcal{V})$.
\end{definition}

\noindent Indeed, by construction, this family is the (unique) solution of
the abstract Cauchy initial value problem%
\begin{equation*}
\forall s,t\in {\mathbb{R}}:\quad \partial _{t}\tau _{t,s}=\tau _{t,s}\circ
\delta _{W_{t}}\ ,\quad \tau _{s,s}=\mathbf{1}_{\mathcal{V}}\ ,
\end{equation*}%
in the strong sense on $(\mathrm{Dom}(\delta ),\left\Vert \cdot \right\Vert
_{\mathcal{V}})$.

\subsection{Time--dependent $C$--Liouvilleans\protect\footnote{%
This section has been partially done in collaboration with M. Westrich
during his PhD \cite[Chapter 5]{Westrich2011}.}\label{Time--dependent
C--Liouvilleans}}

Assume first that the initial state $\omega _{0}$ (\ref{initial state}) is
of the form
\begin{equation*}
\omega _{0}=\mathfrak{g}_{\mathrm{at}}\otimes \omega _{{\mathcal{R}}}\ ,
\end{equation*}%
where $\mathfrak{g}_{\mathrm{at}}$ is the (Gibbs) state corresponding to the
density matrix\ $\rho _{\mathfrak{g}}$ (\ref{Gibbs.init}). Denote by $(%
\mathfrak{H}_{\mathrm{at}},\pi _{\mathrm{at}},\Omega _{\mathrm{at,}\mathfrak{%
g}})$ and $(\mathfrak{H}_{\mathcal{R}},\pi _{\mathcal{R}},\Omega _{\mathcal{R%
}})$ GNS\ representations of respectively $\mathfrak{g}_{\mathrm{at}}$\ and $%
\omega _{{\mathcal{R}}}$. Let $\mathfrak{H}:=\mathfrak{H}_{\mathrm{at}%
}\otimes \mathfrak{H}_{\mathcal{R}}$, $\pi :=\pi _{\mathrm{at}}\otimes \pi _{%
\mathcal{R}}$ and $\Omega _{\mathfrak{g}}:=\Omega _{\mathrm{at,}\mathfrak{g}%
}\otimes \Omega _{\mathcal{R}}$. Observe $\Omega _{\mathfrak{g}}$ is cyclic
because it is the tensor product of cyclic vectors and hence, $(\mathfrak{H}%
,\pi ,\Omega _{\mathfrak{g}})$ is a GNS representation of $\omega _{0}$. An
important property of the initial state $\omega _{0}$ is that it is
faithful. In particular, $\pi $ is injective. So, to simplify notation, $\pi
\left( A\right) $ and $\pi \left( \mathcal{V}\right) $\ are identified with $%
A$ and $\mathcal{V}$, respectively. The weak closure of the $C^{\ast }$%
--algebra $\mathcal{V}\subset \mathcal{B}(\mathfrak{H})$ is the von Neumann
algebra $\mathcal{V}^{\prime \prime }$ denoted by $\mathfrak{M}$. The state $%
\omega _{0}$ is a $(\tau ,\beta )$--KMS state, where $\{\tau _{t}\}_{t\in
\mathbb{R}}$ is the one--parameter group of $\ast $--automorphisms on ${%
\mathcal{V}}$ defined by (\ref{free dynamics}). By \cite[Corollary 5.3.9]%
{BratteliRobinson}, the cyclic vector $\Omega _{\mathfrak{g}}$\ is
separating for $\mathfrak{M}$, i.e., $A\Omega _{\mathfrak{g}}=0$ implies $%
A=0 $ for all $A\in \mathfrak{M}$. The state $\omega _{0}$ on ${\mathcal{V}}$
extends uniquely to a normal state on the von Neumann algebra $\mathfrak{M}$
and $\{\tau _{t}\}_{t\in \mathbb{R}}$ uniquely extends to a $\sigma $%
--weakly continuous $\ast $--automorphism group on $\mathfrak{M}$, see \cite[%
Corollary 5.3.4]{BratteliRobinson}. Both extensions are again denoted by $%
\omega _{0}$ and $\{\tau _{t}\}_{t\in \mathbb{R}}$, respectively. Because $%
\omega _{0}$ is invariant w.r.t. $\{\tau _{t}\}_{t\in \mathbb{R}}$, there is
a unique unitary representation $\{U_{t}\}_{t\in \mathbb{R}}$ of $\{\tau
_{t}\}_{t\in \mathbb{R}}$ by conjugation, i.e.,
\begin{equation*}
\forall t\in \mathbb{R},\ A\in \mathfrak{M}:\qquad \tau _{t}\left( A\right)
=U_{t}AU_{t}^{\ast }\ ,
\end{equation*}%
such that $U_{t}\Omega _{\mathfrak{g}}=\Omega _{\mathfrak{g}}$. As $t\mapsto
\tau _{t}$ is $\sigma $--weakly continuous, the map $t\mapsto U_{t}$ is
strongly continuous. Therefore, there is a a self--adjoint operator $L_{%
\mathfrak{g}}$ with $U_{t}=\mathrm{e}^{itL_{\mathfrak{g}}}$ for all $t\in
\mathbb{R}$. In particular, $\Omega _{\mathfrak{g}}\in \mathrm{Dom}(L_{%
\mathfrak{g}})$ and $L_{\mathfrak{g}}$ annihilates $\Omega _{\mathfrak{g}}$,
i.e.,
\begin{equation}
L_{\mathfrak{g}}\Omega _{\mathfrak{g}}=0\ .  \label{important eq c louvilean}
\end{equation}%
Moreover, $L_{\mathfrak{g}}$ is related to the generator $\delta $ of the
group $\{\tau _{t}\}_{t\in \mathbb{R}}$ by the following relations:
\begin{equation}
\left\{ A\Omega _{\mathfrak{g}}:A\in \mathrm{Dom}\left( \delta \right)
\right\} \subset \mathrm{Dom}\left( L_{\mathfrak{g}}\right) \subset
\mathfrak{H}  \label{domain delata1}
\end{equation}%
and
\begin{equation}
\forall A\in \mathrm{Dom}\left( \delta \right) :\qquad L_{\mathfrak{g}%
}\left( A\Omega _{\mathfrak{g}}\right) =\delta \left( A\right) \Omega _{%
\mathfrak{g}}\ .  \label{domain delata2}
\end{equation}

The (Tomita--Takesaki) modular objects of the pair $\left( \mathfrak{M}%
,\Omega _{\mathfrak{g}}\right) $ are important for our further analysis. We
write $\Delta _{\mathfrak{g}}=\mathrm{e}^{-\beta L_{\mathfrak{g}}}$, $J_{%
\mathfrak{g}}$, and
\begin{equation*}
\mathcal{P}_{\mathfrak{g}}:=\overline{\left\{ AJ_{\mathfrak{g}}AJ_{\mathfrak{%
g}}\Omega _{\mathfrak{g}}\;:\;A\in \mathfrak{M}\right\} }
\end{equation*}%
respectively for the modular operator, the modular conjugation and the
natural positive cone of the pair $\left( \mathfrak{M},\Omega _{\mathfrak{g}%
}\right) $. For a detailed exposition on the use of the Tomita--Takesaki
modular theory in quantum statistical mechanics, see, for instance, the
textbook \cite{BratteliRobinsonI}. The family $\{U_{t}\}_{t\in \mathbb{R}}$
of unitaries representing $\{\tau _{t}\}_{t\in \mathbb{R}}$ by conjugation
also satisfies $U_{t}\mathcal{P}_{\mathfrak{g}}\subset \mathcal{P}_{%
\mathfrak{g}}$ for any $t\in \mathbb{R}$. The self--adjoint operator $L_{%
\mathfrak{g}}$, generator of the strongly continuous group $\{U_{t}\}_{t\in
\mathbb{R}}$ on $\mathfrak{H}$, is named the \emph{standard Liouvillean} of
the $\ast $--automorphism group $\{\tau _{t}\}_{t\in \mathbb{R}}$.

Now, if the faithful state $\omega _{\mathrm{at}}$ is not the Gibbs state $%
\mathfrak{g}_{\mathrm{at}}$ in (\ref{initial state}) then, because $\Omega _{%
\mathrm{at}}$ is cyclic for $\pi _{\mathrm{at}}\left( \mathcal{B}(\mathbb{C}%
^{d})\right) $, a\ GNS\ representation of%
\begin{equation*}
\omega _{0}:=\omega _{\mathrm{at}}\otimes \omega _{{\mathcal{R}}}
\end{equation*}%
is also given by $(\mathfrak{H},\pi ,\Omega )$ where $\Omega =\Omega _{%
\mathrm{at}}\otimes \Omega _{\mathcal{R}}$ for some $\Omega _{\mathrm{at}%
}\in \mathfrak{H}_{\mathrm{at}}$, see (\ref{defnition vaccum at}).

Observe that $\Omega =AJ_{\mathfrak{g}}AJ_{\mathfrak{g}}\Omega _{\mathfrak{g}%
}\in \mathcal{P}_{\mathfrak{g}}$ with
\begin{equation*}
A=\rho _{\mathrm{at}}^{1/4}\rho _{\mathfrak{g}}^{-1/4}\otimes \mathbf{1}_{%
\mathfrak{H}_{\mathcal{R}}}\ ,
\end{equation*}%
where we recall that $\rho _{\mathfrak{g}}$ is the density matrix (\ref%
{Gibbs.init}) of the Gibbs state $\mathfrak{g}_{\mathrm{at}}$. Additionally,
by cyclicity of $\Omega $ and \cite[Proposition 2.5.30 (1)]%
{BratteliRobinsonI}, $\Omega $ is also separating for $\mathfrak{M}$. So, we
can define the modular operator $\Delta $ and the modular conjugation $J$ of
the pair $\left( \mathfrak{M},\Omega \right) $. By \cite[Proposition 2.5.30
(2)]{BratteliRobinsonI}, $J_{\mathfrak{g}}=J$.

Assume that the density matrix $\rho _{\mathrm{at}}$ of the initial atomic
state is block--diagonal in the eigenbasis of the atomic Hamiltonian $H_{%
\mathrm{at}}$, i.e., $\rho _{\mathrm{at}}\in \mathfrak{D}$ (cf. (\ref{ninja0}%
)). Then $\rho _{\mathrm{at}}$ commutes with $H_{\mathrm{at}}$ an it is
straightforward to verify that
\begin{equation}
L_{\mathfrak{g}}\Omega =0\text{ }.  \label{annhilation bis}
\end{equation}

In our setting, however, the free dynamics is perturbed by the pump and the
atom--reservoir interaction. Altogether, this leads to a perturbation $W_{t}$
of $L_{\mathfrak{g}}$. For autonomous perturbations of the generator $\delta
$ of the dynamics $\{\tau _{t}\}_{t\in \mathbb{R}}$ (on ${\mathcal{V}}$) of
the form $i[W,\ \cdot \ ]$ with some self--adjoint $W\in {\mathcal{V}}$, one
has
\begin{equation*}
\forall t\in \mathbb{R},\ A\in {\mathcal{V}}:\qquad \tau _{t}^{W}\left(
A\right) =\mathrm{e}^{it\left( L_{\mathfrak{g}}+W\right) }A\mathrm{e}%
^{-it\left( L_{\mathfrak{g}}+W\right) }\in \mathfrak{M}\ ,
\end{equation*}%
where $\{\tau _{t}^{W}\}_{t\in \mathbb{R}}$ is the strongly continuous $\ast
$--automorphism group on ${\mathcal{V}}$ generated by $\delta +i[W,\ \cdot \
]$. Analogously as above, $\{\tau _{t}^{W}\}_{t\in \mathbb{R}}$ defines a $%
\sigma $--weakly continuous group on whole $\mathfrak{M}$. Nevertheless, in
general, the operator $L_{\mathfrak{g}}+W$ does not annihilate anymore $%
\Omega $. A\ way to get around this problem is presented in \cite[Section 2.2%
]{Pillet-Jacblabla} by introducing the notion of $C$--Liouvilleans $\mathcal{%
L}$, which is constructed such that $\mathcal{L}\Omega =0$.

Observe that, in our case, the dynamics is non--autonomous. Using the $C$%
--Liouvilleans construction of \cite[Section 2.2]{Pillet-Jacblabla} we can
design the time--depending $C$--Liouvillean of the non--autonomous dynamics
such that $\mathcal{L}_{t}\Omega =0$. This is a very useful property for the
analysis of the dynamics. As already done in a few previous works (see for
instance \cite{BMPS, S, FS}), we thus extend the definition of $C$%
--Liouvilleans \cite[Section 2.2]{Pillet-Jacblabla} to non--autonomous
evolutions. Note that we exploit the fact that the density matrix $\rho _{%
\mathrm{at}}$ of the initial atomic state commutes\footnote{%
The mp\_arc preprint of this paper gives this construction without this
assumption.} with the atomic Hamiltonian $H_{\mathrm{at}}$. This simplifies
the explicit expression of the $C$--Liouvillean, see Definition \ref%
{Time--dependent C--Liouvillean} and (\ref{Liouvillean}) below.

Following \cite[Section 2.2]{Pillet-Jacblabla}, we define the linear space%
\begin{equation*}
\mathcal{O}:=\left\{ A\Omega :A\in \mathcal{V}\right\} \subset \mathfrak{H}\
.
\end{equation*}%
Let $\iota $ be the map from $\mathcal{V}$ to $\mathcal{O}$ defined by $%
\iota \left( A\right) :=A\Omega $. This map is an isomorphism of the linear
spaces $\mathcal{V}$ and $\mathcal{O}$ because $\Omega $ is a separating
vector for $\mathfrak{M}$. In particular, $\Vert A\Omega \Vert _{\mathcal{O}%
}:=\Vert A\Vert _{\mathcal{V}}$ defines a norm on the space $\mathcal{O}$, $%
\iota $ is an isometry w.r.t. this norm, and $(\mathcal{O},\Vert \cdot \Vert
_{\mathcal{O}})$ is a Banach space. Any element $A\in \mathcal{V}$ also
defines a bounded operator on $\mathcal{O}$ by left multiplication, i.e., $%
A\left( B\Omega \right) :=\left( AB\right) \Omega $. Moreover, we define a
strongly continuous two--parameter family $\{T_{s,t}\}_{s,t\in {\mathbb{R}}}$
acting on $\mathcal{O}$ by%
\begin{equation}
T_{s,t}:=\iota \circ \mathfrak{W}_{s,t}\circ \iota ^{-1}\ ,\qquad s,t\in {%
\mathbb{R}}\ .  \label{eq idiote2}
\end{equation}%
Note that, since $\{\mathfrak{W}_{s,t}\}_{s,t\in {\mathbb{R}}}$ is a family
of $\ast $--automorphisms, the operator $T_{s,t}$ has a bounded inverse.
Moreover,
\begin{equation}
\forall s,t\in {\mathbb{R}},\ A\in \mathcal{V}:\quad T_{s,t}\left( \Omega
\right) =\Omega \quad \text{and}\quad T_{s,t}AT_{s,t}^{-1}=\mathfrak{W}%
_{s,t}\left( A\right) \ .  \label{eq de genie pillet-etal 2}
\end{equation}%
Here, $A,\mathfrak{W}_{s,t}\left( A\right) \in \mathcal{V}$ are seen as
bounded operators on $\mathcal{O}$ defined by left multiplication, as
explained above. We would like now to extend the two--parameter family $%
\{T_{s,t}\}_{s,t\in {\mathbb{R}}}$ to whole $\mathfrak{H}$. To this end,
similarly as done in \cite[Eq. (2.5)]{Pillet-Jacblabla} for the autonomous
case, we define the time--dependent $C$--Liouvillean as follows:

\begin{definition}[Time--dependent $C$--Liouvilleans]
\label{Time--dependent C--Liouvillean}\mbox{ }\newline
For any self--adjoint family $\{W_{t}\}_{t\in \mathbb{R}}\subset \mathcal{V}$%
, the time--dependent $C$--Liouvillean is the family of operators acting on $%
\mathfrak{H}$ defined by
\begin{equation*}
\mathcal{L}_{t}:=L_{\mathfrak{g}}+W_{t}-J\Delta ^{1/2}W_{t}\Delta ^{-1/2}J\
,\qquad t\in {\mathbb{R}}\ .
\end{equation*}
\end{definition}

\noindent This time--dependent operator is, in general, not anymore
self--adjoint. Note also that the term
\begin{equation*}
V_{t}:=W_{t}-J\Delta ^{1/2}W_{t}\Delta ^{-1/2}J\in \mathcal{B}\left(
\mathcal{O}\right)
\end{equation*}%
implements the commutator $[W_{t},\ \cdot \ ]$ on $\mathcal{O}$ for any $%
t\in \mathbb{R}$. Indeed, for any $A\in \mathcal{V}$,
\begin{equation}
\left[ W_{t},A\right] \Omega =W_{t}A\Omega -\left( W_{t}A^{\ast }\right)
^{\ast }\Omega  \label{commutator}
\end{equation}%
and using $J\Delta ^{1/2}A\Omega =A^{\ast }\Omega $, $\Delta ^{-1/2}=J\Delta
^{1/2}J$, and $J^{2}=1$,
\begin{equation}
J\Delta ^{1/2}W_{t}\Delta ^{-1/2}JA\Omega =\left( W_{t}A^{\ast }\right)
^{\ast }\Omega \ .  \label{commutatorbis}
\end{equation}%
In particular,
\begin{equation}
\Vert J\Delta ^{1/2}W_{t}\Delta ^{-1/2}J\Vert _{\mathcal{B}\left( \mathcal{O}%
\right) }=\Vert W_{t}\Vert _{\mathcal{V}}<\infty
\label{liouvilliean sup0bis}
\end{equation}%
and
\begin{equation}
\forall t\in \mathbb{R}:\qquad \mathcal{L}_{t}\Omega =0\ .
\label{cool c liuouvillean}
\end{equation}

Observe that the norm $\Vert \cdot \Vert _{\mathcal{O}}$ on $\mathcal{O}$ is
not equivalent to the Hilbert space norm on this subspace of $\mathfrak{H}$.
In particular, the boundedness of the operator
\begin{equation*}
J\Delta ^{1/2}W_{t}\Delta ^{-1/2}J
\end{equation*}%
as an operator on $\mathfrak{H}$ is unclear, in spite of (\ref{liouvilliean
sup0bis}). Therefore, for every $t\in \mathbb{R}$, we assume some sufficient
conditions on the operator family $\{V_{t}\}_{t\in \mathbb{R}}$, like the
boundedness of its elements as linear operators on the Hilbert space $%
\mathfrak{H}$, in order to extend the elements of the two--parameter family $%
\{T_{s,t}\}_{s,t\in {\mathbb{R}}}$ to whole $\mathfrak{H}\supset \mathcal{O}$%
.

\begin{proposition}[Extension of $\{T_{s,t}\}_{s,t\in {\mathbb{R}}}$--I]
\label{section extension copy(3)}\mbox{ }\newline
Assume that $\{V_{t}\}_{t\in {\mathbb{R}}}\in C^{1}(\mathbb{R},\mathcal{B}(%
\mathfrak{H}))$. Then, there is a unique evolution family $%
\{U_{s,t}\}_{s,t\in {\mathbb{R}}}\subset \mathcal{B}\left( \mathfrak{H}%
\right) $ with the following properties: \newline
\emph{(i)}\ It satisfies the property%
\begin{equation*}
\forall t,r,s\in \mathbb{R}:\quad U_{s,t}=U_{s,r}U_{r,t}\ .
\end{equation*}%
\emph{(ii)}\ It is the unique fundamental solution of the abstract Cauchy
initial value\ problem
\begin{equation*}
\forall s,t\in {\mathbb{R}}:\quad \partial _{s}U_{s,t}=-i\mathcal{L}%
_{s}U_{s,t},\quad U_{t,t}=\mathbf{1}_{\mathfrak{H}}\ .
\end{equation*}%
\emph{(iii)}\ It solves, in the strong sense on $\mathrm{Dom}(L_{\mathfrak{g}%
})$, the Cauchy initial value\ problem
\begin{equation*}
\forall s,t\in {\mathbb{R}}:\quad \partial _{t}U_{s,t}=iU_{s,t}\mathcal{L}%
_{t},\quad U_{s,s}=\mathbf{1}_{\mathfrak{H}}\ .
\end{equation*}%
\emph{(iv)}\ For any $s,t\in {\mathbb{R}}$, $U_{t,s}$ has a bounded inverse $%
U_{t,s}^{-1}$. \newline
\emph{(v)}\ If $\{W_{t}\}_{t\in \mathbb{R}}$ is periodic with period $T>0$,
then%
\begin{equation*}
\forall s,t\in {\mathbb{R}},\ p\in \mathbb{Z}:\quad U_{s,t}=U_{s+pT,t+pT}\ .
\end{equation*}
\end{proposition}

\noindent \textit{Proof. }This result corresponds to \cite[Lemma 3.1]{BMPS}.
The arguments of its proof are standard and thus only partially indicated in
\cite{BMPS}. We prove the proposition here for completeness. Since $\mathcal{%
L}_{t}=L_{\mathfrak{g}}+V_{t}$, Assertions (i)--(iii) can easily be deduced
from \cite[Sect. 5.4, Theorem 4.8.]{Pazy}. To prove that $U_{t,s}$ has a
bounded inverse $U_{s,t}^{-1}$, it suffices to prove that $U_{s,t}^{-1}\in
\mathcal{B}\left( \mathfrak{H}\right) $ exists for small time differences $%
\left\vert t-s\right\vert $, by the property (i). To this aim, we observe
that
\begin{equation}
\forall s,t\in {\mathbb{R}}:\quad U_{s,t}=\mathrm{e}^{i\left( t-s\right) L_{%
\mathfrak{g}}}+i\int_{s}^{t}U_{s,r}V_{r}\mathrm{e}^{i\left( r-t\right) L_{%
\mathfrak{g}}}\mathrm{d}r\,,  \label{lemma 4.5 pazy}
\end{equation}%
in the strong sense. See, e.g., \cite[Sect. 5.4, Lemma 4.5. and Theorem 4.6.]%
{Pazy}. Using that $\{V_{t}\}_{t\in {\mathbb{R}}}\in C^{1}(\mathbb{R},%
\mathcal{B}(\mathfrak{H}))$, it follows by standard arguments that $%
\left\Vert U_{r,r^{\prime }}\right\Vert _{\mathcal{B}\left( \mathfrak{H}%
\right) }$ is uniformly bounded for $r,r^{\prime }$ in compact subsets of ${%
\mathbb{R}}$. [To show this, $U_{s,t}$ can, for instance, be represented by
Dyson--Phillips series.] Therefore, by unitarity of $\mathrm{e}^{i\left(
t-s\right) L_{\mathfrak{g}}}$, for any $t\in {\mathbb{R}}$ and sufficiently
small $\left\vert t-s\right\vert $ one has
\begin{equation*}
\left\Vert \int_{s}^{t}U_{s,r}V_{r}\mathrm{e}^{i\left( r-t\right) L_{%
\mathfrak{g}}}\mathrm{d}r\right\Vert _{\mathcal{B}\left( \mathfrak{H}\right)
}\leq \frac{1}{2}\ .
\end{equation*}%
By using Neumann series to construct $U_{s,t}^{-1}\in \mathcal{B}\left(
\mathfrak{H}\right) $ for such times, Assertion (iv) holds for all $s,t\in {%
\mathbb{R}}$ because of (i). If $\{W_{t}\}_{t\in \mathbb{R}}$ is $T$%
--periodic, then both $\{U_{s,t}\}_{s,t\in {\mathbb{R}}}$ and $%
\{U_{s+pT,t+pT}\}_{s,t\in {\mathbb{R}}}$ satisfy the integral equation (\ref%
{lemma 4.5 pazy}). By uniqueness of solution of (\ref{lemma 4.5 pazy}) (cf.
\cite[Sect. 5.4, Lemma 4.5.]{Pazy}), it follows that $\{U_{s,t}\}_{s,t\in {%
\mathbb{R}}}$ is $T$--periodic in this case.\hfill $\Box $

Combining this with Proposition \ref{time.dep.trotter}, we deduce that the
evolution family $\{U_{s,t}\}_{s,t\in {\mathbb{R}}}$ is the unique
continuous extension of the two--parameter family $\{T_{s,t}\}_{s,t\in {%
\mathbb{R}}}$ to the Hilbert space $\mathfrak{H}$:

\begin{proposition}[Extension of $\{T_{s,t}\}_{s,t\in {\mathbb{R}}}$--II]
\label{section extension}\mbox{ }\newline
Assume that $\{W_{t}\}_{t\in {\mathbb{R}}}\in C^{1}(\mathbb{R},\mathcal{V})$
and $\{V_{t}\}_{t\in {\mathbb{R}}}\in C^{1}(\mathbb{R},\mathcal{B}(\mathfrak{%
H}))$. Then, the evolution family of Proposition \ref{section extension
copy(3)} satisfies $U_{s,t}\Omega =\Omega $ and
\begin{equation*}
\forall s,t\in {\mathbb{R}},\ A\in \mathcal{V}:\quad \mathfrak{W}%
_{s,t}\left( A\right) =U_{s,t}AU_{s,t}^{-1}\ .
\end{equation*}%
Here, $A\equiv \pi (A)$ and $\mathfrak{W}_{s,t}\left( A\right) \equiv \pi
\left( \mathfrak{W}_{s,t}\left( A\right) \right) $ are seen as bounded
operators on $\mathfrak{H}$, by left multiplication.
\end{proposition}

\noindent \textit{Proof. }This proposition corresponds to \cite[Theorem 3.2]%
{BMPS}. Here we propose an alternative proof which is somewhat simpler. Note
first that $U_{s,t}\Omega =\Omega $ is a direct consequence of (\ref{cool c
liuouvillean}) and Proposition \ref{section extension copy(3)} (iii). Using
the isometry between $\mathcal{O}$ and $\mathcal{V}$,
\begin{equation*}
T_{\cdot ,t}(A\Omega )\in C^{1}(\mathbb{R};\mathcal{O})\ ,\text{\qquad }%
T_{s,\cdot }(A\Omega )\in C^{1}(\mathbb{R};\mathcal{O})
\end{equation*}%
for any $A\in \mathrm{Dom}\left( \delta \right) $ and $s,t\in {\mathbb{R}}$,
by Proposition \ref{time.dep.trotter} (ii). By using Proposition \ref%
{time.dep.trotter} (ii)--(iii), (\ref{derivatiion time dependent}), (\ref%
{domain delata1})--(\ref{domain delata2}) with $\Omega $ replacing $\Omega _{%
\mathfrak{g}}$, (\ref{eq idiote2}), and (\ref{commutator})--(\ref%
{commutatorbis}), it follows that%
\begin{eqnarray}
\partial _{s}T_{s,t}\left( A\Omega \right) &=&-\delta _{W_{s}}\left(
\mathfrak{W}_{s,t}\left( A\right) \right) \Omega =-i\mathcal{L}_{s}\left(
T_{s,t}\left( A\Omega \right) \right) \ ,  \label{derivative wrt s} \\
\partial _{t}T_{s,t}\left( A\Omega \right) &=&\mathfrak{W}_{s,t}\left(
\delta _{W_{t}}\left( A\right) \right) \Omega =iT_{s,t}\left( \mathcal{L}%
_{t}\left( A\Omega \right) \right) \ ,  \label{derivative wrt t}
\end{eqnarray}%
for all $s,t\in {\mathbb{R}}$ and $A\in \mathrm{Dom}\left( \delta \right) $.
Meanwhile, since
\begin{equation}
\forall A\in \mathcal{V}:\quad \left\Vert A\Omega \right\Vert _{\mathfrak{H}%
}\leq \left\Vert A\right\Vert _{\mathcal{V}}=\left\Vert A\Omega \right\Vert
_{\mathcal{O}}\ ,  \label{finner topology}
\end{equation}%
we have%
\begin{equation*}
T_{\cdot ,t}(A\Omega )\in C^{1}(\mathbb{R};(\mathrm{Dom}\left( L_{\mathfrak{g%
}}\right) ,\left\Vert \cdot \right\Vert _{\mathfrak{H}}))\ ,\text{\qquad }%
T_{s,\cdot }(A\Omega )\in C^{1}(\mathbb{R};(\mathrm{Dom}\left( L_{\mathfrak{g%
}}\right) ,\left\Vert \cdot \right\Vert _{\mathfrak{H}}))\ ,
\end{equation*}%
and (\ref{derivative wrt s})--(\ref{derivative wrt t}) also holds in the
sense of $\mathfrak{H}$. By Proposition \ref{section extension copy(3)}
(ii),
\begin{equation}
\forall A\in \mathrm{Dom}\left( \delta \right) :\quad U_{s,t}\left( A\Omega
\right) =T_{s,t}\left( A\Omega \right) :=\mathfrak{W}_{s,t}\left( A\right)
\Omega \ .  \label{blabla}
\end{equation}%
By density of $\mathrm{Dom}\left( \delta \right) $\ in $\mathcal{V}$, for
any $A\in \mathcal{V}$, there is a sequence $\left\{ A_{n}\right\}
_{n=1}^{\infty }\subset \mathrm{Dom}\left( \delta \right) $ converging in $%
\mathcal{V}$ to $A$. We infer from (\ref{finner topology}) that this
sequence $\left\{ A_{n}\right\} _{n=1}^{\infty }$ also converges to $A$ in
the sense of $\mathfrak{H}$. On the one hand, by the boundedness of $U_{s,t}$
in $\mathfrak{H}$,
\begin{equation*}
\underset{n\rightarrow \infty }{\lim }U_{s,t}\left( A_{n}\Omega \right)
=U_{s,t}\left( A\Omega \right) \ .
\end{equation*}%
On the other hand, using (\ref{blabla}) and the boundedness of $T_{t,s}$ in $%
\mathcal{O}$, one gets%
\begin{equation*}
\underset{n\rightarrow \infty }{\lim }U_{s,t}\left( A_{n}\Omega \right) =%
\underset{n\rightarrow \infty }{\lim }T_{s,t}\left( A_{n}\Omega \right)
=T_{s,t}\left( A\Omega \right) \ .
\end{equation*}%
As a consequence, $U_{s,t}|_{\mathcal{O}}=T_{s,t}$. In particular, from the
uniqueness of the inverse, one has $U_{s,t}^{-1}|_{\mathcal{O}}=T_{s,t}^{-1}$%
. We then use (\ref{eq de genie pillet-etal 2}) to deduce that
\begin{equation*}
\forall s,t\in {\mathbb{R}},\ A\in \mathcal{V},\ x\in \mathcal{O}:\quad
\mathfrak{W}_{s,t}\left( A\right) x=U_{s,t}AU_{s,t}^{-1}\left( x\right) \ .
\end{equation*}%
By density of $\mathcal{O}$ in $\mathfrak{H}$, we arrive at the
assertion.\hfill $\Box $

The use of $C$--Liouvilleans is advantageous because of the following
identity:
\begin{equation*}
\forall s,t\in {\mathbb{R}},\ A\in \mathfrak{M}:\quad
U_{s,t}AU_{s,t}^{-1}\Omega =U_{s,t}A\Omega \ .
\end{equation*}

\subsection{Dynamics in an explicit GNS\ representation\label{Section
modular reservoir}}

To obtain the atomic Lindbladians (\ref{lindblad ninja1}) from the
time--dependant $C$--Liouvilleans (Definition \ref{Time--dependent
C--Liouvillean}), we use a convenient explicit GNS\ representation of the
initial state (\ref{initial state}). As already mentioned, this GNS
representation includes the GNS representation defined in Section \ref%
{part.syst} for the atomic initial state $\omega _{\mathrm{at}}$. So, it
remains to give an explicit GNS representation $(\mathfrak{H}_{\mathcal{R}%
},\pi _{\mathcal{R}},\Omega _{\mathcal{R}})$ for the $\left( \tau ^{{%
\mathcal{R}}},\beta \right) $--KMS\ state $\omega _{\mathcal{R}}$ of the
fermionic reservoir at inverse temperature $\beta \in \mathbb{R}^{+}$. As
briefly discussed in \cite[Section 6.1]{BruPedraWestrich2011a}, we use the
so--called \emph{Jak{\v{s}}i{\'{c}}--Pillet glued representation} \cite%
{Pillet-Jacblabla}, because it is well--adapted to the application of
spectral deformation methods, see Section \ref{section spectral deformation}.

Consider the Hilbert space
\begin{equation}
\mathfrak{h}_{2}:=L^{2}(\mathbb{R}\times S^{2},\mathbb{C})\ ,
\label{definition h2}
\end{equation}%
where $S^{2}\subset \mathbb{R}^{3}$ is the two--dimensional unit sphere
centered at the origin and $\mathbb{R}\times S^{2}$ (spherical coordinates
of $\mathbb{R}^{3}\times \mathbb{R}^{3}$) is equipped with the measure d$%
\lambda \otimes $d$^{2}s$. Here, d$^{2}s$ is the usual rotation invariant
measure induced by the Euclidean norm of $\mathbb{R}^{3}$ on $S^{2}$ and d$%
\lambda $ is the Lebesgue measure. The Hilbert space of the Jak\v{s}i\'{c}%
--Pillet glued\emph{\ }representation is the antisymmetric Fock space%
\begin{equation*}
\mathfrak{H}_{\mathcal{R}}:={\mathcal{F}}_{-}(\mathfrak{h}_{2})\ .
\end{equation*}%
The cyclic vector $\Omega _{\mathcal{R}}$ is the vacuum of ${\mathcal{F}}%
_{-}(\mathfrak{h}_{2})$. The representation map $\pi _{\mathcal{R}}$ of the $%
C^{\ast }$--algebra ${\mathcal{V}}_{\mathcal{R}}$ is the $C^{\ast }$%
--homomorphism uniquely defined by%
\begin{equation}
\forall f\in \mathfrak{h}_{1}:\quad \pi _{\mathcal{R}}(\Phi (f))=\tilde{\Phi}%
(g_{f})\ ,  \label{representation}
\end{equation}%
with $\Phi ,\tilde{\Phi}$ being the field operators defined by (\ref{field
operator}) respectively on ${\mathcal{F}}_{-}(\mathfrak{h}_{1})$ and ${%
\mathcal{F}}_{-}(\mathfrak{h}_{2})$, and where $g_{f}\in \mathfrak{h}_{2}$
is given, for $(p,\vartheta )\in \mathbb{R}\times S^{2}$ a.e., by
\begin{equation*}
g_{f}(p,\vartheta ):=|p|\left( 1+\mathrm{e}^{-\beta p}\right) ^{-1/2}\left\{
\begin{array}{lcr}
f(p\vartheta ) & , & p\geq 0\ , \\
&  &  \\
\overline{f(-p\vartheta )} & , & p<0\ .%
\end{array}%
\right.
\end{equation*}%
Compare with (\ref{function gl}).

Note that $\omega _{\mathcal{R}}$ is faithful and we thus identify $\pi _{%
\mathcal{R}}\left( A\right) $ and $\pi _{\mathcal{R}}\left( \mathcal{V}_{%
\mathcal{R}}\right) $ with $A$ and $\mathcal{V}_{\mathcal{R}}$,
respectively. The weak closure of the $C^{\ast }$--algebra $\pi _{\mathcal{R}%
}\left( \mathcal{V}_{\mathcal{R}}\right) \equiv \mathcal{V}_{\mathcal{R}}$
is the von Neumann algebra $\mathfrak{M}_{\mathcal{R}}:=\mathcal{V}_{%
\mathcal{R}}^{\prime \prime }$. The family $\{\tau _{t}^{{\mathcal{R}}%
}\}_{t\in \mathbb{R}}$ of Bogoliubov automorphisms on the algebra ${\mathcal{%
V}}_{{\mathcal{R}}}$ analogously defined as in (\ref{eq:Bogoliubov})
uniquely extends to an automorphism group of the von Neumann algebra $%
\mathfrak{M}_{\mathcal{R}}$, again denoted by $\tau ^{{\mathcal{R}}}:=\{\tau
_{t}^{{\mathcal{R}}}\}_{t\in \mathbb{R}}$. $(\mathfrak{M}_{\mathcal{R}},\tau
^{{\mathcal{R}}})$ defines a $W^{\ast }$--dynamical system and $\omega _{%
\mathcal{R}}$ is a $(\tau ^{{\mathcal{R}}},\beta )$--KMS state. Hence, $%
\Omega _{\mathcal{R}}$ is cyclic and separating for $\mathfrak{M}_{\mathcal{R%
}}$.

As above, there is a unique unitary representation of $\tau ^{{\mathcal{R}}}$
by conjugation, the generator $iL_{\mathcal{R}}$ of which satisfies $L_{%
\mathcal{R}}\Omega _{\mathcal{R}}=0$ and $\tau _{t}^{{\mathcal{R}}}(\cdot )=%
\mathrm{e}^{itL_{\mathfrak{g}}}(\cdot )\mathrm{e}^{-itL_{\mathfrak{g}}}$ for
all $t\in \mathbb{R}$. The standard Liouvillean $L_{\mathcal{R}}$ is the
second quantization%
\begin{equation}
L_{\mathcal{R}}=\mathrm{d}\Gamma (p)  \label{liouvillean reservoir}
\end{equation}%
of the multiplication operator by $p\in \mathbb{R}$, that is, the operator
acting on $\mathfrak{h}_{2}$ as $(pf)(p,\vartheta )=pf(p,\vartheta )$.

An explicit GNS representation of the initial state $\omega _{0}$ of the
composite system is now easy to derive. Using indeed the representations $(%
\mathfrak{H}_{\mathrm{at}},\pi _{\mathrm{at}},\Omega _{\mathrm{at}})$ (see
Section \ref{part.syst}) and $(\mathfrak{H}_{\mathcal{R}},\pi _{\mathcal{R}%
},\Omega _{\mathcal{R}})$ of the states $\omega _{\mathrm{at}}$ and $\omega
_{{\mathcal{R}}}$, a GNS representation $(\mathfrak{H},\pi ,\Omega )$ of $%
\omega _{0}$ is given by
\begin{equation}
\mathfrak{H}:=\mathfrak{H}_{\mathrm{at}}\otimes \mathfrak{H}_{\mathcal{R}}\
,\quad \pi :=\pi _{\mathrm{at}}\otimes \pi _{\mathcal{R}}\ ,\quad \Omega
:=\Omega _{\mathrm{at}}\otimes \Omega _{\mathcal{R}}\ .  \label{definition H}
\end{equation}%
Since $\mathfrak{H}_{\mathrm{at}}$ is finite dimensional and ${\mathcal{V}}:=%
\mathcal{B}(\mathbb{C}^{d})\otimes {\mathcal{V}}_{\mathcal{R}}$, we do not
have to specify the meaning of the tensor product. Recall that, for
simplicity of notation, $\pi \left( A\right) $ and $\pi \left( \mathcal{V}%
\right) $\ are respectively identified with $A$ and $\mathcal{V}$. $%
\mathfrak{M}:=\mathcal{V}^{\prime \prime }$ is the weak closure of the $%
C^{\ast }$--algebra $\pi \left( \mathcal{V}\right) \equiv \mathcal{V}$. In
this explicit representation, $\mathfrak{M}=\mathcal{B}(\mathbb{C}%
^{d})\otimes \mathfrak{M}_{{\mathcal{R}}}$, where $\mathfrak{M}_{\mathcal{R}%
}:=\mathcal{V}_{\mathcal{R}}^{\prime \prime }$.

As explained in\ Section \ref{Time--dependent C--Liouvilleans}, (\ref{free
dynamics}) defines a one--parameter group $\{\tau _{t}\}_{t\in \mathbb{R}}$
of $\ast $--automorphisms on $\mathfrak{M}$. The standard Liouvillean of $%
\{\tau _{t}\}_{t\in \mathbb{R}}$ reads%
\begin{equation}
L_{\mathfrak{g}}=i\mathfrak{L}_{\mathrm{at}}\otimes \mathbf{1}_{\mathfrak{H}%
_{\mathcal{R}}}+\mathbf{1}_{\mathfrak{H}_{\mathrm{at}}}\otimes L_{\mathcal{R}%
}=(\underrightarrow{H_{\mathrm{at}}}-\,\underleftarrow{H_{\mathrm{at}}}%
)\otimes \mathbf{1}_{\mathfrak{H}_{\mathcal{R}}}+\mathbf{1}_{\mathfrak{H}_{%
\mathrm{at}}}\otimes \mathrm{d}\Gamma (p)  \label{liouvillean free0}
\end{equation}%
in the representation given above and satisfies (\ref{important eq c
louvilean}). See also (\ref{Lat}) and (\ref{liouvillean reservoir}). For any
$A\in \mathfrak{H}_{\mathrm{at}}\equiv \mathcal{B}(\mathbb{C}^{d})$ and $%
B\in \mathrm{Dom}(\mathrm{e}^{-\beta L_{\mathcal{R}}/2})\subset \mathfrak{H}%
_{\mathcal{R}}$,
\begin{equation}
\Delta ^{1/2}\left( A\otimes B\right) =(\rho _{\mathrm{at}}^{1/2}A\rho _{%
\mathrm{at}}^{-1/2})\otimes (\mathrm{e}^{-\beta L_{\mathcal{R}}/2}B)\ ,\quad
J\left( A\otimes B\right) =A^{\ast }\otimes \left( J_{\mathcal{R}}B\right)
\label{JR0}
\end{equation}%
with $J_{\mathcal{R}}$ being the modular conjugation associated with the
pair $\left( \mathfrak{M}_{\mathcal{R}},\Omega _{\mathcal{R}}\right) $. In
fact,%
\begin{equation}
J_{\mathcal{R}}\tilde{\Phi}(g_{f})J_{\mathcal{R}}=(-1)^{\mathrm{d}\Gamma (%
\mathbf{1}_{\mathfrak{h}_{2}})}\tilde{\Phi}(g_{f}^{\#})\ ,\qquad f\in
\mathfrak{h}_{1}\ ,  \label{JR}
\end{equation}%
where
\begin{equation*}
g_{f}^{\#}(p,\vartheta ):=i\overline{g_{f}(-p,\vartheta )}\ ,\qquad
(p,\vartheta )\in \mathbb{R}\times S^{2}\text{ (a.e.) .}
\end{equation*}%
Compare with (\ref{g diese}). For more details, see \cite[Theorem 3.3. and
Proposition 3.4.]{Pillet-Jacblabla}.

In the same representation, the time--dependent $C$--Liouvillean $\mathcal{L}%
_{t}$ (Definition \ref{Time--dependent C--Liouvillean}) then equals
\begin{equation}
\mathcal{L}_{t}\equiv \mathcal{L}_{t}^{(\lambda ,\eta )}:=L_{\mathfrak{g}%
}+W_{t}-J\Delta ^{1/2}W_{t}\Delta ^{-1/2}J  \label{Liouvillean}
\end{equation}%
with
\begin{eqnarray}
W_{t} &:&=\eta \cos (\varpi t)\underrightarrow{H_{\mathrm{p}}}\otimes
\mathbf{1}_{{\mathcal{V}}_{\mathcal{R}}}+\lambda \sum_{\kappa =1}^{\mathrm{K}%
}\sum_{\ell _{1},\ldots ,\ell _{\kappa }=1}^{m}\underrightarrow{Q_{_{\ell
_{1},\ldots ,\ell _{\kappa }}}^{(\kappa )}}\otimes  \label{Wt} \\
&&\qquad \qquad \qquad \left( \frac{1}{\sqrt{2}}\right) ^{\kappa }\left(
a^{+}(g_{\ell _{1}}^{(\kappa )})+a(g_{\ell _{1}}^{(\kappa )})\right) \cdots
\left( a^{+}(g_{\ell _{\kappa }}^{(\kappa )})+a(g_{\ell _{\kappa }}^{(\kappa
)})\right)  \notag
\end{eqnarray}%
and, by (\ref{JR0})--(\ref{JR}),
\begin{eqnarray}
J\Delta ^{1/2}W_{t}\Delta ^{-1/2}J &=&\eta \cos (\varpi t)\underleftarrow{%
\rho _{\mathrm{at}}^{-1/2}H_{\mathrm{p}}\rho _{\mathrm{at}}^{1/2}}\otimes
\mathbf{1}_{{\mathcal{V}}_{\mathcal{R}}}-\lambda \sum_{\kappa =1}^{\mathrm{K}%
}\sum_{\ell _{1},\ldots ,\ell _{\kappa }=1}^{m}\underleftarrow{\rho _{%
\mathrm{at}}^{-1/2}Q_{_{\ell _{1},\ldots ,\ell _{\kappa }}}^{(\kappa )}\rho
_{\mathrm{at}}^{1/2}}\otimes  \label{Wt'} \\
&&\left( \frac{-i}{\sqrt{2}}\right) ^{\kappa }(-1)^{\mathrm{d}\Gamma (%
\mathbf{1}_{\mathfrak{h}_{2}})}\left( a\left( i\mathrm{e}^{-\beta p}g_{\ell
_{1}}^{(\kappa )}\right) +a^{+}\left( ig_{\ell _{1}}^{(\kappa )}\right)
\right)  \notag \\
&&\cdots (-1)^{\mathrm{d}\Gamma (\mathbf{1}_{\mathfrak{h}_{2}})}\left(
a\left( i\mathrm{e}^{-\beta p}g_{\ell _{\kappa }}^{(\kappa )}\right)
+a^{+}\left( ig_{\ell _{\kappa }}^{(\kappa )}\right) \right) \ .  \notag
\end{eqnarray}%
Recall that $\rho _{\mathrm{at}}\in \mathcal{B}(\mathbb{C}^{d})$ is a
(invertible) density matrix of the (faithful) state $\omega _{\mathrm{at}}$
with $[H_{\mathrm{at}},\rho _{\mathrm{at}}]=0$ and, for any $\kappa
=\{1,\ldots ,\mathrm{K}\}$ and $\ell \in \{1,\ldots ,m\}$ with $\mathrm{K}%
,m\in \mathbb{N}$,%
\begin{equation}
g_{\ell }^{(\kappa )}(p,\vartheta ):=\mathrm{g}_{\ell }^{(\kappa )}(p)\
,\qquad (p,\vartheta )\in \mathbb{R}\times S^{2}\ ,  \label{function cool}
\end{equation}%
where $\{\mathrm{g}_{\ell }^{(\kappa )}\}_{\ell =1}^{m}$ is the family of
complex--valued functions defined on $\mathbb{R}$ by (\ref{function gl}).
Note that $\mathrm{d}\Gamma (\mathbf{1}_{\mathfrak{h}_{2}})$, the second
quantization of $\mathbf{1}_{\mathfrak{h}_{2}}$, is the particle number
operator acting on the antisymmetric Fock space $\mathfrak{H}_{\mathcal{R}}:=%
{\mathcal{F}}_{-}(\mathfrak{h}_{2})$.

$\{W_{t}\}_{t\in \mathbb{R}}\in C^{\infty }(\mathbb{R},\mathcal{V})$ is $%
2\pi \varpi ^{-1}$--periodic and the operators%
\begin{equation}
V_{t}:=W_{t}-J\Delta ^{1/2}W_{t}\Delta ^{-1/2}J  \label{definition V}
\end{equation}%
defines a smooth family $\{V_{t}\}_{t\in \mathbb{R}}\in C^{\infty }(\mathbb{R%
},\mathcal{B}(\mathfrak{H}))$. Therefore, the assumptions of Proposition \ref%
{section extension} are satisfied for this explicit example. The evolution
family $\{U_{s,t}\}_{s,t\in {\mathbb{R}}}$ is in this case $2\pi \varpi
^{-1} $--periodic, by Proposition \ref{section extension copy(3)} (v).

\subsection{Evolution group of $\{U_{s,t}^{(\protect\lambda ,\protect\eta %
)}\}_{t\geq s}$}

We now represent the non--autonomous evolution family $\{U_{s,t}\equiv
U_{s,t}^{(\lambda ,\eta )}\}_{s,t\in {\mathbb{R}}}$ of Proposition \ref%
{section extension copy(3)} as an \emph{autonomous} dynamics on the enlarged
Hilbert space%
\begin{equation*}
\mathcal{H}:=L^{2}(\left[ 0,2\pi \varpi ^{-1}\right) ,\mathfrak{H})
\end{equation*}%
of $2\pi \varpi ^{-1}$--periodic $\mathfrak{H}$--valued functions. The same
procedure is used in \cite[Section 3.4]{BMPS} and thus there is some overlap
between the latter and the present section. However, as the
time/coupling/frequency regime considered here is completely different to
the one studied in \cite{BMPS} (see Introduction), various important
technical aspects are different. See, for instance, discussion after Lemma %
\ref{lemmalongtime}. In spite of such an overlap with \cite[Section 3.4]%
{BMPS}, for the reader's convenience, we prove all statements we use in this
section, because their proofs are rather short. The scalar product on $%
\mathcal{H}$ is naturally defined by%
\begin{equation*}
\left\langle f,g\right\rangle _{\mathcal{H}}:=\frac{\varpi }{2\pi }\int_{0}^{%
\frac{2\pi }{\varpi }}\left\langle f\left( t\right) ,g\left( t\right)
\right\rangle _{\mathfrak{H}}\mathrm{d}t\ ,\qquad f,g\in \mathcal{H}\ .
\end{equation*}%
In the sequel we identify $\mathfrak{H}$ with the subspace of constant
functions on $\left[ 0,2\pi \varpi ^{-1}\right) $. See also (\ref{definition
H}).

From the strongly continuous two--parameter family $\{U_{s,t}\}_{s,t\in {%
\mathbb{R}}}$ on $\mathfrak{H}$ we define a strongly continuous
one--parameter group $\{\mathfrak{T}_{\alpha }\}_{\alpha \in \mathbb{R}}$ on
$\mathcal{H}$\ by the condition%
\begin{equation}
\forall t\in \left[ 0,2\pi \varpi ^{-1}\right) \ \text{(a.e.)}:\qquad
\mathfrak{T}_{\alpha }\left( f\right) \left( t\right) =U_{t,t+\alpha
}f\left( t+\alpha \right) \ ,  \label{definition howland}
\end{equation}%
for all $\alpha \in \mathbb{R}$ and $f\in \mathcal{H}$. Because of (\ref{Wt}%
) and Proposition \ref{section extension copy(3)} (v), $\mathfrak{T}_{\alpha
}$ is an operator acting on $\mathcal{H}$ for any $\alpha \in \mathbb{R}$.
The strong continuity of $\alpha \mapsto \mathfrak{T}_{\alpha }$ follows
from the strong continuity of $t\mapsto U_{s,t}$, and the group property of $%
\{\mathfrak{T}_{\alpha }\}_{\alpha \in \mathbb{R}}$ from Proposition \ref%
{section extension copy(3)} (i).

The Howland operator of the non--autonomous dynamics $\{U_{s,t}\}_{s,t\in {%
\mathbb{R}}}$ is, by definition, the generator $\mathcal{G}$ of the strongly
continuous group $\{\mathfrak{T}_{\alpha }\}_{\alpha \in \mathbb{R}}$. It is
a closed unbounded operator acting on $\mathcal{H}$. It is not a priori
clear whether the group $\{\mathfrak{T}_{\alpha }\}_{\alpha \in \mathbb{R}}$
is contractive. In fact, it is quasi--contractive. Such a property can be
useful to analyze the domain of generators of semigroups.

We show that $\left( \pm \mathcal{G}-C\right) $ is dissipative, i.e.,
\begin{equation*}
\forall f\in \mathrm{Dom}(\mathcal{G}):\qquad \mathop{\rm Re}\left\{ \langle
f,\left( \pm \mathcal{G}-C\right) f\rangle _{\mathcal{H}}\right\} \leq 0\ ,
\end{equation*}%
for\ a sufficiently large positive constant $C$. By the Lumer--Phillips
theorem, $\left( \pm \mathcal{G}-C\right) $ generate contraction semigroups.

\begin{lemma}[Quasi--Contractivity of $\{\mathfrak{T}_{\protect\alpha }\}_{%
\protect\alpha \in \mathbb{R}}$]
\label{section extension copy(4)}\mbox{ }\newline
There is $C\in \mathbb{R}_{0}^{+}$ such that $\left( \pm \mathcal{G}%
-C\right) $ is dissipative. In particular, $\{\mathrm{e}^{-C\alpha }%
\mathfrak{T}_{\alpha }\}_{\alpha \geq 0}$ and $\{\mathrm{e}^{-C\alpha }%
\mathfrak{T}_{-\alpha }\}_{\alpha \geq 0}$ are contraction semigroups.
\end{lemma}

\noindent \textit{Proof.} For any positive constant $C\in \mathbb{R}_{0}^{+}$%
, we define the operators%
\begin{equation*}
U_{s,t}^{(C)}:=U_{s,t}\mathrm{e}^{\left( s-t\right) C}\ ,\qquad s,t\in {%
\mathbb{R}}\ .
\end{equation*}%
By Proposition \ref{section extension copy(3)} and Equations (\ref%
{liouvillean free0}) and (\ref{Liouvillean}), $\{U_{s,t}^{(C)}\}_{s,t\in {%
\mathbb{R}}}$ is the fundamental solution on $\mathrm{Dom}(L_{\mathfrak{g}})$
of the Cauchy initial value\ problem
\begin{equation*}
\forall s,t\in {\mathbb{R}}:\quad \partial _{s}U_{s,t}^{(C)}=-\left( i%
\mathcal{L}_{s}-C\right) U_{s,t}^{(C)},\quad U_{t,t}^{(C)}=\mathbf{1}_{%
\mathfrak{H}}\ ,
\end{equation*}%
while it solves on $\mathrm{Dom}(L_{\mathfrak{g}})$ the Cauchy initial
value\ problem
\begin{equation*}
\forall s,t\in {\mathbb{R}}:\quad \partial
_{t}U_{s,t}^{(C)}=U_{s,t}^{(C)}\left( i\mathcal{L}_{t}-C\right) ,\quad
U_{s,s}^{(C)}=\mathbf{1}_{\mathfrak{H}}\ .
\end{equation*}%
Choose
\begin{equation*}
C:=\underset{t\in \mathbb{R}}{\max }\left\Vert V_{t}\right\Vert _{\mathcal{B}%
(\mathfrak{H})}<\infty \ ,
\end{equation*}%
where we recall that $V_{t}$ is defined by (\ref{definition V}). As $L_{%
\mathfrak{g}}$ is self--adjoint, $\left( i\mathcal{L}_{t}-C\right) $ is
dissipative for all $t\in \mathbb{R}$ and \cite[Theorem 4.8.]{Pazy} implies
that $\Vert U_{s,t}^{(C)}\Vert _{\mathcal{B}(\mathfrak{H})}\leq 1$ for all $%
t\geq s$. It follows that $\Vert \mathrm{e}^{-C\alpha }\mathfrak{T}_{\alpha
}\Vert _{\mathcal{B}(\mathcal{H})}\leq 1$ for all $\alpha \geq 0$.
Similarly, we show that $\Vert \mathrm{e}^{-C\alpha }\mathfrak{T}_{-\alpha
}\Vert _{\mathcal{B}(\mathcal{H})}\leq 1$ for all $\alpha \geq 0$. To finish
the proof, one uses the fact that $\left( \pm \mathcal{G}-C\right) $ is the
generator of the contraction semigroup $\{$\textrm{e}$^{-C\alpha }\mathfrak{T%
}_{\pm \alpha }\}_{\alpha \geq 0}$ together with the Lumer--Phillips
theorem.\hfill $\Box $

Observe that $\mathcal{H}$ is unitarily equivalent -- via the Fourier
transform $\mathfrak{F}$ -- to the Hilbert space $\mathcal{\hat{H}}:=\ell
^{2}(\mathbb{Z},\mathfrak{H})$ with scalar product
\begin{equation*}
\langle \hat{f},\hat{g}\rangle _{\mathcal{\hat{H}}}:=\sum\limits_{k\in
\mathbb{Z}}\langle \hat{f}(k),\hat{g}(k)\rangle _{\mathfrak{H}}\ ,\qquad
\hat{f},\hat{g}\in \mathcal{\hat{H}}\ .
\end{equation*}%
It is indeed more convenient to analyze the Fourier transform $\mathfrak{F}{%
\mathcal{G\mathfrak{F}}}^{\ast }$ instead of $\mathcal{G}$ directly. To this
end, we define the dense subspace%
\begin{equation*}
\mathcal{\hat{D}}:=\left\{ \hat{f}\in \mathcal{\hat{H}}\;:\;\hat{f}(\mathbb{Z%
})\subset \mathrm{Dom}(L_{\mathfrak{g}}),\;\sum\limits_{k\in \mathbb{Z}%
}\left( k^{2}\Vert \hat{f}(k)\Vert _{\mathfrak{H}}^{2}+\Vert L_{\mathfrak{g}%
}(\hat{f}(k))\Vert _{\mathfrak{H}}^{2}\right) <\infty \right\}
\end{equation*}%
of the Hilbert space $\mathcal{\hat{H}}$ as well as the (unbounded)
operators $k_{\mathcal{\hat{H}}}$ and $L_{\mathfrak{g},\mathcal{\hat{H}}}$
on $\mathcal{\hat{D}}$ by
\begin{equation}
\forall k\in \mathbb{Z}:\quad \big(k_{\mathcal{\hat{H}}}\hat{f}\big)(k):=k%
\hat{f}(k)\ ,\quad \big(L_{\mathfrak{g},\mathcal{\hat{H}}}\hat{f}\big)%
(k):=L_{\mathfrak{g}}\hat{f}(k)\ ,  \label{def trivial1}
\end{equation}%
see (\ref{liouvillean free0}). By abuse of notation, we denote $k_{\mathcal{%
\hat{H}}}$ and $L_{\mathfrak{g},\mathcal{\hat{H}}}$ respectively by $k$ and $%
L_{\mathfrak{g}}$. In the same way, let $V_{\mathcal{H}}\equiv V$ be the
bounded operator acting on $\mathcal{H}$ defined by
\begin{equation}
\forall t\in \left[ 0,2\pi \varpi ^{-1}\right) \ \text{(a.e.)}:\qquad \big(%
V_{\mathcal{H}}f\big)(t):=V_{t}\big(f(t)\big)\ .  \label{important operators}
\end{equation}%
We prove below that the unbounded operator
\begin{equation}
{\mathcal{\hat{G}}}:=i\big(\varpi k+L_{\mathfrak{g}}+\hat{V}\big)
\label{holand fourier}
\end{equation}%
defined on $\mathcal{\hat{D}}$ with $\hat{V}:=\mathfrak{F}V\mathfrak{F}%
^{\ast }$ is the Fourier transform $\mathfrak{F}{\mathcal{G\mathfrak{F}}}%
^{\ast }$ of $\mathcal{G}$:

\begin{satz}[Explicit form of the Howland operator in Fourier space]
\label{section extension copy(5)}\mbox{ }\newline
The Fourier transform of the generator $\mathcal{G}$ of the strongly
continuous group $\{\mathfrak{T}_{\alpha }\}_{\alpha \in \mathbb{R}}$ equals%
\begin{equation*}
\mathfrak{F}{\mathcal{G\mathfrak{F}}}^{\ast }={\mathcal{\hat{G}}}\ .
\end{equation*}
\end{satz}

\noindent \textit{Proof.} The operator $\varpi k+L_{\mathfrak{g}}$ can be
viewed as a tensor sum of self--adjoint operators acting on
\begin{equation*}
\mathcal{\hat{H}}=\underset{k\in \mathbb{Z}}{\bigoplus }\mathfrak{H}_{k}\
,\qquad \mathfrak{H}_{k}\equiv \mathfrak{H}\ .
\end{equation*}%
It is essentially self--adjoint on
\begin{equation}
\mathcal{\hat{D}}_{0}:=\left\{ \hat{f}\in \mathcal{\hat{H}}\;:\;\hat{f}%
\left( k\right) =0\text{ for }k\text{ outside a finite set and }\hat{f}(%
\mathbb{Z})\subset \mathrm{Dom}(L_{\mathfrak{g}})\right\} \subset \mathcal{%
\hat{D}}\   \label{core}
\end{equation}%
and $\mathcal{\hat{D}}$ is the graph norm closure of $\mathcal{\hat{D}}_{0}$
w.r.t. the operator $\varpi k+L_{\mathfrak{g}}$. Hence, $\varpi k+L_{%
\mathfrak{g}}$ is self--adjoint on $\mathcal{\hat{D}}$. Since $%
\{V_{t}\}_{t\in \mathbb{R}}\in C^{\infty }(\mathbb{R},\mathcal{B}(\mathfrak{H%
}))$, the operator $\hat{V}\in \mathcal{B}(\mathcal{\hat{H}})$ is bounded.
Thus, ${\mathcal{\hat{G}}}$ is closed on $\mathcal{\hat{D}}$. The unbounded
part $i(\varpi k+L_{\mathfrak{g}})$ of ${\mathcal{\hat{G}}}$ is dissipative,
as $\varpi k+L_{\mathfrak{g}}$ is self--adjoint. Hence, by adding a
sufficiently large constant $C\geq \Vert \hat{V}\Vert _{\mathcal{B}(\mathcal{%
\hat{H}})}$, the operator $({\mathcal{\hat{G}}}-C)$ defined on the dense set
$\mathcal{\hat{D}}$ is the dissipative generator of a strongly continuous
semigroup. On the other hand, as $\mathfrak{F}{\mathcal{G\mathfrak{F}}}%
^{\ast }={\mathcal{\hat{G}}}$ on the core $\mathcal{\hat{D}}_{0}$ of $({%
\mathcal{\hat{G}}}-C)$, $(\mathfrak{F}{\mathcal{G\mathfrak{F}}}^{\ast }-C)$
is a closed extension of $({\mathcal{\hat{G}}}-C)$ and generates a strongly
continuous semigroup. Choosing $C$ sufficiently large, both generators $({%
\mathcal{\hat{G}}}-C)$ and $(\mathfrak{F}{\mathcal{G\mathfrak{F}}}^{\ast
}-C) $ are dissipative, by Lemma \ref{section extension copy(4)}. Using the
fact that generators of contraction semigroups have no proper dissipative
extensions, it follows that ${\mathcal{\hat{G}}}=\mathfrak{F}{\mathcal{G%
\mathfrak{F}}}^{\ast }$. \hfill $\Box $

We show next that -- at small couplings -- the quantity%
\begin{equation*}
\left\langle \Omega ,U_{0,t}\left( A\otimes \mathbf{1}_{\mathfrak{H}_{%
\mathcal{R}}}\Omega \right) \right\rangle _{\mathfrak{H}}\ ,
\end{equation*}%
which gives the time evolution of the atomic state, are well--approximated
by
\begin{equation*}
\left\langle \Omega ,\mathfrak{T}_{t}\left( A\otimes \mathbf{1}_{\mathfrak{H}%
_{\mathcal{R}}}\Omega \right) \right\rangle _{\mathcal{H}}\ ,
\end{equation*}%
provided the density matrix $\rho _{\mathrm{at}}$ of the initial atomic
state as well as the atomic observable $A$\ are block diagonal (cf. (\ref%
{ninja0})):

\begin{satz}[Effective behavior of $\protect\rho (t)$]
\label{lemmalongtime}\mbox{ }\newline
For any initial faithful state $\omega _{\mathrm{at}}$ with density matrix $%
\rho _{\mathrm{at}}\in \mathfrak{D}\subset $ $\mathcal{B}(\mathbb{C}^{d})$,
any observable $A\in \mathfrak{D}$, and $\alpha \in \mathbb{R}$,%
\begin{equation}
\left\vert \left\langle \Omega ,U_{0,\alpha }\left( A\otimes \mathbf{1}_{%
\mathfrak{H}_{\mathcal{R}}}\Omega \right) \right\rangle _{\mathfrak{H}%
}-\left\langle \Omega ,\mathfrak{T}_{\alpha }\left( A\otimes \mathbf{1}_{%
\mathfrak{H}_{\mathcal{R}}}\Omega \right) \right\rangle _{\mathcal{H}%
}\right\vert \leq C\left\Vert A\right\Vert _{\mathcal{B}(\mathbb{C}%
^{d})}|\lambda |(1+|\lambda |)\varpi ^{-1}\mathrm{e}^{D\left( 1+\left\vert
\lambda \right\vert \right) ^{2}},  \label{longtimebehavior}
\end{equation}%
where $C,D\in \mathbb{R}_{0}^{+}$ are finite constants not depending on $%
\omega _{\mathrm{at}}$, $A$, $\lambda $, $\eta $, $\varpi $, and $\alpha $.
\end{satz}

\noindent \textit{Proof.} Since
\begin{equation}
\left\langle \Omega ,U_{0,\alpha }\left( A\otimes \mathbf{1}_{\mathfrak{H}_{%
\mathcal{R}}}\Omega \right) \right\rangle _{\mathfrak{H}}-\left\langle
\Omega ,\mathfrak{T}_{\alpha }\left( A\otimes \mathbf{1}_{\mathfrak{H}_{%
\mathcal{R}}}\Omega \right) \right\rangle _{\mathcal{H}}=\frac{\varpi }{2\pi
}\int_{0}^{\frac{2\pi }{\varpi }}\left\langle \Omega ,\left( U_{0,\alpha
}-U_{t,t+\alpha }\right) \left( A\otimes \mathbf{1}_{\mathfrak{H}_{\mathcal{R%
}}}\Omega \right) \right\rangle _{\mathfrak{H}}\mathrm{d}t\ ,
\label{super gnack2bis}
\end{equation}%
we need to estimate the difference
\begin{equation*}
\left\langle \Omega ,\left( U_{0,\alpha }-U_{t,t+\alpha }\right) \left(
A\otimes \mathbf{1}_{\mathfrak{H}_{\mathcal{R}}}\Omega \right) \right\rangle
_{\mathfrak{H}}
\end{equation*}%
for any $\alpha \in \mathbb{R}$ and $t\in \left[ 0,2\pi \varpi ^{-1}\right) $%
. Using Proposition \ref{section extension copy(3)} (i), note that
\begin{equation*}
U_{0,\alpha }-U_{t,t+\alpha }=U_{0,\alpha }\left( \mathbf{1}_{\mathfrak{H}%
}-U_{\alpha ,t+\alpha }\right) +\left( U_{0,t}-\mathbf{1}_{\mathfrak{H}%
}\right) U_{t,t+\alpha }\ .
\end{equation*}%
By Proposition \ref{section extension},
\begin{eqnarray}
\left\langle \Omega ,\left( U_{0,\alpha }-U_{t,t+\alpha }\right) \left(
A\otimes \mathbf{1}_{\mathfrak{H}_{\mathcal{R}}}\Omega \right) \right\rangle
_{\mathfrak{H}} &=&\left\langle \Omega ,\mathfrak{W}_{0,\alpha }\left(
A\otimes \mathbf{1}_{\mathfrak{H}_{\mathcal{R}}}-\mathfrak{W}_{\alpha
,t+\alpha }\left( A\otimes \mathbf{1}_{\mathfrak{H}_{\mathcal{R}}}\right)
\right) \Omega \right\rangle _{\mathfrak{H}}  \notag \\
&&+\left\langle \Omega ,\left( U_{0,t}-\mathbf{1}_{\mathfrak{H}}\right)
U_{t,t+\alpha }\left( A\otimes \mathbf{1}_{\mathfrak{H}_{\mathcal{R}}}\Omega
\right) \right\rangle _{\mathfrak{H}}  \label{gnack combine}
\end{eqnarray}%
for any $t\in \left[ 0,2\pi \varpi ^{-1}\right) $. On the one hand, for any $%
B\in \mathcal{V}$,
\begin{equation}
\left\vert \left\langle \Omega ,\mathfrak{W}_{0,\alpha }\left( B-\mathfrak{W}%
_{\alpha ,t+\alpha }\left( B\right) \right) \Omega \right\rangle _{\mathfrak{%
H}}\right\vert \leq \Vert \mathfrak{W}_{0,\alpha }\Vert _{\mathcal{B}(%
\mathcal{V})}\Vert B-\mathfrak{W}_{\alpha ,t+\alpha }\left( B\right) \Vert _{%
\mathcal{V}}=\Vert B-\mathfrak{W}_{\alpha ,t+\alpha }\left( B\right) \Vert _{%
\mathcal{V}}\ .  \label{combine}
\end{equation}%
By (\ref{w fract})--(\ref{w fractbis}),%
\begin{eqnarray}
\Vert B-\mathfrak{W}_{\alpha ,t+\alpha }\left( B\right) \Vert _{\mathcal{V}}
&\leq &\left\Vert \mathfrak{V}_{t+\alpha ,\alpha }-\mathbf{1}_{\mathcal{V}%
}\right\Vert _{\mathcal{V}}\left\Vert B\right\Vert _{\mathcal{V}}\left\Vert
\mathfrak{V}_{t+\alpha ,\alpha }^{\ast }\right\Vert _{\mathcal{V}%
}+\left\Vert B\right\Vert _{\mathcal{V}}\left\Vert \mathfrak{V}_{t+\alpha
,\alpha }^{\ast }-\mathbf{1}_{\mathcal{V}}\right\Vert _{\mathcal{V}}  \notag
\\
&&+\left\Vert \tau _{t}\left( B\right) -B\right\Vert _{\mathcal{V}%
}\left\Vert \mathfrak{V}_{t+\alpha ,\alpha }^{\ast }\right\Vert _{\mathcal{V}%
}  \label{combinebis}
\end{eqnarray}%
with%
\begin{equation}
\mathfrak{V}_{t+\alpha ,\alpha }=\mathbf{1}_{\mathcal{V}}\mathbf{+}%
\sum\limits_{k\in {\mathbb{N}}}i^{k}\int_{0}^{t}\mathrm{d}s_{1}\cdots
\int_{0}^{s_{k-1}}\mathrm{d}s_{k}\tau _{s_{k}+\alpha }\left( \tau _{\alpha
}\left( W_{s_{k}+\alpha }\right) \right) \cdots \tau _{s_{1}}\left( \tau
_{\alpha }\left( W_{s_{1}+\alpha }\right) \right) \ .  \label{combinebisbis}
\end{equation}%
Since $\tau _{t}\left( A\otimes \mathbf{1}_{\mathfrak{H}_{\mathcal{R}%
}}\right) =A\otimes \mathbf{1}_{\mathfrak{H}_{\mathcal{R}}}$ for $A\in
\mathfrak{D}\subset \mathcal{B}(\mathbb{C}^{d})$, it follows from (\ref%
{combine})--(\ref{combinebisbis}) together with (\ref{Wt}) and Assumption %
\ref{assumption important} that
\begin{equation}
\left\vert \left\langle \Omega ,\mathfrak{W}_{0,\alpha }\left( A\otimes
\mathbf{1}_{\mathfrak{H}_{\mathcal{R}}}-\mathfrak{W}_{\alpha ,t+\alpha
}\left( A\otimes \mathbf{1}_{\mathfrak{H}_{\mathcal{R}}}\right) \right)
\Omega \right\rangle _{\mathfrak{H}}\right\vert \leq \frac{|\lambda
|(1+|\lambda |)}{\varpi }C\left\Vert A\right\Vert _{\mathcal{V}}\ .
\label{combine1}
\end{equation}%
On the other hand, for any $B\in \mathcal{V}$,
\begin{equation}
\left\vert \left\langle \Omega ,\left( U_{0,t}-\mathbf{1}_{\mathfrak{H}%
}\right) U_{t,t+\alpha }\left( B\Omega \right) \right\rangle _{\mathfrak{H}%
}\right\vert \leq \left\Vert \left( U_{0,t}^{\ast }-\mathbf{1}_{\mathfrak{H}%
}\right) \Omega \right\Vert _{\mathfrak{H}}\left\Vert B\right\Vert _{%
\mathcal{V}}\ .  \label{combine2}
\end{equation}%
Meanwhile, we deduce from (\ref{lemma 4.5 pazy}) that
\begin{equation}
\forall s,t\in {\mathbb{R}}:\quad U_{s,t}^{\ast }=\mathrm{e}^{i\left(
s-t\right) L_{\mathfrak{g}}}-i\left( \int_{s}^{t}U_{s,r}V_{r}\mathrm{e}%
^{i\left( r-t\right) L_{\mathfrak{g}}}\mathrm{d}r\right) ^{\ast }
\label{combine3}
\end{equation}%
while
\begin{equation}
\forall s,t\in {\mathbb{R}}:\quad \mathrm{e}^{i\left( s-t\right) L_{%
\mathfrak{g}}}\Omega =\Omega +i\int_{t}^{s}\mathrm{e}^{-irL_{\mathfrak{g}%
}}L_{\mathfrak{g}}\Omega \mathrm{d}r\ ,  \label{combine4}
\end{equation}%
because $\Omega \in \mathrm{Dom}\left( L_{\mathfrak{g}}\right) $. If $\rho _{%
\mathrm{at}}\in \mathfrak{D}$ then $L_{\mathfrak{g}}\Omega =0$. So, we can
combine (\ref{combine2}) with (\ref{combine3})--(\ref{combine4}) and the
upper bound \cite[Theorem 4.8.]{Pazy}
\begin{equation*}
\left\Vert U_{t,s}\right\Vert _{\mathcal{B}(\mathfrak{H})}\leq C\mathrm{e}%
^{D\left( t-s\right) }\ ,
\end{equation*}%
to arrive at%
\begin{equation}
\left\vert \left\langle \Omega ,\left( U_{0,t}-\mathbf{1}_{\mathfrak{H}%
}\right) U_{t,t+\alpha }\left( A\otimes \mathbf{1}_{\mathfrak{H}_{\mathcal{R}%
}}\Omega \right) \right\rangle _{\mathfrak{H}}\right\vert \leq C\left\Vert
A\right\Vert _{\mathcal{V}}|\lambda |(1+|\lambda |)\varpi ^{-1}\mathrm{e}%
^{D\left( 1+\left\vert \lambda \right\vert \right) ^{2}}\ .  \label{combine5}
\end{equation}%
Finally, using (\ref{super gnack2bis}), (\ref{gnack combine}), (\ref%
{combine1}), and (\ref{combine5}) we obtain the assertion. \hfill $\Box $

The above estimate is similar to \cite[Lemma 3.3]{BMPS}. But, in contrast to
the latter, the above lemma gives a bound which is uniform in time. Recall
that, as explained in Section \ref{Section intro}, \cite{BMPS} and the
present work consider completely different regimes of couplings and times.
Note also that we use, in an essential way, the equality $[\rho _{\mathrm{at}%
},H_{\mathrm{at}}]=0$ in the proof of Theorem \ref{lemmalongtime}.

\subsection{Resonances of the Howland operator\label{section spectral
deformation}}

Similar to \cite{Pillet-Jacblabla}, we now perform an analytic deformation
of the Howland operator in Fourier space (cf. see (\ref{holand fourier}) and
Theorem \ref{section extension copy(5)}) and study its spectrum after
deformation. Note again that the same technique was also used in \cite{BMPS}
and there are some common issues between the present Section and \cite[%
Section 3.5]{BMPS}. Nevertheless, as already stressed many times, the
considered regimes are very different and the studies of spectral properties
of the deformed Howland operator are, from a technical point of view, not
the same.

Let%
\begin{equation}
{\mathcal{\hat{G}}}_{0}(\theta ):=i\big(\varpi k+L_{\mathfrak{g}}+\theta
\hat{N}\big)\ ,\qquad \theta \in \mathbb{C}\ ,
\label{definition G non perturbe}
\end{equation}%
where, for any $\hat{f}\in \mathcal{\hat{H}}$ and $k\in \mathbb{Z}$,%
\begin{equation}
\hat{N}(\hat{f})(k):=\mathbf{1}_{\mathfrak{H}_{\mathrm{at}}}\otimes \mathrm{d%
}\Gamma (\mathbf{1}_{\mathfrak{h}_{2}})(\hat{f}(k))\ .  \label{def trivial2}
\end{equation}%
Recall that $\mathrm{d}\Gamma (\mathbf{1}_{\mathfrak{h}_{2}})$ is the second
quantization of $\mathbf{1}_{\mathfrak{h}_{2}}$, i.e., the particle number
operator acting on $\mathfrak{H}_{\mathcal{R}}:={\mathcal{F}}_{-}(\mathfrak{h%
}_{2})$. For all $\theta \in \mathbb{C}$ such that $\mathop{\rm Im}\{\theta
\}>0$, ${\mathcal{\hat{G}}}_{0}$ is a normal operator with domain
\begin{equation*}
\mathcal{\hat{D}}\cap \mathrm{Dom}(\hat{N})=\mathrm{Dom}({\mathcal{\hat{G}}}%
_{0}(\theta ))
\end{equation*}%
and spectrum in the left half--plane. In particular, by the spectral theorem
for normal operators, ${\mathcal{\hat{G}}}_{0}(\theta )$ is the generator of
a strongly continuous contraction semigroup for all $\theta \in \mathbb{C}$
such that $\mathop{\rm Im}\{\theta \}\geq 0$. [It cannot be extended to a
group, as the (negative) real part of the spectrum of ${\mathcal{\hat{G}}}%
_{0}(\theta )$ is unbounded.]

Similarly, we define the operator $\hat{V}\left( \theta \right) $ by
replacing in Equations (\ref{Wt})--(\ref{definition V}) the functions $%
g_{\ell }^{(\kappa )}$ with
\begin{equation*}
\tilde{g}_{\ell ,\theta }^{(\kappa )}(p,\vartheta ):=g_{\ell }^{(\kappa
)}(p+\theta ,\vartheta )\ ,\qquad (p,\vartheta )\in \mathbb{R}\times S^{2}\ ,
\end{equation*}%
in the creation operators and with $\tilde{g}_{\ell ,\bar{\theta}}^{(\kappa
)}$ in the annihilation operators for every $\kappa \in \{1,\ldots ,\mathrm{K%
}\}$ and $\ell \in \{1,\ldots ,m\}$, see (\ref{function cool}). Indeed, for
real parameters $\theta \in \mathbb{R\subset C}$, it is easy to see that%
\begin{equation*}
\hat{V}\left( \theta \right) =U(\theta )\hat{V}U(\theta )^{\ast }\text{ },
\end{equation*}%
where $U(\theta )$, $\theta \in \mathbb{R}$, is the unitary operator defined
on $\mathcal{\hat{H}}$ by%
\begin{equation}
\big(U(\theta )\hat{f}\big)(k):=\big(\mathbf{1}_{\mathfrak{H}_{\mathrm{at}%
}}\otimes \Gamma (u(\theta )\big)(\hat{f}(k))  \label{unitary U}
\end{equation}%
for any $\hat{f}\in \mathcal{\hat{H}}$\ and $k\in \mathbb{Z}$. Here, $\Gamma
(u(\theta ))$ is the second quantization of the unitary (translation)
operator $u(\theta )$ from $\mathfrak{h}_{2}$ to $\mathfrak{h}_{2}$ defined
by
\begin{equation*}
\big(u(\theta )f\big)(p,\vartheta ):=f(p+\theta ,\vartheta )
\end{equation*}%
for any $f\in \mathfrak{h}_{2}$ and $(p,\vartheta )\in \mathbb{R}\times
S^{2} $ (a.e.). By Assumption \ref{d Hardy}, there is $\mathrm{r}_{\max }\in
\mathbb{R}^{+}$ such that, for all $\theta \in \mathbb{C}$ such that $|%
\mathop{\rm Im}\{\theta \}|<\mathrm{r}_{\max }$, $\hat{V}\left( \theta
\right) $ is a well--defined bounded operator. The \emph{deformed} Howland
operator
\begin{equation}
{\mathcal{\hat{G}}}(\theta ):={\mathcal{\hat{G}}}_{0}(\theta )+i\hat{V}%
\left( \theta \right) \ ,\qquad \theta \in \mathbb{S}\ ,\qquad \mathbb{S}:=%
\mathbb{R+}i[0,\mathrm{r}_{\max })\ ,  \label{Howland deformed}
\end{equation}%
is thus the generator of a strongly continuous semigroup $\{\mathrm{e}^{{%
\mathcal{\hat{G}}}(\theta )\alpha }\}_{\alpha \geq 0}$. Let $C_{\theta }\in
\mathbb{R}_{0}^{+}$ and $D_{\theta }\in \left[ 1,\infty \right) $ be the
stability constants of ${\mathcal{\hat{G}}}(\theta )$, i.e.,
\begin{equation}
\Vert \mathrm{e}^{{\mathcal{\hat{G}}}(\theta )\alpha }\Vert _{\mathcal{\hat{H%
}}}\leq D_{\theta }\mathrm{e}^{\alpha C_{\theta }}\ .  \label{kato constant}
\end{equation}%
The family $\{{\mathcal{\hat{G}}}(\theta )\}_{\theta \in \mathbb{S}%
\backslash \mathbb{R}}$ of closed operators is of type A (Definition \ref%
{Families of type A}). See also \cite{Pillet-Jacblablabis}. This property is
an obvious consequence of the following lemma:

\begin{lemma}[Analyticity of $\hat{V}(\cdot )$]
\label{analyticity V}\mbox{ }\newline
The map $\theta \mapsto \hat{V}\left( \theta \right) $ from $\mathbb{R+}i(-%
\mathrm{r}_{\max },\mathrm{r}_{\max })\subset \mathbb{C}$ to $\mathcal{B}(%
\mathcal{\hat{H}})$ is analytic.
\end{lemma}

\noindent \textit{Proof. }Define, for all $\kappa =\{1,\ldots ,\mathrm{K}\}$
and $\ell \in \{1,\ldots ,m\}$, the maps
\begin{equation*}
\psi _{\ell }^{(\kappa )},\psi _{\ell ,+}^{(\kappa )},\psi _{\ell
,-}^{(\kappa )}:\mathbb{R+}i(-\mathrm{r}_{\max },\mathrm{r}_{\max
})\rightarrow \mathfrak{h}_{2}:=L^{2}(\mathbb{R}\times S^{2},\mathbb{C})
\end{equation*}%
by%
\begin{eqnarray*}
\left[ \psi _{\ell }^{(\kappa )}(\theta )\right] (p,\vartheta ) &:=&\mathrm{g}_{\ell }^{(\kappa )}(p+\theta )\ , \\
\left[ \psi _{\ell ,+}^{(\kappa )}(\theta )\right] (p,\vartheta ) &:=&\mathrm{e}^{\frac{\beta }{2}(p+\theta )}{\mathrm{g}_{\ell }^{(\kappa )\#}}(p+\theta ,\vartheta )\ , \\
\left[ \psi _{\ell ,-}^{(\kappa )}(\theta )\right] (p,\vartheta ) &:=&\mathrm{e}^{-\frac{\beta }{2}(p+\theta )}{\mathrm{g}_{\ell }^{(\kappa )\#}}(p+\theta ,\vartheta )\text{ }.
\end{eqnarray*}%
These three maps are weakly continuous. Indeed, by analyticity of $\mathrm{g}%
_{\ell }^{(\kappa )}$, for any fixed $\psi \in \mathfrak{h}_{2}$, $\theta
,\theta ^{\prime }\in \mathbb{R+}i(-\mathrm{r}_{\max },\mathrm{r}_{\max })$
and some sufficiently small radius $\mathfrak{r}>0$ with $|\theta ^{\prime
}-\theta |<\mathfrak{r}$,
\begin{equation*}
\left\langle \psi ,\psi _{\ell }^{(\kappa )}(\theta ^{\prime })\right\rangle
_{\mathfrak{h}_{2}}=\frac{1}{2\pi }\int_{\mathbb{R}}\mathrm{d}p\int_{S^{2}}%
\mathrm{d}\vartheta \int_{-\pi }^{\pi }\mathrm{d}\varphi \text{ }\psi
(p,\vartheta )\mathrm{g}_{\ell }^{(\kappa )}(p+\theta +\mathfrak{r}\mathrm{e}%
^{i\varphi },\vartheta )\frac{\mathfrak{r}\mathrm{e}^{i\varphi }}{\mathfrak{r%
}\mathrm{e}^{i\varphi }+\theta -\theta ^{\prime }}\ .
\end{equation*}%
By Assumption \ref{d Hardy} and the Cauchy--Schwarz inequality,
\begin{equation}
\int_{-\pi }^{\pi }\mathrm{d}\varphi \int_{S^{2}}\mathrm{d}\vartheta \int_{%
\mathbb{R}}\mathrm{d}p\ |\psi (p,\vartheta )\mathrm{g}_{\ell }^{(\kappa
)}(p+\theta +\mathfrak{r}\mathrm{e}^{i\varphi },\vartheta )|<\infty \text{ }.
\label{borne citroen c4}
\end{equation}%
Thus, by\ the Fubini theorem and the Lebesgue dominated convergence theorem,
the map $\theta \mapsto \langle \psi ,\psi _{\ell }^{(\kappa )}(\theta
)\rangle _{\mathfrak{h}_{2}}$ is continuous on $\mathbb{R+}i(-\mathrm{r}%
_{\max },\mathrm{r}_{\max })$. The weak continuity of $\psi _{\ell
,+}^{(\kappa )}$ and $\psi _{\ell ,-}^{(\kappa )}$ is shown in the same way.
Let $\gamma $ be any closed contour in the domain $\mathbb{R+}i(-\mathrm{r}%
_{\max },\mathrm{r}_{\max })\subset \mathbb{C}$ (with finite length). For
any fixed $\psi \in \mathfrak{h}_{2}$, we infer from a similar estimate as (%
\ref{borne citroen c4}), the Fubini theorem and the analyticity of $\mathrm{g%
}_{\ell }^{(\kappa )}(\cdot ,\vartheta )$ that
\begin{eqnarray*}
\int_{\gamma }\mathrm{d}z\ \langle \psi ,\psi _{\ell }^{(\kappa )}(z)\rangle
_{\mathfrak{h}_{2}} &=&\int_{\gamma }\mathrm{d}z\int_{\mathbb{R}}\mathrm{d}%
p\int_{S^{2}}\mathrm{d}\vartheta \text{ }\psi (p,\vartheta )\mathrm{g}_{\ell
}^{(\kappa )}(p+z,\vartheta ) \\
&=&\int_{\mathbb{R}}\mathrm{d}p\int_{S^{2}}\mathrm{d}\vartheta \text{ }\psi
(p,\vartheta )\int_{\gamma }\mathrm{d}z\text{ }\mathrm{g}_{\ell }^{(\kappa
)}(p+z,\vartheta )=0\ .
\end{eqnarray*}%
Thus, by Morera's lemma, the map $\theta \mapsto \langle \psi ,\psi _{\ell
}^{(\kappa )}(\theta )\rangle _{\mathfrak{h}_{2}}$ is analytic on $\mathbb{R+%
}i(-\mathrm{r}_{\max },\mathrm{r}_{\max })$. In other words, $\psi _{\ell
}^{(\kappa )}$ is weakly analytic and hence, by \cite[Chap. III, Theorem 1.37%
]{Kato}, it is strongly analytic. The (strong) analyticity of $\psi _{\ell
,+}^{(\kappa )},\psi _{\ell ,-}^{(\kappa )}$ is shown by the same arguments.

Finally, observe that
\begin{equation*}
i\mathrm{e}^{-\beta p}\mathrm{g}_{\ell }^{(\kappa )}\left( p\right) =\mathrm{%
e}^{-\frac{\beta }{2}p}{\mathrm{g}_{\ell }^{(\kappa )\#}}\left( p\right) \ ,%
\text{\qquad }i\mathrm{g}_{\ell }^{(\kappa )}(p)=\mathrm{e}^{\frac{\beta }{2}%
p}{\mathrm{g}_{\ell }^{(\kappa )\#}}(p)\text{ }.
\end{equation*}%
From the analyticity of $\psi _{\ell }^{(\kappa )}$, $\psi _{\ell
,+}^{(\kappa )}$, $\psi _{\ell ,-}^{(\kappa )}$ together with the bounds%
\begin{equation*}
\left\Vert a(f)\right\Vert _{\mathcal{H}},\left\Vert a^{+}(f)\right\Vert _{%
\mathcal{H}}\leq \left\Vert f\right\Vert _{\mathfrak{h}_{2}}\ ,\text{\qquad }%
f\in \mathfrak{h}_{2}\text{ },
\end{equation*}%
the linearity of $f\mapsto a^{+}(f)$, the antilinearity of $f\mapsto a(f)$,
and Equations (\ref{Wt}) and (\ref{Wt'}), it then follows that the map $%
\theta \mapsto \hat{V}\left( \theta \right) $ is analytic on the domain $%
\mathbb{R+}i(-\mathrm{r}_{\max },\mathrm{r}_{\max })$, in the sense of $%
\mathcal{B}(\mathcal{\hat{H}})$. \hfill $\Box $

The subspace $\mathcal{\hat{D}}_{0}\subset \mathrm{Dom}({\mathcal{\hat{G}}}%
_{0}(\theta ))$ defined by (\ref{core}) is a core of ${\mathcal{\hat{G}}}%
_{0}(\theta )$ for all $\theta \in \mathbb{C}$. Hence, for all $\theta \in
\mathbb{S}$, by the boundedness of $\hat{V}\left( \theta \right) $, $%
\mathcal{\hat{D}}_{0}$ is also core of ${\mathcal{\hat{G}}}(\theta )$. This
fact implies the following:

\begin{lemma}[Limit of semigroups]
\label{typeAbis}\mbox{ }\newline
For all $\hat{f}\in \mathcal{\hat{H}}$, $\alpha \in \mathbb{R}_{0}^{+}$, and
$\theta \in \mathbb{S}$,
\begin{equation*}
\mathrm{e}^{{\mathcal{\hat{G}}}(\theta )\alpha }\hat{f}=\lim\limits_{\theta
^{\prime }\rightarrow \theta }\left\{ \mathrm{e}^{{\mathcal{\hat{G}}}(\theta
^{\prime })\alpha }\hat{f}\right\} \ .
\end{equation*}%
In particular, for all $\hat{f}\in \mathcal{\hat{H}}$ and $\zeta \in \mathbb{%
C}$ with $\mathrm{\mathop{\rm Re}}\{\zeta \}>C_{\theta }$,
\begin{equation*}
(\zeta -{\mathcal{\hat{G}}}(\theta ))^{-1}\hat{f}=\lim\limits_{\theta
^{\prime }\rightarrow \theta }\left\{ (\zeta -{\mathcal{\hat{G}}}(\theta
))^{-1}\hat{f}\right\} \ .
\end{equation*}
\end{lemma}

\noindent \textit{Proof.} The first assertion follows from the Trotter--Kato
approximation theorem \cite[Chap. III, Theorem 4.8]{Nagel-engel} as ${%
\mathcal{\hat{G}}}(\theta ^{\prime })$ converges strongly on the common core
$\mathcal{\hat{D}}_{0}$\ when $\theta ^{\prime }\rightarrow \theta $, by
Lemma \ref{analyticity V}. The second assertion results from the integral
representation of the resolvent \cite[Chap. III, Eq. (5.18)]{Nagel-engel}:
\begin{equation}
(\zeta -{\mathcal{\hat{G}}}(\theta ))^{-1}\hat{f}=\int_{0}^{\infty }\mathrm{e%
}^{{\mathcal{\hat{G}}}(\theta )\alpha }\mathrm{e}^{-\zeta \alpha }\hat{f}\
\mathrm{d}\alpha \ ,  \label{Laplace transform}
\end{equation}%
for all $\zeta \in \mathbb{C}$ with $\mathop{\rm Re}\{\zeta \}>C_{\theta }$.
\hfill $\Box $

\noindent Recall that $\mathfrak{H}_{\mathrm{at}}$ is defined by (\ref{H at}%
), while $\Omega _{\mathcal{R}}$ is the vacuum of $\mathfrak{H}_{\mathcal{R}%
}:={\mathcal{F}}_{-}(\mathfrak{h}_{2})$ and $\mathfrak{F}$\ denotes the
Fourier transform. Using Lemma \ref{typeAbis}, we prove now that the
dynamics given by $\{\mathrm{e}^{{\mathcal{\hat{G}}}(\theta )\alpha
}\}_{\alpha \geq 0}$ restricted to the atomic space
\begin{equation}
\mathfrak{\hat{H}}_{\mathrm{at}}:=\mathfrak{F}\left( \mathfrak{H}_{\mathrm{at%
}}\otimes \Omega _{\mathcal{R}}\right) \subset \mathcal{\hat{H}}
\label{atomic space fourier}
\end{equation}%
does not depend on the choice of $\theta \in \mathbb{S}$, in the following
sense:

\begin{satz}[Invariance of the evolution semigroup under analytic
translations]
\label{invariance picasseuse}\mbox{ }\newline
For all $\alpha \in \mathbb{R}_{0}^{+}$, $\hat{\psi}_{1},\hat{\psi}_{2}\in
\mathfrak{\hat{H}}_{\mathrm{at}}$ and $\theta ,\theta ^{\prime }\in \mathbb{S%
}$,
\begin{equation*}
\langle \hat{\psi}_{1},\mathrm{e}^{{\mathcal{\hat{G}}}(\theta )\alpha }\hat{%
\psi}_{2}\rangle _{\mathcal{\hat{H}}}=\langle \hat{\psi}_{1},\mathrm{e}^{{%
\mathcal{\hat{G}}}(\theta ^{\prime })\alpha }\hat{\psi}_{2}\rangle _{%
\mathcal{\hat{H}}}\ .
\end{equation*}
\end{satz}

\noindent \textit{Proof}. Using (\ref{Laplace transform}) applied to any
vector $\hat{f}\in \mathfrak{\hat{H}}_{\mathrm{at}}$ and the injectivity of
the Laplace transform, we only need to show that%
\begin{equation}
\langle \hat{\psi}_{1},(\zeta -{\mathcal{\hat{G}}}(\theta ))^{-1}\hat{\psi}%
_{2}\rangle _{\mathcal{\hat{H}}}=\langle \hat{\psi}_{1},(\zeta -{\mathcal{%
\hat{G}}}(\theta ^{\prime }))^{-1}\hat{\psi}_{2}\rangle _{\mathcal{\hat{H}}}
\label{assertion}
\end{equation}%
for any $\hat{\psi}_{1},\hat{\psi}_{2}\in \mathfrak{\hat{H}}_{\mathrm{at}}$,
$\theta ,\theta ^{\prime }\in \mathbb{S}$, and $\zeta \in \left( D,\infty
\right) $ with
\begin{equation*}
D>\sup_{\theta \in \mathbb{S}}C_{\theta }\ ,
\end{equation*}
see (\ref{kato constant}). Indeed,
\begin{equation*}
\sup_{\theta \in \mathbb{S}}C_{\theta }<\infty \ .
\end{equation*}

For any real parameter $\theta \in \mathbb{R}$, let the unitary operator $%
U(\theta )$ be defined by (\ref{unitary U}). Clearly $U(\theta )=U(-\theta
)^{\ast }$ for all $\theta \in \mathbb{R}$ and
\begin{equation}
\forall \theta \in \mathbb{R}:\quad (\zeta -{\mathcal{\hat{G}}}(\theta
))^{-1}=U(\theta )(\zeta -{\mathcal{\hat{G}}}(0))^{-1}U(\theta )^{\ast }\ ,
\label{sympa1}
\end{equation}%
while
\begin{equation}
\forall \theta \in \mathbb{R},\ \hat{\psi}\in \mathfrak{\hat{H}}_{\mathrm{at}%
}:\qquad U(\theta )(\hat{\psi})=\hat{\psi}\ .  \label{sympa2}
\end{equation}%
It follows that the function
\begin{equation*}
g\left( \theta \right) :=\langle \hat{\psi}_{1},(\zeta -{\mathcal{\hat{G}}}%
(\theta ))^{-1}\hat{\psi}_{2}\rangle _{\mathcal{\hat{H}}}
\end{equation*}%
is constant on $\mathbb{R}$, i.e.,
\begin{equation*}
\forall \theta \in \mathbb{R}:\qquad g\left( \theta \right) =g\left(
0\right) \ .
\end{equation*}%
By Lemma \ref{analyticity V}, the family $\{{\mathcal{\hat{G}}}(\theta
)\}_{\theta \in \mathbb{S}\backslash \mathbb{R}}$ of closed operators is of
type A, see Definition \ref{Families of type A}. Therefore, we infer from
Lemma \ref{typeA} that the function $g\ $is analytic on $\mathbb{S}%
\backslash \mathbb{R}$. Finally, using the Schwarz reflection principle, we
deduce that $g$ is constant on $\mathbb{S}$. \hfill $\Box $

Therefore, as soon as the restricted dynamics on the atom is concerned, we
can analyze the evolution given by the strongly continuous semigroup $\{%
\mathrm{e}^{{\mathcal{\hat{G}}}(ir)\alpha }\}_{\alpha \geq 0}$ at a fixed $%
r\in (0,\mathrm{r}_{\max })$. The main advantage of studying $\{\mathrm{e}^{{%
\mathcal{\hat{G}}}(ir)\alpha }\}_{\alpha \geq 0}$ instead of $\{\mathrm{e}^{{%
\mathcal{\hat{G}}}(0)\alpha }=\mathrm{e}^{{\mathcal{\hat{G}}}\alpha
}\}_{\alpha \geq 0}$ is that the continuous spectrum of ${\mathcal{\hat{G}}}$
(coming from the reservoir) is shifted to the left half plane. Indeed, in
contrast to ${\mathcal{\hat{G}}}$, if $r>0$ is sufficiently large, the
generator ${\mathcal{\hat{G}}}(ir)$ has discrete spectrum, as explained
below. The effective atomic Lindbladian defined in (\ref{Lindblad gol}) is
related to Kato's perturbation theory at second order for the discrete
spectrum of ${\mathcal{\hat{G}}}(ir)$ near the origin.

From now on, let $r\in (0,\mathrm{r}_{\max })$. For $\lambda =\eta =0$,
i.e., in absence of pump and atom--reservoir interaction, the discrete
spectrum of the operator ${\mathcal{\hat{G}}}_{0}(ir)$ equals%
\begin{equation*}
\sigma _{d}({\mathcal{\hat{G}}}_{0}(ir))=i\left( \varpi \mathbb{Z+}\sigma (i%
\mathfrak{L}_{\mathrm{at}})\right) \ ,
\end{equation*}%
see (\ref{definition spectre atomique}) and (\ref{definition G non perturbe}%
). The full spectrum of ${\mathcal{\hat{G}}}_{0}(ir)$ is
\begin{equation*}
\sigma ({\mathcal{\hat{G}}}_{0}(ir))=\sigma _{d}({\mathcal{\hat{G}}}%
_{0}(ir))\cup \left\{ i\mathbb{R-}r\mathbb{N}\right\} \ .
\end{equation*}%
In particular, there is a strictly positive gap between the discrete and
essential spectra of ${\mathcal{\hat{G}}}_{0}(ir)$:%
\begin{equation*}
\mathrm{dist}\left( \sigma _{d}({\mathcal{\hat{G}}}_{0}(ir)),\left\{ i%
\mathbb{R-}r\mathbb{N}\right\} \right) =r>0\ .
\end{equation*}

For $r\in \lbrack 0,\mathrm{r}_{\max })$, let $\mathcal{J}\in \mathcal{B}(%
\mathcal{\hat{H}})$ and $\mathcal{I}(ir)\in \mathcal{B}(\mathcal{\hat{H}})$
be such that%
\begin{equation}
{\mathcal{\hat{G}}}(ir):={\mathcal{\hat{G}}}_{0}(ir)+i\hat{V}\left(
ir\right) ={\mathcal{\hat{G}}}_{0}(ir)+\eta \mathcal{J}+\lambda \mathcal{I}%
(ir)  \label{Howland deformedbis}
\end{equation}%
for any $\lambda ,\eta \in \mathbb{R}$, see (\ref{Howland deformed}). I.e., $%
\eta \mathcal{J}:=\hat{V}\left( ir\right) |_{\lambda =0}$ and $\lambda
\mathcal{I}(ir):=\hat{V}\left( ir\right) |_{\eta =0}$ are the interaction
parts of ${\mathcal{\hat{G}}}(ir)$ respectively related to the pump and the
atom--reservoir interaction. By Kato's perturbation theory for the discrete
spectrum, if $\left\vert \lambda \right\vert ,\left\vert \eta \right\vert $
are small, the deformed Howland generator ${\mathcal{\hat{G}}}(ir)$ has
discrete spectrum.

Indeed, for $r\in (0,\mathrm{r}_{\max })$, let $\mathfrak{r}\in (0,1/2)\cap
(0,r)$ be such that
\begin{equation}
\forall p\in \mathbb{Z},\ \epsilon \in \sigma (i\mathfrak{L}_{\mathrm{at}%
}):\qquad \mathrm{dist}\left( i(p\varpi +\epsilon ),\sigma ({\mathcal{\hat{G}%
}}_{0}(ir))\backslash \{i(p\varpi +\epsilon )\}\right) \geq 2\mathfrak{r}
\label{gamma oo}
\end{equation}%
and, for $p\in \mathbb{Z}$, $\epsilon \in \sigma (i\mathfrak{L}_{\mathrm{at}%
})$, the contour $\gamma _{p,\epsilon }$ be defined by
\begin{equation}
\gamma _{p,\epsilon }\left( y\right) :=i\left( p\varpi +\epsilon \right) +%
\mathfrak{r}\mathrm{e}^{2\pi iy}\in \mathbb{C}\ ,\qquad y\in \lbrack 0,1]\ .
\label{gammak0}
\end{equation}%
Then, for every $\eta \in \mathbb{R}$, $p\in \mathbb{Z}$, $\epsilon \in
\sigma (i\mathfrak{L}_{\mathrm{at}})$, and sufficiently small $\left\vert
\lambda \right\vert $, the operator
\begin{equation}
\mathcal{P}_{p,\epsilon }^{\left( \lambda ,\eta \right) }:=\frac{1}{2\pi i}%
\oint\limits_{\gamma _{p,\epsilon }}(\zeta -{\mathcal{\hat{G}}}(ir))^{-1}%
\mathrm{d}\zeta  \label{Kato projection1}
\end{equation}%
is the well--known Riesz projection associated with ${\mathcal{\hat{G}}}(ir)$
and the discrete eigenvalue $i\left( p\varpi +\epsilon \right) $ of ${%
\mathcal{\hat{G}}}_{0}(ir)$. Define also%
\begin{equation*}
\mathcal{\bar{P}}_{p,\epsilon }^{\left( \lambda ,\eta \right) }:=\mathbf{1}_{%
\mathcal{\hat{H}}}-\mathcal{P}_{p,\epsilon }^{\left( \lambda ,\eta \right)
}\ ,\qquad \lambda ,\eta \in \mathbb{R},\ p\in \mathbb{Z},\ \epsilon \in
\sigma (i\mathfrak{L}_{\mathrm{at}})\ .
\end{equation*}

If $\zeta \in \mathbb{C}$ is in the spectrum of ${\mathcal{\hat{G}}}_{0}(ir)$
and thus $(\zeta -{\mathcal{\hat{G}}}_{0}(ir))$ is not a bijective map from $%
\mathrm{Dom}({\mathcal{\hat{G}}}_{0}(ir))$ to $\mathcal{\hat{H}}$, the
inverse $(\zeta -{\mathcal{\hat{G}}}_{0}(ir))^{-1}$ will be understood below
in the sense of multi--valued functions. In fact, observe that with this
convention expressions of the form%
\begin{equation*}
\mathcal{\bar{P}}_{p,\epsilon }^{\left( 0,0\right) }(i(p\varpi +\epsilon )-{%
\mathcal{\hat{G}}}_{0}(ir))^{-1}\mathcal{\bar{P}}_{p,\epsilon }^{\left(
0,0\right) }
\end{equation*}%
with $p\in \mathbb{Z}$, $\epsilon \in \sigma (i\mathfrak{L}_{\mathrm{at}})$
define single--valued linear maps. Recall that $p\in \mathbb{Z}$, $\epsilon
\in \sigma (i\mathfrak{L}_{\mathrm{at}})$ are eigenvalues of ${\mathcal{\hat{%
G}}}_{0}(ir)$ and thus $(i(p\varpi +\epsilon )-{\mathcal{\hat{G}}}%
_{0}(ir))^{-1}$ is not single--valued in this case.

\begin{lemma}[Perturbative expansions of the deformed Howland operator]
\label{pert G}\mbox{ }\newline
Let $r\in (0,\mathrm{r}_{\max })$. For all $p\in \mathbb{Z}$, $\epsilon \in
\sigma (i\mathfrak{L}_{\mathrm{at}})$ and sufficiently small $\left\vert
\lambda \right\vert $,%
\begin{eqnarray*}
\mathcal{P}_{p,\epsilon }^{\left( \lambda ,\eta \right) }{\mathcal{\hat{G}}}%
(ir)\mathcal{P}_{p,\epsilon }^{\left( \lambda ,\eta \right) } &=&i\left(
p\varpi +\epsilon \right) \mathcal{P}_{p,\epsilon }^{\left( 0,0\right)
}+\eta \ \mathcal{P}_{p,\epsilon }^{\left( 0,0\right) }\ \mathcal{J}\
\mathcal{P}_{p,\epsilon }^{\left( 0,0\right) } \\
&&+\lambda ^{2}\ \mathcal{P}_{p,\epsilon }^{\left( 0,0\right) }\ \mathcal{I}%
(ir)\ \mathcal{\bar{P}}_{p,\epsilon }^{\left( 0,0\right) }\big(i(p\varpi
+\epsilon )-{\mathcal{\hat{G}}}_{0}(ir)\big)^{-1}\mathcal{\bar{P}}%
_{p,\epsilon }^{\left( 0,0\right) }\ \mathcal{I}(ir)\ \mathcal{P}%
_{p,\epsilon }^{\left( 0,0\right) } \\
&&+\lambda ^{3}\ \mathrm{R}
\end{eqnarray*}%
with $\mathrm{R}\equiv \mathrm{R}^{(p,\epsilon ,\lambda ,\eta )}$ being an
operator with norm $\left\Vert \mathrm{R}\right\Vert _{\mathcal{B}(\mathcal{%
\hat{H}})}\leq C$ for some finite constant $C\in \mathbb{R}^{+}$ not
depending on $p$, $\epsilon $, $\lambda $ and $\eta $.
\end{lemma}

\noindent \textit{Proof.} We fix w.l.o.g. $p=\epsilon =0$. To simplify
notation, in all the proof we denote by $\mathrm{R}\equiv \mathrm{R}%
^{(\lambda ,\eta )}$ any operator with norm $\Vert \mathrm{R}\Vert \leq C$
for some fixed constant $C\in \mathbb{R}^{+}$ not depending on $\lambda $, $%
\eta $. Note that the operator $\mathrm{R}$ does not need to be the same
from one statement to another.

Assumption \ref{assumption important} yields, at small $|\lambda |$, $\Vert
\hat{V}\left( ir\right) \Vert \leq C|\lambda |$, see (\ref{Howland
deformedbis}). Hence, if $\left\vert \lambda \right\vert $ is sufficiently
small then the resolvent $(\zeta -{\mathcal{\hat{G}}}(ir))^{-1}$ equals the
absolutely convergent Neumann series
\begin{equation}
(\zeta -{\mathcal{\hat{G}}}(ir))^{-1}=\underset{n=0}{\overset{\infty }{\sum }%
}(\zeta -{\mathcal{\hat{G}}}_{0}(ir))^{-1}\left\{ \left( \eta \mathcal{J}%
+\lambda \mathcal{I}(ir)\right) (\zeta -{\mathcal{\hat{G}}}%
_{0}(ir))^{-1}\right\} ^{n}  \label{Neumann series1}
\end{equation}%
for all $\zeta \in \gamma _{0,0}$. By (\ref{Kato projection1}) and
Assumption \ref{assumption important}, it follows that%
\begin{equation}
\mathcal{P}_{0,0}^{\left( \lambda ,\eta \right) }=\mathcal{P}_{0,0}^{\left(
0,0\right) }+\eta P_{\mathrm{p}}^{(1)}+\lambda P_{\mathrm{at},\mathcal{R}%
}^{(1)}+\lambda ^{2}P_{\mathrm{at},\mathcal{R}}^{(2)}+\lambda ^{3}\mathrm{R}%
\ ,  \label{Neumann series2}
\end{equation}%
where
\begin{eqnarray*}
P_{\mathrm{p}}^{(1)} &:=&\frac{1}{2\pi i}\oint\limits_{\gamma _{0,0}}(\zeta -{\mathcal{\hat{G}}}_{0}(ir))^{-1}\mathcal{J}(\zeta -{\mathcal{\hat{G}}}_{0}(ir))^{-1}\mathrm{d}\zeta \ , \\
P_{\mathrm{at},\mathcal{R}}^{(1)} &:=&\frac{1}{2\pi i}\oint\limits_{\gamma
_{0,0}}(\zeta -{\mathcal{\hat{G}}}_{0}(ir))^{-1}\mathcal{I}(ir)(\zeta -{\mathcal{\hat{G}}}_{0}(ir))^{-1}\mathrm{d}\zeta \ , \\
P_{\mathrm{at},\mathcal{R}}^{(2)} &:=&\frac{1}{2\pi i}\oint\limits_{\gamma
_{0,0}}(\zeta -{\mathcal{\hat{G}}}_{0}(ir))^{-1}\left\{ \mathcal{I}(ir)(\zeta -{\mathcal{\hat{G}}}_{0}(ir))^{-1}\right\} ^{2}\mathrm{d}\zeta \ .
\end{eqnarray*}%
We infer from (\ref{Neumann series1})--(\ref{Neumann series2}) that
\begin{eqnarray}
&&\mathcal{P}_{0,0}^{\left( \lambda ,\eta \right) }{\mathcal{\hat{G}}}(ir)%
\mathcal{P}_{0,0}^{\left( \lambda ,\eta \right) }  \label{Neumann series3} \\
&=&\left( \mathcal{P}_{0,0}^{\left( 0,0\right) }+\eta P_{\mathrm{p}%
}^{(1)}+\lambda P_{\mathrm{at},\mathcal{R}}^{(1)}+\lambda ^{2}P_{\mathrm{at},%
\mathcal{R}}^{(2)}\right) {\mathcal{\hat{G}}}(ir)\left( \mathcal{P}%
_{0,0}^{\left( 0,0\right) }+\eta P_{\mathrm{p}}^{(1)}+\lambda P_{\mathrm{at},%
\mathcal{R}}^{(1)}+\lambda ^{2}P_{\mathrm{at},\mathcal{R}}^{(2)}\right)
\notag \\
&&+\lambda ^{3}\mathrm{R}\ .  \notag
\end{eqnarray}%
Note that, by Assumption \ref{assumption important}, $\eta =O(\lambda ^{2})$%
. Thus, using the equality $\mathcal{P}_{0,0}^{\left( 0,0\right) }{\mathcal{%
\hat{G}}}_{0}(ir)=0$ and (\ref{Howland deformedbis}) we obtain
\begin{eqnarray}
\left( \mathcal{P}_{0,0}^{\left( 0,0\right) }+\eta P_{\mathrm{p}%
}^{(1)}+\lambda P_{\mathrm{at},\mathcal{R}}^{(1)}+\lambda ^{2}P_{\mathrm{at},%
\mathcal{R}}^{(2)}\right) {\mathcal{\hat{G}}}(ir) &=&\mathcal{P}%
_{0,0}^{\left( 0,0\right) }(\eta \mathcal{J}+\lambda \mathcal{I}(ir))  \notag
\\
&&+(\eta P_{\mathrm{p}}^{(1)}+\lambda P_{\mathrm{at},\mathcal{R}%
}^{(1)}+\lambda ^{2}P_{\mathrm{at},\mathcal{R}}^{(2)}){\mathcal{\hat{G}}}%
_{0}(ir)  \notag \\
&&+\lambda ^{2}P_{\mathrm{at},\mathcal{R}}^{(1)}\mathcal{I}(ir)+\lambda ^{3}%
\mathrm{R}\ .  \label{Neumann series4}
\end{eqnarray}%
Furthermore, from Assumption \ref{assumption important 0} and similar
analyticity arguments as used in the proof of Theorem \ref{invariance
picasseuse}, one gets that
\begin{equation}
{\mathcal{\hat{G}}}_{0}(ir)\mathcal{P}_{0,0}^{\left( 0,0\right) }=0\quad
\text{and}\quad \mathcal{P}_{0,0}^{\left( 0,0\right) }\mathcal{I}(ir)%
\mathcal{P}_{0,0}^{\left( 0,0\right) }=0\ .  \label{equality sympa}
\end{equation}%
We deduce from (\ref{Neumann series3})--(\ref{equality sympa}) that%
\begin{eqnarray}
\mathcal{P}_{0,0}^{\left( \lambda ,\eta \right) }{\mathcal{\hat{G}}}(ir)%
\mathcal{P}_{0,0}^{\left( \lambda ,\eta \right) } &=&\eta \mathcal{P}%
_{0,0}^{\left( 0,0\right) }\mathcal{JP}_{0,0}^{\left( 0,0\right) }+\lambda
^{2}\mathcal{P}_{0,0}^{\left( 0,0\right) }\mathcal{I}(ir)P_{\mathrm{at},%
\mathcal{R}}^{(1)}  \notag \\
&&+\lambda ^{2}P_{\mathrm{at},\mathcal{R}}^{(1)}\mathcal{I}(ir)\mathcal{P}%
_{0,0}^{\left( 0,0\right) }+\lambda ^{2}P_{\mathrm{at},\mathcal{R}}^{(1)}{%
\mathcal{\hat{G}}}_{0}(ir)P_{\mathrm{at},\mathcal{R}}^{(1)}  \notag \\
&&+\lambda ^{3}\mathrm{R}\ .  \label{Neumann series5}
\end{eqnarray}%
Since $\mathcal{P}_{0,0}^{\left( \lambda ,\eta \right) }$ is a projection,
obviously,%
\begin{equation*}
\mathcal{P}_{0,0}^{\left( \lambda ,\eta \right) }{\mathcal{\hat{G}}}(ir)%
\mathcal{P}_{0,0}^{\left( \lambda ,\eta \right) }=\mathcal{P}_{0,0}^{\left(
\lambda ,\eta \right) }\left( \mathcal{P}_{0,0}^{\left( \lambda ,\eta
\right) }{\mathcal{\hat{G}}}(ir)\mathcal{P}_{0,0}^{\left( \lambda ,\eta
\right) }\right) \mathcal{P}_{0,0}^{\left( \lambda ,\eta \right) }
\end{equation*}%
and, by (\ref{Neumann series2}) and (\ref{Neumann series5}),
\begin{eqnarray}
\mathcal{P}_{0,0}^{\left( \lambda ,\eta \right) }{\mathcal{\hat{G}}}(ir)%
\mathcal{P}_{0,0}^{\left( \lambda ,\eta \right) } &=&\eta \mathcal{P}%
_{0,0}^{\left( 0,0\right) }\mathcal{JP}_{0,0}^{\left( 0,0\right) }+\lambda
^{2}\mathcal{P}_{0,0}^{\left( 0,0\right) }\mathcal{I}(ir)P_{\mathrm{at},%
\mathcal{R}}^{(1)}\mathcal{P}_{0,0}^{\left( 0,0\right) }+\lambda ^{2}%
\mathcal{P}_{0,0}^{\left( 0,0\right) }P_{\mathrm{at},\mathcal{R}}^{(1)}%
\mathcal{I}(ir)\mathcal{P}_{0,0}^{\left( 0,0\right) }  \notag \\
&&+\lambda ^{2}\mathcal{P}_{0,0}^{\left( 0,0\right) }P_{\mathrm{at},\mathcal{%
R}}^{(1)}{\mathcal{\hat{G}}}_{0}(ir)P_{\mathrm{at},\mathcal{R}}^{(1)}%
\mathcal{P}_{0,0}^{\left( 0,0\right) }+\lambda ^{3}\mathrm{R}\ .
\label{Neumann series5bis}
\end{eqnarray}%
Now, using again $\mathcal{P}_{0,0}^{\left( 0,0\right) }{\mathcal{\hat{G}}}%
_{0}(ir)=0$ and (\ref{equality sympa}) we observe that%
\begin{eqnarray}
&&\mathcal{P}_{0,0}^{\left( 0,0\right) }P_{\mathrm{at},\mathcal{R}}^{(1)}{%
\mathcal{\hat{G}}}_{0}(ir)P_{\mathrm{at},\mathcal{R}}^{(1)}\mathcal{P}%
_{0,0}^{\left( 0,0\right) }  \label{Neumann series6} \\
&=&\frac{1}{2\pi i}\oint\limits_{\gamma _{0,0}}\Big\{\zeta _{1}^{-1}\mathcal{%
P}_{0,0}^{\left( 0,0\right) }\mathcal{I}(ir)(\zeta _{1}-{\mathcal{\hat{G}}}%
_{0}(ir))^{-1}\frac{1}{2\pi i}\oint\limits_{\gamma _{0,0}}{\mathcal{\hat{G}}}%
_{0}(ir)(\zeta _{2}-{\mathcal{\hat{G}}}_{0}(ir))^{-1}\mathcal{I}(ir)\mathcal{%
P}_{0,0}^{\left( 0,0\right) }\zeta _{2}^{-1}\mathrm{d}\zeta _{2}\Big\}%
\mathrm{d}\zeta _{1}\ ,  \notag
\end{eqnarray}%
while, by (\ref{equality sympa}),
\begin{equation}
\mathcal{I}(ir)\mathcal{P}_{0,0}^{\left( 0,0\right) }=(\mathbf{1}_{\mathcal{%
\hat{H}}}-\mathcal{P}_{0,0}^{\left( 0,0\right) })\mathcal{I}(ir)\mathcal{P}%
_{0,0}^{\left( 0,0\right) }\quad \text{and}\quad \mathcal{P}_{0,0}^{\left(
0,0\right) }\mathcal{I}(ir)=\mathcal{P}_{0,0}^{\left( 0,0\right) }\mathcal{I}%
(ir)(\mathbf{1}_{\mathcal{\hat{H}}}-\mathcal{P}_{0,0}^{\left( 0,0\right) })\
.  \label{Neumann series7}
\end{equation}%
By analyticity of the map%
\begin{equation*}
\zeta _{2}\mapsto {\mathcal{\hat{G}}}_{0}(ir)(\zeta _{2}-{\mathcal{\hat{G}}}%
_{0}(ir))^{-1}(\mathbf{1}_{\mathcal{\hat{H}}}-\mathcal{P}_{0,0}^{\left(
0,0\right) })\mathcal{I}(ir)\mathcal{P}_{0,0}^{\left( 0,0\right) }\ ,
\end{equation*}%
Equations (\ref{Neumann series6}) and (\ref{Neumann series7}) together imply
that%
\begin{eqnarray}
&&\mathcal{P}_{0,0}^{\left( 0,0\right) }P_{\mathrm{at},\mathcal{R}}^{(1)}{%
\mathcal{\hat{G}}}_{0}(ir)P_{\mathrm{at},\mathcal{R}}^{(1)}\mathcal{P}%
_{0,0}^{\left( 0,0\right) }  \label{Neumann series8} \\
&=&-\frac{1}{2\pi i}\oint\limits_{\gamma _{0,0}}\left\{ \zeta ^{-1}\mathcal{P%
}_{0,0}^{\left( 0,0\right) }\mathcal{I}(ir)(\mathbf{1}_{\mathcal{\hat{H}}}-%
\mathcal{P}_{0,0}^{\left( 0,0\right) })(\zeta -{\mathcal{\hat{G}}}%
_{0}(ir))^{-1}(\mathbf{1}_{\mathcal{\hat{H}}}-\mathcal{P}_{0,0}^{\left(
0,0\right) })\mathcal{I}(ir)\mathcal{P}_{0,0}^{\left( 0,0\right) }\right\}
\mathrm{d}\zeta \ .  \notag
\end{eqnarray}%
We also remark that (\ref{Neumann series7}) together with $\mathcal{P}%
_{0,0}^{\left( 0,0\right) }{\mathcal{\hat{G}}}_{0}(ir)=0$ and (\ref{equality
sympa}) yields%
\begin{eqnarray}
\mathcal{P}_{0,0}^{\left( 0,0\right) }P_{\mathrm{at},\mathcal{R}}^{(1)}%
\mathcal{I}(ir)\mathcal{P}_{0,0}^{\left( 0,0\right) } &=&\frac{1}{2\pi i}%
\oint\limits_{\gamma _{0,0}}\left\{ \zeta ^{-1}\mathcal{P}_{0,0}^{\left(
0,0\right) }\mathcal{I}(ir)(\mathbf{1}_{\mathcal{\hat{H}}}-\mathcal{P}%
_{0,0}^{\left( 0,0\right) })(\zeta -{\mathcal{\hat{G}}}_{0}(ir))^{-1}(%
\mathbf{1}_{\mathcal{\hat{H}}}-\mathcal{P}_{0,0}^{\left( 0,0\right) })%
\mathcal{I}(ir)\mathcal{P}_{0,0}^{\left( 0,0\right) }\right\} \mathrm{d}\zeta
\notag \\
&=&\mathcal{P}_{0,0}^{\left( 0,0\right) }\mathcal{I}(ir)P_{\mathrm{at},%
\mathcal{R}}^{(1)}\mathcal{P}_{0,0}^{\left( 0,0\right) }.
\label{Neumann series9}
\end{eqnarray}%
By analyticity of the map%
\begin{equation*}
\zeta \mapsto \mathcal{P}_{0,0}^{\left( 0,0\right) }\mathcal{I}(ir)(\mathbf{1%
}_{\mathcal{\hat{H}}}-\mathcal{P}_{0,0}^{\left( 0,0\right) })(\zeta -{%
\mathcal{\hat{G}}}_{0}(ir))^{-1}(\mathbf{1}_{\mathcal{\hat{H}}}-\mathcal{P}%
_{0,0}^{\left( 0,0\right) })\mathcal{I}(ir)\mathcal{P}_{0,0}^{\left(
0,0\right) },
\end{equation*}%
we then infer from (\ref{Neumann series8})--(\ref{Neumann series9}) that%
\begin{eqnarray*}
&&\mathcal{P}_{0,0}^{\left( 0,0\right) }\mathcal{I}(ir)P_{\mathrm{at},%
\mathcal{R}}^{(1)}\mathcal{P}_{0,0}^{\left( 0,0\right) }+\mathcal{P}%
_{0,0}^{\left( 0,0\right) }P_{\mathrm{at},\mathcal{R}}^{(1)}\mathcal{I}(ir)%
\mathcal{P}_{0,0}^{\left( 0,0\right) }+\mathcal{P}_{0,0}^{\left( 0,0\right)
}P_{\mathrm{at},\mathcal{R}}^{(1)}{\mathcal{\hat{G}}}_{0}(ir)P_{\mathrm{at},%
\mathcal{R}}^{(1)}\mathcal{P}_{0,0}^{\left( 0,0\right) } \\
&=&\mathcal{P}_{0,0}^{\left( 0,0\right) }\mathcal{I}(ir)(\mathbf{1}_{%
\mathcal{\hat{H}}}-\mathcal{P}_{0,0}^{\left( 0,0\right) })(i0-{\mathcal{\hat{%
G}}}_{0}(ir))^{-1}(\mathbf{1}_{\mathcal{\hat{H}}}-\mathcal{P}_{0,0}^{\left(
0,0\right) })\mathcal{I}(ir)\mathcal{P}_{0,0}^{\left( 0,0\right) }\ .
\end{eqnarray*}%
Using this and (\ref{Neumann series5bis}) we arrive at the assertion for $%
p=\epsilon =0$.

Up to some obvious changes in the above arguments, the general case with $%
p\in \mathbb{Z}$ and $\epsilon \in \sigma (i\mathfrak{L}_{\mathrm{at}})$ is
proven in the same way. Note only that $\mathrm{R}$ is an operator with norm
$\left\Vert \mathrm{R}\right\Vert _{\mathcal{B}(\mathcal{\hat{H}})}\leq C$
for some finite constant $C$ that does not depend on $p$, $\epsilon $
because of (\ref{gamma oo})--(\ref{gammak0}).\hfill $\Box $

Similar to the atom--reservoir Lindbladian $\mathfrak{L}_{\mathcal{R}%
}^{(\varepsilon )}\in \mathcal{B}(\mathfrak{H}_{\mathrm{at}})$ (\ref%
{Lindblad gol}), we define the operator $\mathcal{L}_{\mathcal{R}%
}^{(\varepsilon ,r,p)}\in \mathcal{B}(\mathcal{\hat{H}})$ by%
\begin{equation}
\mathcal{L}_{\mathcal{R}}^{(\varepsilon ,r,p)}:=\sum_{\epsilon \in \sigma (i%
\mathfrak{L}_{\mathrm{at}})}\left[ \mathcal{P}_{p,\epsilon }^{\left(
0,0\right) }\ \mathcal{I}(ir)\ \mathcal{\bar{P}}_{p,\epsilon }^{\left(
0,0\right) }\big(\varepsilon +i(p\varpi +\epsilon )-{\mathcal{\hat{G}}}%
_{0}(ir)\big)^{-1}\mathcal{\bar{P}}_{p,\epsilon }^{\left( 0,0\right) }\
\mathcal{I}(ir)\ \mathcal{P}_{p,\epsilon }^{\left( 0,0\right) }\right]
^{\ast }  \label{atom reservoir louvi}
\end{equation}%
for any $\varepsilon \in \mathbb{R}_{0}^{+}$, $r\in \lbrack 0,\mathrm{r}%
_{\max })$ and $p\in \mathbb{Z}$. This operator has the following important
properties:

\begin{lemma}[Properties of the operator $\mathcal{L}_{\mathcal{R}}^{(%
\protect\varepsilon ,r,p)}$]
\label{pert G copy(1)}\mbox{ }\newline
For any $p\in \mathbb{Z}$, $\varepsilon \in \mathbb{R}_{0}^{+}$ and $%
r_{1},r_{2}\in \lbrack 0,\mathrm{r}_{\max })$,
\begin{equation*}
\mathcal{L}_{\mathcal{R}}^{(\varepsilon ,r_{1},p)}=\mathcal{L}_{\mathcal{R}%
}^{(\varepsilon ,r_{2},p)} \ .
\end{equation*}%
Moreover, in the sense of $\mathcal{B}(\mathcal{\hat{H}})$,
\begin{equation*}
\mathcal{L}_{\mathcal{R}}^{(p)}:=\mathcal{L}_{\mathcal{R}}^{(0,0,p)}=\lim_{%
\varepsilon \searrow 0}\mathcal{L}_{\mathcal{R}}^{(\varepsilon ,0,p)}\ .
\end{equation*}
\end{lemma}

\noindent \textit{Proof.} Fix $p\in \mathbb{Z}$ and $\varepsilon \in \mathbb{%
R}_{0}^{+}$. By using similar analyticity arguments to those used in the
proof of Theorem \ref{invariance picasseuse}, one shows that $(\mathcal{L}_{%
\mathcal{R}}^{(\varepsilon ,r_{1},p)})^{\ast }=(\mathcal{L}_{\mathcal{R}%
}^{(\varepsilon ,r_{2},p)})^{\ast }$ for all $r_{1},r_{2}\in \lbrack 0,%
\mathrm{r}_{\max })$, which is equivalent to the first assertion. Choose now
$r\in (0,\mathrm{r}_{\max })$. Then, by the first part of the lemma, $%
\mathcal{L}_{\mathcal{R}}^{(\varepsilon ,0,p)}=\mathcal{L}_{\mathcal{R}%
}^{(\varepsilon ,r,p)}$ for all $\varepsilon \in \mathbb{R}_{0}^{+}$. For $%
r>0$,%
\begin{equation*}
\lim_{\varepsilon \searrow 0}\mathcal{\bar{P}}_{p,\epsilon }^{\left(
0,0\right) }\big(\varepsilon +i(p\varpi +\epsilon )-{\mathcal{\hat{G}}}%
_{0}(ir)\big)^{-1}\mathcal{\bar{P}}_{p,\epsilon }^{\left( 0,0\right) }=%
\mathcal{\bar{P}}_{p,\epsilon }^{\left( 0,0\right) }\big(i(p\varpi +\epsilon
)-{\mathcal{\hat{G}}}_{0}(ir)\big)^{-1}\mathcal{\bar{P}}_{p,\epsilon
}^{\left( 0,0\right) }\ ,
\end{equation*}%
in the sense of $\mathcal{B}(\mathcal{\hat{H}})$ and the second assertion
follows.\hfill $\Box $

\noindent We show below that $\mathcal{L}_{\mathcal{R}}^{(p)}$ acts -- up to
an equivalence transformation -- as the Lindbladian $\mathfrak{L}_{\mathcal{R%
}}$ defined in (\ref{Lindblad gol}).

Recall that the eigenspaces of the atomic Hamiltonian $H_{\mathrm{at}}\in
\mathcal{B}(\mathbb{C}^{d})$ associated with the eigenvalues $E_{k}$, for $%
k\in \{1,\ldots ,N\}$, and their dimensions are denoted by ${\mathcal{H}}%
_{k}\subset \mathbb{C}^{d}$ and $n_{k}\in \mathbb{N}$, respectively. By
taking any arbitrary orthonormal basis $\{e_{n}^{\left( k\right)
}\}_{n=1}^{n_{k}}$ of ${\mathcal{H}}_{k}$ for each $k\in \{1,\ldots ,N\}$,
we define the elements
\begin{equation*}
W_{\left( k,n\right) }^{\left( k^{\prime },n^{\prime }\right) }\in \mathfrak{%
H}_{\mathrm{at}}\equiv \mathcal{B}(\mathbb{C}^{d})
\end{equation*}%
for any $k,k^{\prime }\in \{1,\ldots ,N\}$, $n\in \{1,\ldots ,n_{k}\}$ and $%
n^{\prime }\in \{1,\ldots ,n_{k^{\prime }}\}$ by the condition%
\begin{equation}
\forall k^{\prime \prime }\in \{1,\ldots ,N\},\ n^{\prime \prime }\in
\{1,\ldots ,n_{k^{\prime \prime }}\}:\qquad W_{\left( k,n\right) }^{\left(
k^{\prime },n^{\prime }\right) }(e_{n^{\prime \prime }}^{\left( k^{\prime
\prime }\right) })=\delta _{n,n^{\prime \prime }}\delta _{k,k^{\prime \prime
}}e_{n^{\prime }}^{\left( k^{\prime }\right) }\ .  \label{def sup1}
\end{equation}%
Then, for any $p\in \mathbb{Z}$ and $\epsilon \in \sigma (i\mathfrak{L}_{%
\mathrm{at}})$, straightforward computations show that%
\begin{equation}
\mathrm{ran}(\mathcal{P}_{p,\epsilon }^{\left( 0,0\right) })=\mathrm{span}%
\Big\{\mathfrak{F}%
\big(%
\mathrm{e}^{it\varpi m}(W_{\left( k,n\right) }^{\left( k^{\prime },n^{\prime
}\right) }\rho _{\mathrm{at}}^{1/2})\otimes \Omega _{\mathcal{R}}%
\big)%
:(m,k,k^{\prime })\in \mathbb{O}_{p,\epsilon },\ n\in \{1,\ldots
,n_{k}\},\;n^{\prime }\in \{1,\ldots ,n_{k^{\prime }}\}\Big\}
\label{eq penible}
\end{equation}%
with
\begin{equation*}
\mathbb{O}_{p,\epsilon }:=\left\{ (m,k,k^{\prime })\in \mathbb{Z}\times
\{1,\ldots ,N\}^{2}:m\varpi +E_{k^{\prime }}-E_{k}=\varpi p+\epsilon
\right\} \ .
\end{equation*}%
Recall that $\rho _{\mathrm{at}}\in \mathcal{B}(\mathbb{C}^{d})$ is the
density matrix of the initial state of the atom. The range $\mathrm{ran}(%
\mathcal{P}_{p,\epsilon }^{\left( 0,0\right) })$ of the Riesz projection $%
\mathcal{P}_{p,\epsilon }^{\left( 0,0\right) }$ does obviously not belong to
the atomic space $\mathfrak{\hat{H}}_{\mathrm{at}}$ (\ref{atomic space
fourier}). We can remove oscillating terms by using a unitary map $\mathrm{U}%
_{p,\epsilon }$ from $\mathrm{ran}(\mathcal{P}_{p,\epsilon }^{\left(
0,0\right) })$ to the atomic subspace
\begin{eqnarray}
\mathfrak{H}_{\mathrm{at}}^{(p,\epsilon )} &:=&\mathrm{span}\Big\{W_{\left(
k,n\right) }^{\left( k^{\prime },n^{\prime }\right) }\rho _{\mathrm{at}}^{1/2}:(k,k^{\prime })\in \{1,\ldots ,N\}^{2},\ n\in \{1,\ldots
,n_{k}\},\;n^{\prime }\in \{1,\ldots ,n_{k^{\prime }}\},  \notag \\
&&\qquad \qquad \qquad \qquad \qquad \qquad \qquad \qquad \qquad \exists m \text{ with }(m,k,k^{\prime
}) \in \mathbb{O}_{p,\epsilon }\Big\}\subset \mathfrak{H}_{\mathrm{at}}
\label{space 0}
\end{eqnarray}%
for $p\in \mathbb{Z}$ and $\epsilon \in \sigma (i\mathfrak{L}_{\mathrm{at}})$
as follows:
\begin{equation}
\mathrm{U}_{p,\epsilon }\Big(\mathfrak{F}%
\big(%
\mathrm{e}^{it\varpi m}(W_{\left( k,n\right) }^{\left( k^{\prime },n^{\prime
}\right) }\rho _{\mathrm{at}}^{1/2})\otimes \Omega _{\mathcal{R}}%
\big)%
\Big):=W_{\left( k,n\right) }^{\left( k^{\prime },n^{\prime }\right) }\rho _{%
\mathrm{at}}^{1/2}\in \mathfrak{H}_{\mathrm{at}}^{(p,\epsilon )}
\label{unitary transform}
\end{equation}%
for any $k,k^{\prime }\in \{1,\ldots ,N\}$, $n\in \{1,\ldots ,n_{k}\}$ and $%
n^{\prime }\in \{1,\ldots ,n_{k^{\prime }}\}$. Let $\mathfrak{S}$ be the
bounded self--adjoint operator on $\mathcal{B}(\mathcal{\hat{H}})$ defined by%
\begin{equation}
\big(\mathfrak{S}\hat{f}\big)(k):=\frac{1}{2}(\hat{f}(k+1)+\hat{f}(k-1))\
,\qquad k\in \mathbb{Z}\ .  \label{sigma}
\end{equation}

Using the identification above of $\mathrm{ran}(\mathcal{P}_{p,\epsilon
}^{\left( 0,0\right) })$ and $\mathfrak{H}_{\mathrm{at}}^{(p,\epsilon
)}\subset \mathfrak{H}_{\mathrm{at}}$, we can establish a relation between
the deformed Howland operator ${\mathcal{\hat{G}}}(ir)$ and the Lindbladians
$\mathfrak{L}_{\mathrm{p}}$ (\ref{L pump}) and $\mathfrak{L}_{\mathcal{R}}$ (%
\ref{Lindblad gol}) via the adjoints $\mathfrak{L}_{\mathrm{p}}^{\ast }$, $%
\mathfrak{L}_{\mathcal{R}}^{\ast }$ w.r.t. the scalar product of $\mathfrak{H%
}_{\mathrm{at}}\equiv \mathcal{B}(\mathbb{C}^{d})$ defined by (\ref{trace h
at}) and the operator $\mathfrak{Z}_{\mathrm{at}}$ on $\mathfrak{H}_{\mathrm{%
at}}$,
\begin{equation}
\mathfrak{Z}_{\mathrm{at}}\left( A\right) :=A\rho _{\mathrm{at}}^{1/2}\
,\qquad A\in \mathfrak{H}_{\mathrm{at}}\equiv \mathcal{B}(\mathbb{C}^{d})\ .
\label{eq cool g0bis}
\end{equation}%
Observe that, because $\rho _{\mathrm{at}}$ is the density matrix of a
faithful state, $\mathfrak{Z}_{\mathrm{at}}$ has an inverse. By abuse of
notation, $\mathfrak{L}_{\mathrm{p}}^{\ast }$ and $\mathfrak{L}_{\mathcal{R}%
}^{\ast }$ can be seen as operators defined on $\mathrm{ran}(\mathcal{P}%
_{p,\epsilon }^{\left( 0,0\right) })$ like in (\ref{def trivial1}) or (\ref%
{def trivial2}): Let
\begin{equation}
\mathfrak{\hat{H}}_{\mathrm{im}}:=\left\{ \hat{f}:=\{\hat{f}(k)\}_{k\in
\mathbb{Z}}:\hat{f}(k)\in \mathfrak{H}_{\mathrm{at}}\otimes \Omega _{%
\mathcal{R}},\ \sum_{k=-\infty }^{\infty }\Vert \hat{f}(k)\Vert _{\mathfrak{H%
}_{\mathrm{at}}}^{2}<\infty \right\} \subset \mathcal{\hat{H}\ }.
\label{def new label1}
\end{equation}%
For any linear operator $\mathfrak{L}\in \mathcal{B}(\mathfrak{\hat{H}}_{%
\mathrm{im}})$ and $\hat{f}:=\hat{u}\otimes \Omega _{\mathcal{R}}\in
\mathfrak{\hat{H}}_{\mathrm{im}}$, we define $\mathfrak{\tilde{L}}\equiv
\mathfrak{L}\in \mathcal{B}(\mathfrak{\hat{H}}_{\mathrm{im}})$\ by%
\begin{equation}
\big(\mathfrak{\tilde{L}}\hat{f}\big)(k):=\big(\mathfrak{L}\hat{u}\left(
k\right) \big)\otimes \Omega _{\mathcal{R}}\ ,\qquad k\in \mathbb{Z}\ .
\label{def new label2}
\end{equation}%
Note that $\mathrm{ran}(\mathcal{P}_{p,\epsilon }^{\left( 0,0\right)
})\subset \mathfrak{\hat{H}}_{\mathrm{im}}$ is an invariant subspace of $%
\mathfrak{L}_{\mathcal{R}}^{\ast }\equiv \mathfrak{\tilde{L}}_{\mathcal{R}%
}^{\ast }$.

\begin{satz}[Effective microscopic dynamics]
\label{corolario legal}\mbox{ }\newline
For all $p\in \mathbb{Z}$, $\epsilon \in \sigma (i\mathfrak{L}_{\mathrm{at}%
}) $, $A\in \mathfrak{H}_{\mathrm{at}}^{(p,\epsilon )}$ and sufficiently
small $\left\vert \lambda \right\vert $,
\begin{eqnarray*}
\mathcal{P}_{p,\epsilon }^{\left( \lambda ,\eta \right) }{\mathcal{\hat{G}}}%
(ir)\mathcal{P}_{p,\epsilon }^{\left( \lambda ,\eta \right) }\mathrm{U}%
_{p,\epsilon }^{\ast }(A) &=&i\left( p\varpi +\epsilon \right) \mathrm{U}%
_{p,\epsilon }^{\ast }(A)+\eta \mathcal{P}_{p,\epsilon }^{\left( 0,0\right) }%
\mathfrak{SZ}_{\mathrm{at}}\mathfrak{L}_{\mathrm{p}}^{\ast }\mathfrak{Z}_{%
\mathrm{at}}^{-1}\mathrm{U}_{p,\epsilon }^{\ast }(A) \\
&&+\lambda ^{2}\mathfrak{Z}_{\mathrm{at}}\mathfrak{L}_{\mathcal{R}}^{\ast }%
\mathfrak{Z}_{\mathrm{at}}^{-1}\mathrm{U}_{p,\epsilon }^{\ast }(A)+\mathrm{RU%
}_{p,\epsilon }^{\ast }(A)\ ,
\end{eqnarray*}%
where $\mathrm{R}\equiv \mathrm{R}^{(p,\epsilon ,\lambda ,\eta )}$ is an
operator with norm $\left\Vert \mathrm{R}\right\Vert _{\mathcal{B}(\mathcal{%
\hat{H}})}\leq C|\lambda |^{3}$ for some finite constant $C\in \mathbb{R}%
^{+} $ not depending on $p$, $\epsilon $, $\lambda $, $\eta $ and $A$.
\end{satz}

\noindent \textit{Proof}. By Lemma \ref{pert G copy(1)}, if $A\in \mathfrak{H%
}_{\mathrm{at}}^{(p,\epsilon )}$ then%
\begin{eqnarray}
&&\mathcal{P}_{p,\epsilon }^{\left( 0,0\right) }\ \mathcal{I}(ir)\ \mathcal{%
\bar{P}}_{p,\epsilon }^{\left( 0,0\right) }\big(i(p\varpi +\epsilon )-{%
\mathcal{\hat{G}}}_{0}(ir)\big)^{-1}\mathcal{\bar{P}}_{p,\epsilon }^{\left(
0,0\right) }\ \mathcal{I}(ir)\ \mathcal{P}_{p,\epsilon }^{\left( 0,0\right) }%
\mathrm{U}_{p,\epsilon }^{\ast }(A)  \notag \\
&=&\left( \mathcal{L}_{\mathcal{R}}^{(0,r,p)}\right) ^{\ast }\mathrm{U}%
_{p,\epsilon }^{\ast }(A)=\lim_{\varepsilon \searrow 0}\left( \mathcal{L}_{%
\mathcal{R}}^{(\varepsilon ,0,p)}\right) ^{\ast }\mathrm{U}_{p,\epsilon
}^{\ast }(A)\ ,  \label{totocool}
\end{eqnarray}%
see (\ref{atom reservoir louvi}). Straightforward computations show that,
for all $A\in \mathfrak{H}_{\mathrm{at}}^{(p,\epsilon )}$ and $\varepsilon
\in \mathbb{R}^{+}$
\begin{equation*}
\mathcal{L}_{\mathcal{R}}^{(\varepsilon ,0,p)\ast }\mathrm{U}_{p,\epsilon
}^{\ast }(A)=\mathfrak{Z}_{\mathrm{at}}\mathfrak{L}_{\mathcal{R}%
}^{(\varepsilon )\ast }\mathfrak{Z}_{\mathrm{at}}^{-1}\mathrm{U}_{p,\epsilon
}^{\ast }(A)\text{ }.
\end{equation*}%
Hence, by the definition (\ref{Lindblad gol}) of $\mathfrak{L}_{\mathcal{R}}$
together with (\ref{totocool}),
\begin{equation*}
\mathcal{P}_{p,\epsilon }^{\left( 0,0\right) }\ \mathcal{I}(ir)\ \mathcal{%
\bar{P}}_{p,\epsilon }^{\left( 0,0\right) }\big(i(p\varpi +\epsilon )-{%
\mathcal{\hat{G}}}_{0}(ir)\big)^{-1}\mathcal{\bar{P}}_{p,\epsilon }^{\left(
0,0\right) }\ \mathcal{I}(ir)\ \mathcal{P}_{p,\epsilon }^{\left( 0,0\right) }%
\mathrm{U}_{p,\epsilon }^{\ast }(A)=\mathfrak{Z}_{\mathrm{at}}\mathfrak{L}_{%
\mathcal{R}}^{\ast }\mathfrak{Z}_{\mathrm{at}}^{-1}\mathrm{U}_{p,\epsilon
}^{\ast }(A)
\end{equation*}%
for all $A\in \mathfrak{H}_{\mathrm{at}}^{(p,\epsilon )}$. Similarly, from
straightforward computations, one gets that, for all $A\in \mathfrak{H}_{%
\mathrm{at}}^{(p,\epsilon )}$,%
\begin{equation}
\mathcal{P}_{p,\epsilon }^{\left( 0,0\right) }\ \mathcal{J}\ \mathcal{P}%
_{p,\epsilon }^{\left( 0,0\right) }\mathrm{U}_{p,\epsilon }^{\ast }(A)=%
\mathcal{P}_{p,\epsilon }^{\left( 0,0\right) }\mathfrak{SZ}_{\mathrm{at}}%
\mathfrak{L}_{\mathrm{p}}^{\ast }\mathfrak{Z}_{\mathrm{at}}^{-1}\mathrm{U}%
_{p,\epsilon }^{\ast }(A)\ .  \label{totocoolbis}
\end{equation}%
The theorem then follows from Lemma \ref{pert G} combined with (\ref%
{totocool}) and (\ref{totocoolbis}). \hfill $\Box $

We are now in position to analyze the resonances of the Howland operator:

\begin{satz}[Resonances of the Howland operator]
\label{Theorem pas le temps}\mbox{ }\newline
Let $r\in (0,\mathrm{r}_{\max })$. There are constants $\lambda
_{0},C_{1},C_{2},C_{3},C_{4},C_{5}\in \mathbb{R}^{+}$ such that the
following properties holds for all $\left\vert \lambda \right\vert \leq
\lambda _{0}$:\newline
\emph{(i)} For any $p\in \mathbb{Z}$, $ip\varpi $ is a non--degenerated
eigenvalue of ${\mathcal{\hat{G}}}(ir)$ with eigenvector $\hat{f}_{p}$
defined by%
\begin{equation}
\hat{f}_{p}\left( p^{\prime }\right) :=\delta _{p,p^{\prime }}\Omega \
,\qquad p^{\prime }\in \mathbb{Z}\ ,  \label{vecteur propre}
\end{equation}%
where $\delta _{p,p^{\prime }}$ is the Kronecker symbol. \newline
\emph{(ii)} $\left\{ i\mathbb{R}+\left( -C_{1}\lambda ^{2},\infty \right)
\right\} \backslash i\varpi \mathbb{Z}\subset \varrho ({\mathcal{\hat{G}}}%
(ir))$. Here, $\varrho ({\mathcal{\hat{G}}}(ir))$ stands for the resolvent
set of ${\mathcal{\hat{G}}}(ir)$. \newline
\emph{(iii)} The spectrum of ${\mathcal{\hat{G}}}(ir)$ in $i\mathbb{R}%
+\left( -C_{2}\left\vert \lambda \right\vert ,-C_{1}\lambda ^{2}\right) $ is
discrete with algebraic multiplicity at most $d^{2}$, where $%
C_{2}>C_{1}\left\vert \lambda \right\vert $.\newline
\emph{(iv)} $i\mathbb{R}+\left( -C_{3},-C_{2}\left\vert \lambda \right\vert
\right) \subset \varrho ({\mathcal{\hat{G}}}(ir))$, where $%
C_{3}>C_{2}\left\vert \lambda \right\vert $.\newline
\emph{(v)} For all $\zeta \in i\mathbb{R}+\left( -C_{3},-C_{4}\right) $,
where $C_{3}>C_{4}$,
\begin{equation*}
\Vert (\zeta -{\mathcal{\hat{G}}}(ir))^{-1}\Vert _{\mathcal{B}(\mathcal{\hat{%
H}})}\leq C_{5}\ .
\end{equation*}%
\emph{(vi)} For any $D\in \mathbb{R}^{+}$, there is $C_{D}\in \mathbb{R}^{+}$
such that
\begin{equation*}
\underset{\mathop{\rm Re}\zeta >D}{\sup }\Vert (\zeta -{\mathcal{\hat{G}}}%
(ir))^{-1}\Vert _{\mathcal{B}(\mathcal{\hat{H}})}\leq C_{D}\ .
\end{equation*}
\end{satz}

\noindent \textit{Proof}. (i) For $p_{1}\in \mathbb{Z}$, assume that $%
ip_{1}\varpi $ is an eigenvalue of ${\mathcal{\hat{G}}}(ir)$ with
eigenvector $\hat{f}_{p_{1}}\in \mathcal{\hat{H}}$. Then, for $p_{2}\in
\mathbb{Z}$, the vector $\hat{f}_{p_{2}}$ defined by
\begin{equation}
\hat{f}_{p_{2}}\left( p\right) :=\hat{f}_{p_{1}}\left( p-p_{2}+p_{1}\right)
\ ,\qquad p\in \mathbb{Z}\ ,  \label{eq blbilb}
\end{equation}%
is an eigenvector of ${\mathcal{\hat{G}}}(ir)$ with eigenvalue $ip_{2}\varpi
$. In fact, the map $\hat{f}\left( p\right) \mapsto \hat{f}\left(
p-p_{2}+p_{1}\right) $ is a unitary transformation on $\mathcal{\hat{H}}$,
whose conjugation with ${\mathcal{\hat{G}}}(ir)$ is ${\mathcal{\hat{G}}}%
(ir)+i(p_{2}-p_{1})$. Consequently, it suffices to prove that ${\mathcal{%
\hat{G}}}(ir)\hat{f}_{0} =0$ and $0$ is a non--degenerated eigenvalue of ${%
\mathcal{\hat{G}}}(ir)$.

Remark that the vector $\hat{f}_{0}$ defined by (\ref{vecteur propre}) is an
eigenvector of ${\mathcal{\hat{G}}}(0)$ with eigenvalue $0$. Using (\ref%
{sympa1})--(\ref{sympa2}) we then deduce that ${\mathcal{\hat{G}}}(\theta )%
\hat{f}_{0}=0$ for all $\theta \in \mathbb{R}$. Recall that $\{{\mathcal{%
\hat{G}}}(\theta )\}_{\theta \in \mathbb{S}\backslash \mathbb{R}}$ is of
type A (cf. Lemma \ref{analyticity V}) and $\hat{f}_{0}\in \mathrm{Dom}({%
\mathcal{\hat{G}}}(0))$. It follows that, for all $\hat{\psi}\in \mathcal{%
\hat{H}}$, the continuous function
\begin{equation*}
g\left( \theta \right) :=\langle \hat{\psi},{\mathcal{\hat{G}}}(\theta ))%
\hat{f}_{0}\rangle _{\mathcal{\hat{H}}}\ ,\qquad \theta \in \mathbb{S}\ ,
\end{equation*}%
is analytic on $\mathbb{S}\backslash \mathbb{R}$ and is zero on the real
line. We thus infer from the Schwarz reflection principle that $g=0$ on $%
\mathbb{S}$. In other words, ${\mathcal{\hat{G}}}(ir) \hat{f}_{0}=0 \in
\mathcal{\hat{H}}$ for $r\in (0,\mathrm{r}_{\max })$ and so, the vectors $%
\hat{f}_{p}$ defined by (\ref{vecteur propre}) are, for any $p\in \mathbb{Z}$%
, eigenvector of ${\mathcal{\hat{G}}}(ir)$ with eigenvalue $ip\varpi $. It
remains to prove that $0$ is a non--degenerated eigenvalue.

By Assumption \ref{assumption important}, $|\eta |\leq C\lambda ^{2}$ for
some fixed constant $C\in \mathbb{R}^{+}$ and we can infer from (\ref{Wt}), (%
\ref{Wt'}) and (\ref{definition V}) that, at small $|\lambda |$,
\begin{equation}
\forall \theta \in \mathbb{S}:\qquad \Vert \hat{V}\left( \theta \right)
\Vert _{\mathcal{B}(\mathcal{\hat{H}})}\leq C\left\vert \lambda \right\vert
\ .  \label{norm perturbation V deformed}
\end{equation}%
By (\ref{liouvillean free0}) and (\ref{definition G non perturbe}), we
observe that $0$ is also an isolated eigenvalue of the unperturbed Howland
operator ${\mathcal{\hat{G}}}_{0}(ir)$ and by Kato's perturbation theory, it
is an isolated eigenvalue of ${\mathcal{\hat{G}}}(ir)$ for sufficiently
small $\left\vert \lambda \right\vert $ at any $r\in (0,\mathrm{r}_{\max })$%
. However, $0$ is still a degenerated eigenvalue of ${\mathcal{\hat{G}}}%
_{0}(ir)$. The interaction part of ${\mathcal{\hat{G}}}(ir)$ removes this
degeneracy.

Indeed, by (\ref{definition G non perturbe}),
\begin{equation*}
\forall \hat{f}\in \mathrm{Dom}({\mathcal{\hat{G}}}(ir)):\quad \langle \hat{f%
},{\mathcal{\hat{G}}}(ir)\hat{f}\rangle _{\mathcal{\hat{H}}}=i\left\{
\langle \hat{f},\varpi k\ \hat{f}\rangle _{\mathcal{\hat{H}}}+\langle \hat{f}%
,L_{\mathfrak{g}}\hat{f}\rangle _{\mathcal{\hat{H}}}+\langle \hat{f},\hat{V}%
\left( ir\right) \hat{f}\rangle \right\} -r\langle \hat{f},\hat{N}\hat{f}%
\rangle _{\mathcal{\hat{H}}}\ .
\end{equation*}%
The operators $\varpi k$, $L_{\mathfrak{g}}$, $\mathcal{J}$ (cf. (\ref%
{Howland deformedbis})) are self--adjoint and thus,%
\begin{equation*}
i\left\{ \langle \hat{f},\varpi k\ \hat{f}\rangle _{\mathcal{\hat{H}}%
}+\langle \hat{f},L_{\mathfrak{g}}\hat{f}\rangle _{\mathcal{\hat{H}}%
}+\langle \hat{f},\mathcal{J}\hat{f}\rangle _{\mathcal{\hat{H}}}\right\} \in
i\mathbb{R}\text{ }.
\end{equation*}%
Then, using Assumption \ref{assumption important 0}, we conclude that, for
some finite constant $C$ not depending on $\lambda $ and $r\in \lbrack
0,r_{\max })$,
\begin{equation*}
\mathop{\rm Re}\left\{ \lambda \langle \hat{f},\mathcal{I}\left( ir\right)
\hat{f}\rangle -r\langle \hat{f},\hat{N}\hat{f}\rangle _{\mathcal{\hat{H}}%
}\right\} <0\text{ },
\end{equation*}%
whenever $r>C|\lambda |$ and $\hat{f}(k)\notin \mathfrak{H}_{\mathrm{at}%
}\otimes \Omega _{\mathcal{R}}$ for some $k\in \mathbb{Z}$. This implies
that, for fixed $r\in (0,r_{\max })$ and sufficiently small $|\lambda |$,
each eigenvector of ${\mathcal{\hat{G}}}(ir)$ associated with a purely
imaginary eigenvalue $y\in i\mathbb{R}$ must be an element of the subspace $%
\mathfrak{\hat{H}}_{\mathrm{im}}$ defined by (\ref{def new label1}).

Now, we need several definitions. Let
\begin{equation*}
\mathcal{\hat{D}}_{\mathrm{im}}:=\left\{ \hat{f}\in \mathfrak{\hat{H}}_{%
\mathrm{im}}:\sum_{k=-\infty }^{\infty }k^{2}\Vert \hat{f}(k)\Vert _{%
\mathfrak{H}_{\mathrm{at}}}^{2}<\infty \right\}
\end{equation*}%
and the (unbounded) operator $k_{\mathrm{im}}$ be defined on $\mathcal{\hat{D%
}}_{\mathrm{im}}$ by%
\begin{equation*}
\big(k_{\mathrm{im}}\hat{f}\big)(k):=k\hat{f}(k)\ ,\qquad k\in \mathbb{Z}\ .
\end{equation*}%
To simplify notation, we denote the operator $k_{\mathrm{im}}$ by $k$.
Define further the bounded operator $\mathcal{\hat{L}}_{\mathrm{im}}^{\left(
\lambda ,\eta \right) }\in \mathcal{B}(\mathfrak{\hat{H}}_{\mathrm{im}})$ by%
\begin{equation}
\mathcal{\hat{L}}_{\mathrm{im}}^{\left( \lambda ,\eta \right) }:=\mathfrak{Z}%
_{\mathrm{at}}(\mathfrak{L}_{\mathrm{at}}^{\ast }+\lambda ^{2}\mathfrak{L}_{%
\mathcal{R}}^{\ast })\mathfrak{Z}_{\mathrm{at}}^{-1}+\eta \mathfrak{S}%
\mathfrak{Z}_{\mathrm{at}}\mathfrak{L}_{\mathrm{p}}^{\ast }\mathfrak{Z}_{%
\mathrm{at}}^{-1}  \label{ninja bound semibisbisbis}
\end{equation}%
where $\mathfrak{L}_{\mathrm{at}}^{\ast }$, $\mathfrak{L}_{\mathrm{p}}^{\ast
}$ and $\mathfrak{L}_{\mathcal{R}}^{\ast }$ the adjoints of $\mathfrak{L}_{%
\mathrm{at}}$, $\mathfrak{L}_{\mathrm{p}}$ and $\mathfrak{L}_{\mathcal{R}}$,
respectively. See (\ref{Lat}), (\ref{L pump}), (\ref{L R}) and (\ref{eq cool
g0bis}) as well as (\ref{def new label1})--(\ref{def new label2}).

Similar to \cite[Eq. (4.8)]{BruPedraWestrich2011a}, define
\begin{equation}
\hat{G}_{\mathrm{im}}^{\left( \lambda ,\eta \right) }:=i\varpi k+\mathcal{%
\hat{L}}_{\mathrm{im}}^{\left( \lambda ,\eta \right) }
\label{ninja bound semibis}
\end{equation}%
with dense domain $\mathcal{\hat{D}}_{\mathrm{im}}\subset \mathfrak{\hat{H}}%
_{\mathrm{im}}$. We denote by
\begin{equation}
P_{0}^{\left( \lambda ,\eta \right) }:=\frac{1}{2\pi i}\oint\limits_{\gamma
_{0,0}}\big(z-\hat{G}_{\mathrm{im}}^{\left( \lambda ,\eta \right) }\big)^{-1}%
\mathrm{d}z  \label{Kato projection}
\end{equation}%
the Riesz projection associated with the closed operator $\hat{G}_{\mathrm{im%
}}^{\left( \lambda ,\eta \right) }$ defined by (\ref{ninja bound semibis}),
for a sufficiently small enough parameter $\mathfrak{r}>0$ (cf. (\ref{gamma
oo})--(\ref{gammak0})).

From Theorem \ref{corolario legal},
\begin{equation*}
\left\Vert \mathcal{P}_{0,0}^{\left( \lambda ,\eta \right) }{\mathcal{\hat{G}%
}}(ir)\mathcal{P}_{0,0}^{\left( \lambda ,\eta \right) }\mathcal{P}%
_{0,0}^{\left( 0,0\right) }-\mathcal{P}_{0,0}^{\left( 0,0\right) }\hat{G}_{%
\mathrm{im}}^{\left( \lambda ,\eta \right) }\mathcal{P}_{0,0}^{\left(
0,0\right) }\right\Vert _{\mathcal{B}(\mathfrak{\hat{H}})}\leq C\left\vert
\lambda \right\vert ^{3}\ .
\end{equation*}%
Note that $\mathrm{ran}(\mathcal{P}_{0,0}^{\left( 0,0\right) })\subset
\mathfrak{\hat{H}}_{\mathrm{im}}$. Define
\begin{equation*}
\Lambda _{\mathrm{im}}^{\left( \lambda ,\eta \right) }:=P_{0}^{\left(
0,0\right) }\hat{G}_{\mathrm{im}}^{\left( \lambda ,\eta \right)
}P_{0}^{\left( 0,0\right) }\ .
\end{equation*}%
By observing that
\begin{equation}
\mathcal{P}_{0,0}^{\left( 0,0\right) }=P_{0}^{\left( 0,0\right) }\mathcal{P}%
_{0,0}^{\left( 0,0\right) },\quad \mathcal{P}_{0,0}^{\left( 0,0\right) }|_{%
\mathfrak{\hat{H}}_{\mathrm{im}}}=\mathcal{P}_{0,0}^{\left( 0,0\right)
}P_{0}^{\left( 0,0\right) }  \label{P calP}
\end{equation}%
and using the bounds
\begin{equation*}
\left\Vert \mathcal{P}_{0,0}^{\left( \lambda ,\eta \right) }-\mathcal{P}%
_{0,0}^{\left( 0,0\right) }\right\Vert _{\mathcal{B}(\mathcal{\hat{H}})}\leq
C|\lambda |\ ,\quad \left\Vert \mathcal{P}_{0,0}^{\left( \lambda ,\eta
\right) }{\mathcal{\hat{G}}}(ir)\mathcal{P}_{0,0}^{\left( \lambda ,\eta
\right) }\right\Vert _{\mathcal{B}(\mathcal{\hat{H}})}\leq C\lambda ^{2}%
\text{ },
\end{equation*}%
we deduce that%
\begin{equation}
\left\Vert \mathcal{P}_{0,0}^{\left( \lambda ,\eta \right) }{\mathcal{\hat{G}%
}}(ir)\mathcal{P}_{0,0}^{\left( \lambda ,\eta \right) }-\mathcal{P}%
_{0,0}^{\left( 0,0\right) }\Lambda _{\mathrm{im}}^{\left( \lambda ,\eta
\right) }\mathcal{P}_{0,0}^{\left( 0,0\right) }\right\Vert _{\mathcal{B}(%
\mathcal{\hat{H}})}\leq C\left\vert \lambda \right\vert ^{3}\text{ }.
\label{eq sympa}
\end{equation}%
If Assumption \ref{assumption3} holds then, exactly like in \cite[Lemma 6.3]%
{BruPedraWestrich2011a}, for all $(\lambda ,\eta )\in \mathbb{R}\times
\mathbb{R}$,
\begin{equation}
\min \left\{ \ \left\vert \mathop{\rm Re}\left\{ p\right\} \right\vert :p\in
\sigma (\Lambda _{\mathrm{im}}^{\left( \lambda ,\eta \right) })\backslash
\{0\}\right\} \geq C_{\varpi }\lambda ^{2}  \label{bound Gim importante}
\end{equation}%
with $C_{\varpi }\in \mathbb{R}^{+}$ being a constant depending on $\varpi $
but not on $\lambda $, $\eta $. Similar to \cite[Lemma 6.1]%
{BruPedraWestrich2011a} one proves, moreover, that $0$ is a non--degenerated
eigenvalue of $\Lambda _{\mathrm{im}}^{\left( \lambda ,\eta \right) }$.
Using this, the relations (\ref{P calP}) and Kato's perturbation theory
together with the bound (\ref{bound Gim importante}), for small enough $%
\left\vert \lambda \right\vert $, $\mathcal{\hat{G}}(ir)$ has a
non--degenerated eigenvalue $\varepsilon \left( \lambda ,\eta \right) =%
\mathcal{O}\left( \lambda ^{3}\right) $ with
\begin{equation}
\min \left\{ \ \left\vert \mathrm{Re}\{p\}\right\vert :p\in \sigma (\mathcal{%
\hat{G}}(ir))\backslash \{\varepsilon \left( \lambda ,\eta \right)
\}\right\} \geq C_{\varpi }\lambda ^{2}\ .  \label{estimate super hatG}
\end{equation}%
[Lemma \ref{typeAbis copy(5)} applied to the pair of projections $\mathcal{P}%
_{0,0}^{\left( 0,0\right) }$ and $\mathcal{P}_{0,0}^{\left( \lambda ,\eta
\right) }$ can be useful in this context.] As $0$ is an eigenvalue of $%
\mathcal{\hat{G}}(ir)$, it follows that $\varepsilon \left( \lambda ,\eta
\right) =0$ and $0$ is a non--degenerated eigenvalue of ${\mathcal{\hat{G}}}%
(ir)$, provided $|\lambda |$ is sufficiently small. As a consequence, the
vectors $\hat{f}_{p}$ defined by (\ref{vecteur propre}) for $p\in \mathbb{Z}$
are eigenvectors of ${\mathcal{\hat{G}}}(ir)$ associated with the
non--degenerate eigenvalues $ip\varpi $, for small enough $|\lambda |$.

(ii) As explained in the beginning of the proof, it suffices to prove this
statement for the spectrum near $0$. Therefore, the assertion is a direct
consequence of the estimate (\ref{estimate super hatG}).

(iii) Recall that $0$ is an isolated eigenvalue of the unperturbed Howland
operator ${\mathcal{\hat{G}}}_{0}(ir)$ with algebraic multiplicity $n\in
\mathbb{N}$, $n\leq d^{2}$. By Kato's perturbation theory and (\ref{norm
perturbation V deformed}), there are at most $n$ eigenvalues of ${\mathcal{%
\hat{G}}}(ir)$ within a ball of radius $\left\vert \lambda \right\vert $
with algebraic multiplicity at most $d^{2}$. Therefore, we arrive at the
third assertion by combining this observation with (i)--(ii).

(iv) and (v) are easy to verify for the case $V=0$ because the operator ${%
\mathcal{\hat{G}}}_{0}(ir)$ is equivalent to a normal operator with
explicitly known spectrum. The general case is proved by using simple power
expansion for the resolvents of ${\mathcal{\hat{G}}}(ir)$ as $\hat{V}$ is a
bounded operator of order $\mathcal{O}(\lambda )$.

(vi) To prove the last assertion, use the fact that the eigenvalues\ $%
ip\varpi $, $p\in \mathbb{Z}$, are non--degenerated. By Kato's perturbation
theory, its resolvent near such spectral points behaves in the limit $\zeta
\rightarrow ip\varpi $ as%
\begin{equation*}
\Vert (\zeta -ip\varpi -{\mathcal{\hat{G}}}(ir))^{-1}\Vert _{\mathcal{B}(%
\mathcal{\hat{H}})}\leq C\left\vert \zeta -ip\varpi \right\vert ^{-1}\ ,
\end{equation*}%
where $C\in \mathbb{R}^{+}$ is a constant not depending on $p\in \mathbb{Z}$%
. The uniformity of this last estimate is related to the fact that the
spectral spaces associated with the eigenvalues $ip\varpi $ are all
unitarily equivalent, see Equation (\ref{eq blbilb}). \hfill $\Box $

\subsection{Time--uniform approximations of $\{\mathrm{e}^{\protect\alpha {%
\mathcal{\hat{G}}}(ir)}\}_{\protect\alpha \geq 0}$}

Let $r\in (0,\mathrm{r}_{\max })$ (see Section \ref{ninja thm cool}). The
strongly continuous one--parameter semigroup $\{\mathrm{e}^{\alpha {\mathcal{%
\hat{G}}}(ir)}\}_{\alpha \geq 0}$ can be represented as the inverse Laplace
transform of the resolvent of ${\mathcal{\hat{G}}}(ir)$. Indeed, by Theorem %
\ref{Theorem pas le temps} (vi) and the Gearhart--Pr\"{u}ss--Greiner Theorem
\cite[Chap. V, 1.11]{Nagel-engel}, for any $D\in \mathbb{R}^{+}$, there is $%
C_{D}\in \mathbb{R}^{+}$ such that%
\begin{equation*}
\forall \alpha \in \mathbb{R}_{0}^{+}:\qquad \Vert \mathrm{e}^{\alpha {%
\mathcal{\hat{G}}}(ir)}\Vert _{\mathcal{B}(\mathcal{\hat{H}})}\leq C_{D}%
\mathrm{e}^{D\alpha }\ ,
\end{equation*}%
i.e., the growth bound of the semigroup $\{\mathrm{e}^{\alpha {\mathcal{\hat{%
G}}}(ir)}\}_{\alpha \geq 0}$ is zero. Hence, by Lemma \ref{typeAbis copy(1)}%
,
\begin{equation}
\mathrm{e}^{\alpha {\mathcal{\hat{G}}}(ir)}\hat{f}=\underset{N\rightarrow
\infty }{\lim }\left\{ \frac{1}{2\pi i}\int_{w-iN}^{w+iN}\mathrm{e}^{\alpha
\zeta }(\zeta -{\mathcal{\hat{G}}}(ir))^{-1}\hat{f}\ \mathrm{d}\zeta \right\}
\label{impoetant equation laplace transform inverse}
\end{equation}%
for all $\hat{f}\in \mathrm{Dom}({\mathcal{\hat{G}}}(ir))$, $w,\alpha \in
\mathbb{R}^{+}$. Next, we modify the contour of integration to make Riesz
projections appear.

To this end, define%
\begin{equation}
\mathcal{\tilde{P}}_{p}^{\left( \lambda ,\eta \right) }:=\frac{1}{2\pi i}%
\underset{\tilde{\gamma}_{p}}{\oint }(\zeta -{\mathcal{\hat{G}}}(ir))^{-1}%
\mathrm{d}\zeta \ ,\qquad p\in \mathbb{Z}\ ,  \label{Kato projection2}
\end{equation}%
where, for any $p\in \mathbb{Z}$, the contour $\tilde{\gamma}_{p}$ is
defined by
\begin{equation}
\tilde{\gamma}_{p}\left( y\right) :=\left\{
\begin{array}{lll}
i\left( p\varpi +\mathfrak{r}\right) +i\left( \varpi -2\mathfrak{r}\right) y
& \text{for} & y\in \lbrack 0,1]\ . \\
i\left( \left( p+1\right) \varpi -\mathfrak{r}\right) -C_{4}\left( y-1\right)
& \text{for} & y\in \lbrack 1,2]\ . \\
-C_{4}+i\left( \left( p+1\right) \varpi -\mathfrak{r}\right) -i\left( \varpi
-2\mathfrak{r}\right) \left( y-2\right) & \text{for} & y\in \lbrack 2,3]\ .
\\
-C_{4}+i\left( p\varpi +\mathfrak{r}\right) +C_{4}\left( y-3\right) & \text{%
for} & y\in \lbrack 3,4]\ .%
\end{array}%
\right.  \label{gammak}
\end{equation}%
Here, $\mathfrak{r}\in (0,1/2)\cap (0,\mathrm{r}_{\max })$ is a sufficiently
small parameter, see (\ref{gamma oo})--(\ref{Kato projection1}), while $%
C_{4}\in \mathbb{R}^{+}$ is the constant of Theorem \ref{Theorem pas le
temps} (v). For all $\hat{f}\in \mathrm{Dom}({\mathcal{\hat{G}}}(ir))$, $%
w,\alpha \in \mathbb{R}^{+}$, and any negative real number $v\in \left(
-C_{3},-C_{4}\right) $ (cf. Theorem \ref{Theorem pas le temps}) we now
observe that, for $N\in \varpi \mathbb{N+}\mathfrak{r}$ and sufficiently
small $|\lambda |$,
\begin{align}
\frac{1}{2\pi i}\int_{w-iN}^{w+iN}\mathrm{e}^{\alpha \zeta }(\zeta -{%
\mathcal{\hat{G}}}(ir))^{-1}\hat{f}\text{ }\mathrm{d}\zeta & =\underset{p\in
\mathbb{Z}:\text{ }\left\vert p\right\vert \varpi <N}{\sum }\mathrm{e}%
^{\alpha \mathcal{K}_{p}^{\left( \lambda ,\eta \right) }}\mathcal{P}%
_{p,0}^{\left( \lambda ,\eta \right) }\hat{f}  \notag \\
& \text{ \ \ }+\underset{p\in \mathbb{Z}:\text{ }p\varpi ,(p+1)\varpi \in
\lbrack -N,N]}{\sum }\mathrm{e}^{\alpha \mathcal{\tilde{K}}_{p}^{\left(
\lambda ,\eta \right) }}\mathcal{\tilde{P}}_{p}^{\left( \lambda ,\eta
\right) }\hat{f}  \notag \\
& \text{ \ \ }-\frac{1}{2\pi i}\int_{w+iN}^{v+iN}\mathrm{e}^{\alpha \zeta
}(\zeta -{\mathcal{\hat{G}}}(ir))^{-1}\hat{f}\mathrm{d}\zeta  \notag \\
& \text{ \ \ }-\frac{1}{2\pi i}\int_{v+iN}^{v-iN}\mathrm{e}^{\alpha \zeta
}(\zeta -{\mathcal{\hat{G}}}(ir))^{-1}\hat{f}\mathrm{d}\zeta  \notag \\
& \text{ \ \ }-\frac{1}{2\pi i}\int_{v-iN}^{w-iN}\mathrm{e}^{\alpha \zeta
}(\zeta -{\mathcal{\hat{G}}}(ir))^{-1}\hat{f}\mathrm{d}\zeta \ ,
\label{important equationbis}
\end{align}%
where
\begin{eqnarray}
\mathcal{K}_{p}^{\left( \lambda ,\eta \right) } &:=&\mathcal{P}_{p,0}^{\left( \lambda ,\eta \right) }{\mathcal{\hat{G}}}(ir)\mathcal{P}_{p,0}^{\left( \lambda ,\eta \right) }\ ,\qquad p\in \mathbb{Z}\ ,
\label{Kato projection3} \\
\mathcal{\tilde{K}}_{p}^{\left( \lambda ,\eta \right) } &:=&\mathcal{\tilde{P}}_{p}^{\left( \lambda ,\eta \right) }{\mathcal{\hat{G}}}(ir)\mathcal{\tilde{P}}_{p}^{\left( \lambda ,\eta \right) }\ \ ,\qquad p\in \mathbb{Z}\ .
\label{Kato projection3bis}
\end{eqnarray}%
We analyze in the three next lemmata each term of the right hand side
(r.h.s.) of Equation (\ref{important equationbis}), in the limit $%
N\rightarrow \infty $.

\begin{lemma}
\label{typeAbis copy(4)}\mbox{ }\newline
For all $\hat{\psi},\hat{\psi}^{\prime }\in \mathfrak{\hat{H}}_{\mathrm{at}%
}\subset \mathrm{Dom}({\mathcal{\hat{G}}}(ir))$, $w,\alpha \in \mathbb{R}%
^{+} $, and $v\in \left( -C_{3},-C_{4}\right) $,%
\begin{equation*}
\underset{N\rightarrow \infty }{\lim }\int_{w\pm iN}^{v\pm iN}\mathrm{e}%
^{\alpha \zeta }(\zeta -{\mathcal{\hat{G}}}(ir))^{-1}\hat{\psi}\mathrm{d}%
\zeta =0\ ,
\end{equation*}%
while
\begin{equation*}
\limsup\limits_{N\rightarrow \infty }\left\vert \left\langle \hat{\psi}%
,\int_{v+iN}^{v-iN}\mathrm{e}^{\alpha \zeta }(\zeta -{\mathcal{\hat{G}}}%
(ir))^{-1}\hat{\psi}^{\prime }\mathrm{d}\zeta \right\rangle _{\mathcal{\hat{H%
}}}\right\vert \leq C\lambda ^{2}\mathrm{e}^{v\alpha }\Vert \hat{\psi}\Vert
_{\mathcal{\hat{H}}}\Vert \hat{\psi}^{\prime }\Vert _{\mathcal{\hat{H}}}\ .
\end{equation*}
\end{lemma}

\noindent \textit{Proof}. Using the equality
\begin{equation}
\forall \hat{f}\in \mathrm{Dom}({\mathcal{\hat{G}}}(ir)),\ \zeta \in \varrho
({\mathcal{\hat{G}}}(ir)):\quad (\zeta -{\mathcal{\hat{G}}}(ir))^{-1}\hat{f}%
=\zeta ^{-1}\left( (\zeta -{\mathcal{\hat{G}}}(ir))^{-1}{\mathcal{\hat{G}}}%
(ir)\hat{f}+\hat{f}\right) \ ,  \label{Equality trivial}
\end{equation}%
we deduce that, for any $\hat{\psi}\in \mathfrak{\hat{H}}_{\mathrm{at}}$,
\begin{equation*}
\int_{w\pm iN}^{v\pm iN}\mathrm{e}^{\alpha \zeta }(\zeta -{\mathcal{\hat{G}}}%
(ir))^{-1}\hat{\psi}\mathrm{d}\zeta =\int_{w\pm iN}^{v\pm iN}\zeta ^{-1}%
\mathrm{e}^{\alpha \zeta }(\zeta -{\mathcal{\hat{G}}}(ir))^{-1}{\mathcal{%
\hat{G}}}(ir)\hat{\psi}\mathrm{d}\zeta +\int_{w\pm iN}^{v\pm iN}\zeta ^{-1}%
\mathrm{e}^{\alpha \zeta }\hat{\psi}\mathrm{d}\zeta \ .
\end{equation*}%
Recall that $\varrho ({\mathcal{\hat{G}}}(ir))$ stands for the resolvent set
of ${\mathcal{\hat{G}}}(ir)$. In the limit $N\rightarrow \infty $, both
integrals in the r.h.s. of this last equality vanish.

We prove now the second inequality. Observe first that, for any $\zeta \in
\varrho ({\mathcal{\hat{G}}}(ir))$ and sufficiently small $|\lambda |$,
\begin{eqnarray*}
(\zeta -{\mathcal{\hat{G}}}(ir))^{-1} &=&(\zeta -{\mathcal{\hat{G}}}%
_{0}(ir))^{-1}+\lambda (\zeta -{\mathcal{\hat{G}}}_{0}(ir))^{-1}\mathcal{I}%
(ir)(\zeta -{\mathcal{\hat{G}}}_{0}(ir))^{-1} \\
&&+\eta (\zeta -{\mathcal{\hat{G}}}_{0}(ir))^{-1}\mathcal{J}(\zeta -{%
\mathcal{\hat{G}}}_{0}(ir))^{-1} \\
&&+(\zeta -{\mathcal{\hat{G}}}_{0}(ir))^{-1}\left( \eta \mathcal{J}+\lambda
\mathcal{I}(ir)\right) (\zeta -{\mathcal{\hat{G}}}_{0}(ir))^{-1}\left( \eta
\mathcal{J}+\lambda \mathcal{I}(ir)\right) (\zeta -{\mathcal{\hat{G}}}%
(ir))^{-1}.
\end{eqnarray*}%
Because of Assumption \ref{assumption important 0}, for all $\hat{\psi},\hat{%
\psi}^{\prime }\in \mathfrak{\hat{H}}_{\mathrm{at}}$, $\zeta \in \varrho ({%
\mathcal{\hat{G}}}(ir))$ and sufficiently small $|\lambda |$,
\begin{equation*}
\left\langle \hat{\psi},(\zeta -{\mathcal{\hat{G}}}_{0}(ir))^{-1}\mathcal{I}%
(ir)(\zeta -{\mathcal{\hat{G}}}_{0}(ir))^{-1}\hat{\psi}^{\prime
}\right\rangle _{\mathcal{\hat{H}}}=0\ ,
\end{equation*}%
while, by using (\ref{def sup1}),%
\begin{eqnarray*}
&&\underset{N\rightarrow \infty }{\lim }\left\langle \hat{\psi}%
,\int_{v+iN}^{v-iN}(\zeta -{\mathcal{\hat{G}}}_{0}(ir))^{-1}\hat{\psi}%
^{\prime }\mathrm{d}\zeta \right\rangle _{\mathcal{\hat{H}}} \\
&=&\sum\limits_{(n,k),(n^{\prime },k^{\prime })}\left\langle \hat{\psi},%
\mathfrak{F}%
\big(%
(W_{\left( k,n\right) }^{\left( k^{\prime },n^{\prime }\right) }\rho _{%
\mathrm{at}}^{1/2})\otimes \Omega _{\mathcal{R}}%
\big)%
\right\rangle _{\mathcal{\hat{H}}}\left\langle \mathfrak{F}%
\big(%
(W_{\left( k,n\right) }^{\left( k^{\prime },n^{\prime }\right) }\rho _{%
\mathrm{at}}^{1/2})\otimes \Omega _{\mathcal{R}}%
\big)%
,\hat{\psi}^{\prime }\right\rangle _{\mathcal{\hat{H}}} \\
&&\times \underset{N\rightarrow \infty }{\lim }\int_{v+iN}^{v-iN}\mathrm{e}%
^{\alpha \zeta }(\zeta -i(E_{k^{\prime }}-E_{k}))^{-1}\mathrm{d}\zeta \\
&=&0
\end{eqnarray*}%
for all $\alpha \in \mathbb{R}^{+}$ and $v<0$. Clearly, for $\zeta \in v+i%
\mathbb{R}$ and $\hat{\psi}^{\prime }\in \mathfrak{\hat{H}}_{\mathrm{at}}$,
\begin{equation*}
\left\Vert (\zeta -{\mathcal{\hat{G}}}_{0}(ir))^{-1}\mathcal{J}(\zeta -{%
\mathcal{\hat{G}}}_{0}(ir))^{-1}\hat{\psi}^{\prime }\right\Vert _{\mathcal{%
\hat{H}}}\leq \frac{C}{|\zeta |^{2}}\Vert \hat{\psi}^{\prime }\Vert _{%
\mathcal{\hat{H}}}
\end{equation*}%
with $C\in \mathbb{R}^{+}$ being some finite constant that only depends on $%
v $. Thus, using Assumption \ref{assumption important},%
\begin{equation*}
\left\vert \left\langle \hat{\psi},\int_{v+iN}^{v-iN}\mathrm{e}^{\alpha
\zeta }\eta (\zeta -{\mathcal{\hat{G}}}_{0}(ir))^{-1}\mathcal{J}(\zeta -{%
\mathcal{\hat{G}}}_{0}(ir))^{-1}\hat{\psi}^{\prime }\mathrm{d}\zeta
\right\rangle _{\mathcal{\hat{H}}}\right\vert \leq C\lambda ^{2}\mathrm{e}%
^{\alpha v}\Vert \hat{\psi}\Vert _{\mathcal{\hat{H}}}\Vert \hat{\psi}%
^{\prime }\Vert _{\mathcal{\hat{H}}}
\end{equation*}%
for some constant $C\in \mathbb{R}^{+}$depending only on $v$.

Again by Assumption \ref{assumption important}, note that, for all $\zeta
\in v+i\mathbb{R}$ and $\hat{\psi}\in \mathfrak{\hat{H}}_{\mathrm{at}}$,%
\begin{equation*}
\left\Vert \left[ (\zeta -{\mathcal{\hat{G}}}_{0}(ir))^{-1}\left( \eta
\mathcal{J}+\lambda \mathcal{I}(ir)\right) (\zeta -{\mathcal{\hat{G}}}%
_{0}(ir))^{-1}\left( \eta \mathcal{J}+\lambda \mathcal{I}(ir)\right) \right]
^{\ast }\hat{\psi}\right\Vert _{\mathcal{\hat{H}}}\leq \frac{C\lambda ^{2}}{%
|\zeta |}\Vert \hat{\psi}\Vert _{\mathcal{\hat{H}}}
\end{equation*}%
with $C\in \mathbb{R}^{+}$ being some finite constant only depending on $v$.
On the other hand, by using Equality (\ref{Equality trivial}) together with
Theorem \ref{Theorem pas le temps} (v), we obtain%
\begin{equation*}
\left\Vert (\zeta -{\mathcal{\hat{G}}}(ir))^{-1}\hat{\psi}^{\prime
}\right\Vert _{\mathcal{\hat{H}}}\leq \frac{C}{|\zeta |}\Vert \hat{\psi}%
^{\prime }\Vert _{\mathcal{\hat{H}}}
\end{equation*}%
for some constant $C\in \mathbb{R}^{+}$ that does not depend on $\zeta \in
v+i\mathbb{R}$ and $\hat{\psi}^{\prime }\in \mathfrak{\hat{H}}_{\mathrm{at}}$%
. Hence,%
\begin{eqnarray*}
&&\left\vert \left\langle \hat{\psi},\int_{v+iN}^{v-iN}\mathrm{e}^{\alpha
\zeta }(\zeta -{\mathcal{\hat{G}}}_{0}(ir))^{-1}\left( \eta \mathcal{J}%
+\lambda \mathcal{I}(ir)\right) (\zeta -{\mathcal{\hat{G}}}%
_{0}(ir))^{-1}\left( \eta \mathcal{J}+\lambda \mathcal{I}(ir)\right) (\zeta -%
{\mathcal{\hat{G}}}(ir))^{-1}\hat{\psi}^{\prime }\mathrm{d}\zeta
\right\rangle _{\mathcal{\hat{H}}}\right\vert \\
&\leq &C\lambda ^{2}\mathrm{e}^{\alpha v}\Vert \hat{\psi}\Vert _{\mathcal{%
\hat{H}}}\Vert \hat{\psi}^{\prime }\Vert _{\mathcal{\hat{H}}}
\end{eqnarray*}%
for all $\hat{\psi},\hat{\psi}^{\prime }\in \mathfrak{\hat{H}}_{\mathrm{at}}$%
, $\alpha \in \mathbb{R}^{+}$ and some constant $C\in \mathbb{R}^{+}$ that
does not depend on $\hat{\psi},\hat{\psi}^{\prime }$ and $\alpha $. \hfill $%
\Box $

\begin{lemma}
\label{typeAbis copy(2)}\mbox{ }\newline
There is a constant $C\in \mathbb{R}^{+}$ such that, for all $\hat{\psi}\in
\mathfrak{\hat{H}}_{\mathrm{at}}\subset \mathcal{\hat{H}}$,%
\begin{equation*}
\begin{array}{lll}
\forall p\in \mathbb{Z}\backslash \left\{ 0\right\} & : & \Vert \mathcal{P}%
_{p,0}^{\left( \lambda ,\eta \right) }\hat{\psi}\Vert _{\mathcal{\hat{H}}%
}\leq Cp^{-2}\varpi ^{-2}\Vert \hat{\psi}\Vert _{\mathcal{\hat{H}}}\ , \\
\forall p\in \mathbb{Z}\backslash \left\{ 0,-1\right\} & : & \Vert \mathcal{%
\tilde{P}}_{p}^{\left( \lambda ,\eta \right) }\hat{\psi}\Vert _{\mathcal{%
\hat{H}}}\leq Cp^{-2}\varpi ^{-2}\Vert \hat{\psi}\Vert _{\mathcal{\hat{H}}}\
.%
\end{array}%
\end{equation*}%
Moreover, if $\hat{\psi}\in \mathfrak{F}(\mathfrak{D}\otimes \Omega _{%
\mathcal{R}})$ then
\begin{equation*}
\forall p\in \mathbb{Z}:\quad \Vert \mathcal{\tilde{P}}_{p}^{\left( \lambda
,\eta \right) }\hat{\psi}\Vert _{\mathcal{\hat{H}}}\leq C\left\vert \lambda
\right\vert \left( \frac{1}{\varpi ^{2}p^{2}+1}+\frac{1}{\varpi ^{2}\left(
p+1\right) ^{2}+1}\right) \Vert \hat{\psi}\Vert _{\mathcal{\hat{H}}}\ .
\end{equation*}
\end{lemma}

\noindent \textit{Proof}. Using two times Equality (\ref{Equality trivial}),
for any $p\in \mathbb{Z}\backslash \left\{ 0\right\} $ and $\hat{\psi}\in
\mathfrak{\hat{H}}_{\mathrm{at}}\subset \mathrm{Dom}(({\mathcal{\hat{G}}}%
(ir))^{2})$, in the definition (\ref{Kato projection1}) of the Riesz
projection $\mathcal{P}_{p,0}^{\left( \lambda ,\eta \right) }$, we obtain
that%
\begin{eqnarray*}
\mathcal{P}_{p,0}^{\left( \lambda ,\eta \right) }\hat{\psi} &=&\frac{1}{2\pi
i}\underset{\gamma _{p,0}}{\oint }\zeta ^{-2}(\zeta -{\mathcal{\hat{G}}}%
(ir))^{-1}({\mathcal{\hat{G}}}(ir))^{2}\hat{\psi}\mathrm{d}\zeta +\frac{1}{%
2\pi i}\underset{\gamma _{p,0}}{\oint }\zeta ^{-2}{\mathcal{\hat{G}}}(ir)%
\hat{\psi}\mathrm{d}\zeta +\frac{1}{2\pi i}\underset{\gamma _{p,0}}{\oint }%
\zeta ^{-1}\hat{\psi}\mathrm{d}\zeta \\
&=&\frac{1}{2\pi i}\underset{\gamma _{p,0}}{\oint }\zeta ^{-2}(\zeta -{%
\mathcal{\hat{G}}}(ir))^{-1}({\mathcal{\hat{G}}}(ir))^{2}\hat{\psi}\mathrm{d}%
\zeta \ ,
\end{eqnarray*}%
by analyticity. There is $C\in \mathbb{R}^{+}$ such that, for $p\in \mathbb{Z%
}\backslash \left\{ 0\right\} $, $\zeta \in \gamma _{p,0}$ and sufficiently
small $|\lambda |$,
\begin{equation*}
\Vert (\zeta -{\mathcal{\hat{G}}}(ir))^{-1}\Vert _{\mathcal{B}(\mathcal{\hat{%
H}})}\leq C\quad \text{and}\quad \Vert ({\mathcal{\hat{G}}}(ir))^{2}\hat{\psi%
}\Vert _{\mathcal{\hat{H}}}\leq C\Vert \hat{\psi}\Vert _{\mathcal{\hat{H}}}\
,
\end{equation*}%
see (\ref{holand fourier}) and (\ref{Howland deformed}), while%
\begin{equation}
|\zeta |^{-2}\leq Cp^{-2}\varpi ^{-2}\ .  \label{new label}
\end{equation}%
We thus arrive at the assertion%
\begin{equation*}
\Vert \mathcal{P}_{p,0}^{\left( \lambda ,\eta \right) }\hat{\psi}\Vert _{%
\mathcal{\hat{H}}}\leq Cp^{-2}\varpi ^{-2}\Vert \hat{\psi}\Vert _{\mathcal{%
\hat{H}}}
\end{equation*}%
with a constant $C\in \mathbb{R}^{+}$ not depending on $p\in \mathbb{Z}%
\backslash \left\{ 0\right\} $.

To prove the second inequality, we proceed in the same way by using the fact
that%
\begin{equation*}
\Vert (\zeta -{\mathcal{\hat{G}}}(ir))^{-1}\Vert _{\mathcal{B}(\mathcal{\hat{%
H}})}\leq C
\end{equation*}%
for all $p\in \mathbb{Z}$, $\zeta \in \tilde{\gamma}_{p}$ and sufficiently
small $|\lambda |$, by Theorem \ref{Theorem pas le temps}. Note also that (%
\ref{new label}) also holds for all $p\in \mathbb{Z}\backslash \left\{
0,-1\right\} $ and $\zeta \in \tilde{\gamma}_{p}$.

To prove the last assertion, observe that if $\hat{\psi}\in \mathfrak{F}(%
\mathfrak{D}\otimes \Omega _{\mathcal{R}})$ then
\begin{equation*}
\Vert ({\mathcal{\hat{G}}}(ir))^{2}\hat{\psi}\Vert _{\mathcal{\hat{H}}}\leq
C\left\vert \lambda \right\vert \Vert \hat{\psi}\Vert _{\mathcal{\hat{H}}}\
\end{equation*}%
for some constant $C\in \mathbb{R}^{+}$. \hfill $\Box $

\begin{lemma}
\label{typeAbis copy(3)}\mbox{ }\newline
There is a constant $C\in \mathbb{R}^{+}$ such that, for all $p\in \mathbb{Z}
$ and $\alpha \in \mathbb{R}_{0}^{+}$,
\begin{equation*}
\left\Vert \mathrm{e}^{\alpha \mathcal{K}_{p}}\right\Vert _{\mathcal{B}(%
\mathcal{\hat{H}})}\leq C\quad \text{and}\quad \Vert \mathrm{e}^{\alpha
\mathcal{\tilde{K}}_{p}}\Vert _{\mathcal{B}(\mathcal{\hat{H}})}\leq C\ .
\end{equation*}
\end{lemma}

\noindent \textit{Proof}. By Kato's perturbation theory and Assumption \ref%
{assumption important}, which implies (\ref{norm perturbation V deformed}),
the operator
\begin{equation*}
\mathcal{R}_{p}^{\left( \lambda ,\eta \right) }:=\left( \mathcal{P}%
_{p,0}^{\left( 0,0\right) }-\mathcal{P}_{p,0}^{\left( \lambda ,\eta \right)
}\right) ^{2}\ ,\qquad p\in \mathbb{Z}\ ,
\end{equation*}%
is bounded by%
\begin{equation}
\Vert \mathcal{R}_{p}^{\left( \lambda ,\eta \right) }\Vert _{\mathcal{B}(%
\mathcal{\hat{H}})}\leq C\lambda ^{2}<1\   \label{borne 1}
\end{equation}%
for $p\in \mathbb{Z}$, sufficiently small $\left\vert \lambda \right\vert $
and some constant $C\in \mathbb{R}^{+}$ not depending on $p,\lambda $ (cf. (%
\ref{gamma oo})--(\ref{gammak0})). Therefore, we can infer from Lemma \ref%
{typeAbis copy(5)} for $P=\mathcal{P}_{p,0}^{\left( 0,0\right) }$ and $Q=%
\mathcal{P}_{p,0}^{\left( \lambda ,\eta \right) }$ that there are two
invertible, bounded operators
\begin{equation*}
U_{p}^{\left( \lambda ,\eta \right) }:=\left( \mathcal{P}_{p,0}^{\left(
\lambda ,\eta \right) }\mathcal{P}_{p,0}^{\left( 0,0\right) }+\left( 1-%
\mathcal{P}_{p,0}^{\left( \lambda ,\eta \right) }\right) \left( 1-\mathcal{P}%
_{p,0}^{\left( 0,0\right) }\right) \right) \left( 1-\mathcal{R}_{p}^{\left(
\lambda ,\eta \right) }\right) ^{-1/2}
\end{equation*}%
and
\begin{equation*}
V_{p}^{\left( \lambda ,\eta \right) }:=\left( \mathcal{P}_{p,0}^{\left(
0,0\right) }\mathcal{P}_{p,0}^{\left( \lambda ,\eta \right) }+\left( 1-%
\mathcal{P}_{p,0}^{\left( 0,0\right) }\right) \left( 1-\mathcal{P}%
_{p,0}^{\left( \lambda ,\eta \right) }\right) \right) \left( 1-\mathcal{R}%
_{p}^{\left( \lambda ,\eta \right) }\right) ^{-1/2}=\left( U_{p}^{\left(
\lambda ,\eta \right) }\right) ^{-1}
\end{equation*}%
such that
\begin{equation*}
\mathcal{P}_{p,0}^{\left( \lambda ,\eta \right) }=U_{p}^{\left( \lambda
,\eta \right) }\mathcal{P}_{p,0}^{\left( 0,0\right) }V_{p}^{\left( \lambda
,\eta \right) }\ .
\end{equation*}%
By Kato's perturbation theory and (\ref{norm perturbation V deformed}),
there is $C\in \mathbb{R}^{+}$ such that, for all $p\in \mathbb{Z}$ and
sufficiently small $\left\vert \lambda \right\vert $,%
\begin{equation}
\Vert \mathcal{P}_{p,0}^{\left( \lambda ,\eta \right) }-\mathcal{P}%
_{p,0}^{\left( 0,0\right) }\Vert _{\mathcal{B}(\mathcal{\hat{H}})}\leq
C\left\vert \lambda \right\vert \ .  \label{104bis}
\end{equation}%
Combined with (\ref{borne 1}), $\Vert \mathcal{P}_{p,0}^{\left( 0,0\right)
}\Vert =1$ and the absolutely convergent binomial series%
\begin{equation*}
(1-\mathcal{R}_{p}^{\left( \lambda ,\eta \right) })^{-1/2}=\mathbf{1}_{%
\mathcal{\hat{H}}}+\overset{\infty }{\sum_{n=1}}\left(
\begin{array}{c}
-1/2 \\
n%
\end{array}%
\right) (-\mathcal{R}_{p}^{\left( \lambda ,\eta \right) })^{n}\ ,
\end{equation*}%
Inequality (\ref{104bis}) implies that%
\begin{eqnarray}
\Vert U_{p}^{\left( \lambda ,\eta \right) }\mathcal{P}_{p,0}^{\left(
0,0\right) }-\mathcal{P}_{p,0}^{\left( 0,0\right) }\Vert _{\mathcal{B}(%
\mathcal{\hat{H}})} &\leq &C\left\vert \lambda \right\vert \ ,
\label{borne 2} \\
\Vert \mathcal{P}_{p,0}^{\left( 0,0\right) }V_{p}^{\left( \lambda ,\eta
\right) }-\mathcal{P}_{p,0}^{\left( 0,0\right) }\Vert _{\mathcal{B}(\mathcal{%
\hat{H}})} &\leq &C\left\vert \lambda \right\vert \ ,  \label{borne 2bis}
\end{eqnarray}%
for some constant $C\in \mathbb{R}^{+}$ not depending on $p,\lambda $ ($%
|\lambda |$ sufficiently small). By (\ref{Kato projection3}), it follows
that
\begin{eqnarray*}
\Vert \mathrm{e}^{\alpha \mathcal{K}_{p}}\Vert _{\mathcal{B}(\mathcal{\hat{H}%
})} &=&\Vert U_{p}^{\left( \lambda ,\eta \right) }\mathrm{e}^{\alpha
\mathcal{P}_{p,0}^{\left( 0,0\right) }V_{p}^{\left( \lambda ,\eta \right) }%
\mathcal{P}_{p,0}^{\left( \lambda ,\eta \right) }{\mathcal{\hat{G}}}(ir)%
\mathcal{P}_{p,0}^{\left( \lambda ,\eta \right) }U_{p}^{\left( \lambda ,\eta
\right) }\mathcal{P}_{p,0}^{\left( 0,0\right) }}V_{p}^{\left( \lambda ,\eta
\right) }\Vert _{\mathcal{B}(\mathcal{\hat{H}})} \\
&\leq &C\Vert \mathrm{e}^{\alpha \mathcal{P}_{p,0}^{\left( 0,0\right)
}V_{p}^{\left( \lambda ,\eta \right) }\mathcal{P}_{p,0}^{\left( \lambda
,\eta \right) }{\mathcal{\hat{G}}}(ir)\mathcal{P}_{p,0}^{\left( \lambda
,\eta \right) }U_{p}^{\left( \lambda ,\eta \right) }\mathcal{P}%
_{p,0}^{\left( 0,0\right) }}\Vert _{\mathcal{B}(\mathcal{\hat{H}})}
\end{eqnarray*}%
for all $p\in \mathbb{Z}$ and sufficiently small $\left\vert \lambda
\right\vert $, with some constant $C\in \mathbb{R}^{+}$ not depending on $%
p,\lambda $. Meanwhile, by using (\ref{borne 2})--(\ref{borne 2bis}) as well
as the Neumann series (\ref{Neumann series1}) and (\ref{equality sympa})
extended to all $p\in \mathbb{Z}$ as it is done in Lemma \ref{pert G} one
verifies that
\begin{eqnarray}
&&\Vert \mathcal{P}_{p,0}^{\left( 0,0\right) }V_{p}^{\left( \lambda ,\eta
\right) }\mathcal{P}_{p,0}^{\left( \lambda ,\eta \right) }{\mathcal{\hat{G}}}%
(ir)\mathcal{P}_{p,0}^{\left( \lambda ,\eta \right) }U_{p}^{\left( \lambda
,\eta \right) }\mathcal{P}_{p,0}^{\left( 0,0\right) }-\mathcal{P}%
_{p,0}^{\left( 0,0\right) }\mathcal{P}_{p,0}^{\left( \lambda ,\eta \right) }{%
\mathcal{\hat{G}}}(ir)\mathcal{P}_{p,0}^{\left( \lambda ,\eta \right) }%
\mathcal{P}_{p,0}^{\left( 0,0\right) }\Vert _{\mathcal{B}(\mathcal{\hat{H}})}
\notag \\
&\leq &C|\lambda |^{3}  \label{pas verifie2}
\end{eqnarray}%
for some finite constant $C\in \mathbb{R}^{+}$ not depending on $p$, $%
\lambda $, $\eta $. Using the operator $\mathfrak{Z}_{\mathrm{at}}\in
\mathcal{B}(\mathfrak{H}_{\mathrm{at}})$ (\ref{eq cool g0bis}) and Theorem %
\ref{corolario legal}, it follows from (\ref{pas verifie2}) that, for all $%
p\in \mathbb{Z}$, $A\in \mathcal{\hat{H}}$ and sufficiently small $%
\left\vert \lambda \right\vert $,%
\begin{eqnarray*}
\mathcal{P}_{p,0}^{\left( 0,0\right) }V_{p}^{\left( \lambda ,\eta \right)
}\left( \mathcal{P}_{p,0}^{\left( \lambda ,\eta \right) }{\mathcal{\hat{G}}}%
(ir)\mathcal{P}_{p,0}^{\left( \lambda ,\eta \right) }\right) U_{p}^{\left(
\lambda ,\eta \right) }\mathcal{P}_{p,0}^{\left( 0,0\right) }(A) &=&ip\varpi
\mathcal{P}_{p,0}^{\left( 0,0\right) }(A) \\
&&+\eta \mathcal{P}_{p,0}^{\left( 0,0\right) }\mathfrak{S\mathfrak{Z}}_{%
\mathrm{at}}\mathfrak{L}_{\mathrm{p}}^{\ast }\mathfrak{Z}_{\mathrm{at}}^{-1}%
\mathcal{P}_{p,0}^{\left( 0,0\right) }(A) \\
&&+\lambda ^{2}\mathfrak{Z}_{\mathrm{at}}\mathfrak{L}_{\mathcal{R}}^{\ast }%
\mathfrak{Z}_{\mathrm{at}}^{-1}\mathcal{P}_{p,0}^{\left( 0,0\right) }(A) \\
&&+\mathcal{P}_{p,0}^{\left( 0,0\right) }\mathrm{R}\mathcal{P}_{p,0}^{\left(
0,0\right) }(A)
\end{eqnarray*}%
with $\mathrm{R}\equiv \mathrm{R}^{(p,\lambda ,\eta )}$ being an operator
with norm $\left\Vert \mathrm{R}\right\Vert _{\mathcal{B}(\mathcal{\hat{H}}%
)}\leq C|\lambda |^{3}$ for some finite constant $C\in \mathbb{R}^{+}$ not
depending on $p$, $\lambda $, $\eta $ and $A$. Hence, since, by (\ref{sigma}%
),
\begin{equation}
\mathrm{U}_{p,0}\mathcal{P}_{p,0}^{\left( 0,0\right) }\mathfrak{S\mathfrak{Z}%
}_{\mathrm{at}}\mathfrak{L}_{\mathrm{p}}^{\ast }\mathfrak{Z}_{\mathrm{at}%
}^{-1}\mathcal{P}_{p,0}^{\left( 0,0\right) }\mathrm{U}_{p,0}^{\ast }=\frac{1%
}{2}\mathfrak{\mathfrak{Z}}_{\mathrm{at}}\mathfrak{L}_{\mathrm{p}}^{\ast }%
\mathfrak{Z}_{\mathrm{at}}^{-1}|_{\mathfrak{H}_{\mathrm{at}}^{(p,0)}}\
,\qquad \mathrm{U}_{p,0}\mathfrak{Z}_{\mathrm{at}}\mathfrak{L}_{\mathcal{R}%
}^{\ast }\mathfrak{Z}_{\mathrm{at}}^{-1}\mathrm{U}_{p,0}^{\ast }=\mathfrak{Z}%
_{\mathrm{at}}\mathfrak{L}_{\mathcal{R}}^{\ast }\mathfrak{Z}_{\mathrm{at}%
}^{-1}|_{\mathfrak{H}_{\mathrm{at}}^{(p,0)}}\ ,  \label{ident Z importante}
\end{equation}%
(see (\ref{eq cool g0bis})), it suffices to bound the group on $\mathfrak{H}%
_{\mathrm{at}}^{(p,0)}$ (\ref{space 0}) generated by%
\begin{equation*}
\mathcal{M}_{p}:=ip\varpi +\frac{\eta }{2}\mathfrak{L}_{\mathrm{p}}^{\ast
}+\lambda ^{2}\mathfrak{L}_{\mathcal{R}}^{\ast }+\Theta _{p}
\end{equation*}%
with
\begin{equation*}
\Theta _{p}:=\mathfrak{Z}_{\mathrm{at}}^{-1}\mathrm{U}_{p,0}\mathcal{P}%
_{p,0}^{\left( 0,0\right) }\mathrm{R}\mathcal{P}_{p,0}^{\left( 0,0\right) }%
\mathrm{U}_{p,0}^{\ast }\mathfrak{Z}_{\mathrm{at}}\ .
\end{equation*}%
Note that $i\varpi p$ is a non--degenerated eigenvalue of $\mathcal{M}_{p}$
with associated Riesz projection $P$ satisfying $\left\Vert P\right\Vert _{%
\mathcal{B}(\mathfrak{H}_{\mathrm{at}}^{(p,0)})}\leq 2$ for sufficiently
small $|\lambda |$. Hence,%
\begin{equation}
\mathrm{e}^{\alpha \mathcal{M}_{p}}=\mathrm{e}^{ip\varpi \alpha }P+(\mathbf{1%
}_{\mathfrak{H}_{\mathrm{at}}^{(p,0)}}-P)\mathrm{e}^{\alpha \mathcal{M}%
_{p}^{\prime }}(\mathbf{1}_{\mathfrak{H}_{\mathrm{at}}^{(p,0)}}-P)
\label{equacao super legal}
\end{equation}%
with%
\begin{equation*}
\mathcal{M}_{p}^{\prime }:=(\mathbf{1}_{\mathfrak{H}_{\mathrm{at}%
}^{(p,0)}}-P)\mathcal{M}_{p}(\mathbf{1}_{\mathfrak{H}_{\mathrm{at}%
}^{(p,0)}}-P)\ .
\end{equation*}%
Using spectral properties of the operator
\begin{equation*}
\left( i\varpi p+\lambda ^{2}\mathfrak{L}_{\mathcal{R}}^{\ast }+\frac{\eta }{%
2}\mathfrak{L}_{\mathrm{p}}^{\ast }\right) \in \mathcal{B}(\mathfrak{H}_{%
\mathrm{at}}^{(p,0)})
\end{equation*}%
and Kato's perturbation theory we deduce that
\begin{equation*}
\sigma \left( \mathcal{M}_{p}^{\prime }|_{\mathrm{ran}(\mathbf{1}_{\mathfrak{%
H}_{\mathrm{at}}^{(p,0)}}-P)}\right) \subset (-\infty ,C\lambda ^{2})+i%
\mathbb{R}
\end{equation*}%
for some constant $C\in \mathbb{R}^{+}$ and sufficiently small $|\lambda |$.
[It is similar to (\ref{bound Gim importante}) or \cite[Lemma 6.3]%
{BruPedraWestrich2011a} and it results from Assumption \ref{assumption3}.]
Since the generator $\mathcal{M}_{p}^{\prime }|_{\mathrm{ran}(\mathbf{1}_{%
\mathfrak{H}_{\mathrm{at}}^{(p,0)}}-P)}$ acts on a finite dimensional space,
this spectral condition implies that the corresponding semigroup is bounded.
By (\ref{equacao super legal}), the semigroup $\left\{ \mathrm{e}^{\alpha
\mathcal{M}_{p}}\right\} _{\alpha \geq 0}$ is thus bounded.

The proof of $\Vert \mathrm{e}^{\alpha \mathcal{\tilde{K}}_{p}}\Vert _{%
\mathcal{B}(\mathcal{\hat{H}})}\leq C$ is performed in the same way. It is
even simpler because the real part of the spectrum of $\mathcal{\tilde{K}}%
_{p}$ is strictly negative, by Theorem \ref{Theorem pas le temps} (ii).
\hfill $\Box $

We now study the semigroup $\{\mathrm{e}^{\alpha {\mathcal{\hat{G}}}%
(ir)}\}_{\alpha \geq 0}$ via (\ref{impoetant equation laplace transform
inverse}) and (\ref{important equationbis}):

\begin{satz}[Time--Uniform Approximations of $\{\mathrm{e}^{\protect\alpha {%
\mathcal{\hat{G}}}(ir)}\}_{\protect\alpha \geq 0}$]
\label{typeAbis copy(6)}\mbox{ }\newline
There is a constant $C\in \mathbb{R}^{+}$ such that, for all $\alpha \in
\mathbb{R}_{0}^{+}$, $\hat{\psi},\hat{\psi}^{\prime }\in \mathfrak{\hat{H}}_{%
\mathrm{at}}$ and sufficiently small $|\lambda |$,%
\begin{equation*}
\left\vert \left\langle \hat{\psi},(\mathrm{e}^{\alpha {\mathcal{\hat{G}}}%
(ir)}-\mathrm{e}^{\alpha \mathcal{K}_{0}}\mathcal{P}_{0,0}^{\left( \lambda
,\eta \right) }-\mathrm{e}^{\alpha \mathcal{\tilde{K}}_{0}}\mathcal{\tilde{P}%
}_{0}^{\left( \lambda ,\eta \right) }-\mathrm{e}^{\alpha \mathcal{\tilde{K}}%
_{-1}}\mathcal{\tilde{P}}_{-1}^{\left( \lambda ,\eta \right) })\hat{\psi}%
^{\prime }\right\rangle _{\mathcal{\hat{H}}}\right\vert \leq C\left( \varpi
^{-2}+\lambda ^{2}\mathrm{e}^{v\alpha }\right) \Vert \hat{\psi}\Vert _{%
\mathcal{\hat{H}}}\Vert \hat{\psi}^{\prime }\Vert _{\mathcal{\hat{H}}}\ .
\end{equation*}%
Moreover, if $\hat{\psi}^{\prime }\in \mathfrak{F}(\mathfrak{D}\otimes
\Omega _{\mathcal{R}})$ then%
\begin{equation*}
\left\vert \left\langle \hat{\psi},(\mathrm{e}^{\alpha {\mathcal{\hat{G}}}%
(ir)}-\mathrm{e}^{\alpha \mathcal{K}_{0}})\hat{\psi}^{\prime }\right\rangle
_{\mathcal{\hat{H}}}\right\vert \leq C\left( \left( 1+\varpi ^{-2}\right)
\left\vert \lambda \right\vert +\lambda ^{2}\mathrm{e}^{v\alpha }\right)
\Vert \hat{\psi}\Vert _{\mathcal{\hat{H}}}\Vert \hat{\psi}^{\prime }\Vert _{%
\mathcal{\hat{H}}}\ .
\end{equation*}
\end{satz}

\noindent \textit{Proof}. The assertion is a direct consequence of (\ref%
{impoetant equation laplace transform inverse}), (\ref{important equationbis}%
) and (\ref{104bis}) combined with Lemmata \ref{typeAbis copy(4)}--\ref%
{typeAbis copy(3)}. \hfill $\Box $

\subsection{Proof of Theorem \protect\ref{ninja thm cool}\label{thm final
proof}}

\noindent \textbf{1.} As the estimates leading to Theorem \ref{ninja thm
cool} are uniform w.r.t. the initial atomic state and the dynamical problems
involved are well--posed, we consider, w.l.o.g, only the case of a faithful
initial atomic state. This implies, in particular, that the density matrix $%
\rho _{\mathrm{at}}$, and hence the operator $\mathfrak{Z}_{\mathrm{at}}$,
has an inverse. Let%
\begin{equation*}
\Lambda ^{\left( \lambda ,\eta \right) }:=\left( \mathcal{P}_{0,0}^{\left(
0,0\right) }\left( \eta \mathfrak{SZ}_{\mathrm{at}}\mathfrak{L}_{\mathrm{p}%
}^{\ast }\mathfrak{Z}_{\mathrm{at}}^{-1}+\lambda ^{2}\mathfrak{Z}_{\mathrm{at%
}}\mathfrak{L}_{\mathcal{R}}^{\ast }\mathfrak{Z}_{\mathrm{at}}^{-1}\right)
\mathcal{P}_{0,0}^{\left( 0,0\right) }\right) ^{\ast }\in \mathcal{B}(%
\mathcal{\hat{H}})\ .
\end{equation*}%
By (\ref{eq penible}), observe that $\mathrm{ran}(\mathcal{P}_{0,0}^{\left(
0,0\right) })$ is equivalent to the space $\mathfrak{H}_{0}^{\left(
0,0\right) }$ of \cite[Sections 4.4, 6.2]{BruPedraWestrich2011a}. In
particular, $\Lambda ^{\left( \lambda ,\eta \right) }$ can in this way be
identified with the operator $\Lambda ^{\left( \lambda ,\eta \right) }\equiv
\Lambda _{0}^{\left( \lambda ,\eta \right) }$ of \cite[Section 6.2]%
{BruPedraWestrich2011a}. See also \cite[Eq. (6.6)]{BruPedraWestrich2011a}.
By Theorem \ref{corolario legal} and (\ref{ident Z importante}),
\begin{equation}
\Vert (\Lambda ^{\left( \lambda ,\eta \right) })^{\ast }-\mathcal{K}%
_{0}\Vert _{\mathcal{B}(\mathcal{\hat{H}})}\leq C|\lambda |^{3}
\label{estimate to improve}
\end{equation}%
for some finite constant $C\in \mathbb{R}^{+}$ not depending on $\lambda $, $%
\eta $ ($|\lambda |$ sufficiently small). Using exactly the same arguments
as in the proofs of \cite[Lemma 6.4, Theorem 4.4]{BruPedraWestrich2011a},
one verifies that, for any $\varepsilon \in (0,1)$,
\begin{equation}
\left\Vert \mathrm{e}^{\alpha (\Lambda ^{\left( \lambda ,\eta \right)
})^{\ast }}-\mathrm{e}^{\alpha \mathcal{K}_{0}}\right\Vert _{\mathcal{B}(%
\mathcal{\hat{H}})}\leq C\min \left\{ \alpha |\lambda |^{3},|\lambda |+%
\mathrm{e}^{-D\alpha \lambda ^{2}}\right\} \leq C|\lambda |^{1-\varepsilon }
\label{eq fini}
\end{equation}%
for some constants $C,D\in \mathbb{R}^{+}$ not depending on $\lambda $, $%
\eta $ ($|\lambda |$ sufficiently small). \smallskip

\noindent \textbf{2.} Now, we combine (\ref{eq fini}) with Theorems \ref%
{lemmalongtime}, \ref{invariance picasseuse}, \ref{typeAbis copy(6)} to
deduce that, for any initial faithful state $\omega _{\mathrm{at}}$ with
density matrix $\rho _{\mathrm{at}}\in \mathfrak{D}\subset $ $\mathcal{B}(%
\mathbb{C}^{d})$, any observable $A\in \mathfrak{D}\subset \mathcal{B}(%
\mathbb{C}^{d})$, $\varepsilon \in (0,1)$ and sufficiently small $|\lambda |$%
,%
\begin{equation}
\left\vert \left\langle \Omega ,U_{0,\alpha }\left( A\otimes \mathbf{1}_{%
\mathfrak{H}_{\mathcal{R}}}\Omega \right) \right\rangle _{\mathfrak{H}%
}-\left\langle \Omega ,\mathrm{e}^{\alpha (\Lambda ^{\left( \lambda ,\eta
\right) })^{\ast }}\left( A\otimes \mathbf{1}_{\mathfrak{H}_{\mathcal{R}%
}}\Omega \right) \right\rangle _{\mathcal{H}}\right\vert \leq C_{\varpi
,\varepsilon }\left\Vert A\right\Vert _{\mathcal{B}(\mathbb{C}%
^{d})}\left\vert \lambda \right\vert ^{1-\varepsilon }\ ,  \label{eq final1}
\end{equation}%
where $C_{\varpi ,\varepsilon }\in \mathbb{R}_{0}^{+}$ is a finite constant
depending on $\varpi ,\varepsilon $ but not on $\omega _{\mathrm{at}}$, $A$,
$\lambda $, $\eta $, and $\alpha $. Recall that $\{U_{s,t}\equiv
U_{s,t}^{(\lambda ,\eta )}\}_{s,t\in {\mathbb{R}}}$ is the non--autonomous
evolution family of Proposition \ref{section extension copy(3)}. By (\ref%
{ninja1}),
\begin{equation}
\left\langle \Omega ,U_{0,\alpha }\left( A\otimes \mathbf{1}_{\mathfrak{H}_{%
\mathcal{R}}}\Omega \right) \right\rangle _{\mathfrak{H}}=\omega _{\mathrm{at%
}}\left( \alpha \right) \left( A\right) \ .  \label{eq final2}
\end{equation}%
Moreover, for any observable $A\in \mathfrak{D}\subset \mathcal{B}(\mathbb{C}%
^{d})$,
\begin{equation}
\omega _{\mathrm{at}}\left( \alpha \right) \left( A\right) =\left\langle
\rho _{\mathrm{at}}\left( \alpha \right) ,A\right\rangle _{\mathfrak{H}_{%
\mathrm{at}}}=\left\langle P_{\mathfrak{D}}(\rho _{\mathrm{at}}\left( \alpha
\right) ),A\right\rangle _{\mathfrak{H}_{\mathrm{at}}}\ ,  \label{eq final3}
\end{equation}%
see (\ref{trace h at}). Recall also that $\rho _{\mathrm{at}}\left( \alpha
\right) $ is, by definition, the density matrix of $\omega _{\mathrm{at}%
}\left( \alpha \right) $ and $P_{\mathfrak{D}}$ is the orthogonal projection
on the subspace $\mathfrak{D}$ (\ref{ninja0}) of block--diagonal matrices.
Meanwhile, we infer from (\ref{ident Z importante}) that
\begin{equation}
\Lambda ^{\left( \lambda ,\eta \right) }=\left( \mathcal{P}_{0,0}^{\left(
0,0\right) }\mathrm{U}_{0,0}^{\ast }\mathfrak{Z}_{\mathrm{at}}\left( \frac{%
\eta }{2}\mathfrak{L}_{\mathrm{p}}^{\ast }+\lambda ^{2}\mathfrak{L}_{%
\mathcal{R}}^{\ast }\right) \mathfrak{Z}_{\mathrm{at}}^{-1}\mathrm{U}_{0,0}%
\mathcal{P}_{0,0}^{\left( 0,0\right) }\right) ^{\ast }\ .
\label{ptt eq cool}
\end{equation}%
\cite[Theorem 4.6 (i)]{BruPedraWestrich2011a} implies that $\mathfrak{H}_{%
\mathrm{at}}^{(0,0)}$ is an invariant space of%
\begin{equation*}
\mathfrak{Z}_{\mathrm{at}}\left( \frac{\eta }{2}\mathfrak{L}_{\mathrm{p}%
}^{\ast }+\lambda ^{2}\mathfrak{L}_{\mathcal{R}}^{\ast }\right) \mathfrak{Z}%
_{\mathrm{at}}^{-1}\in \mathcal{B}(\mathfrak{H}_{\mathrm{at}})
\end{equation*}%
and we deduce from (\ref{ptt eq cool}) that%
\begin{equation*}
\mathrm{e}^{\alpha (\Lambda ^{\left( \lambda ,\eta \right) })^{\ast }}%
\mathcal{P}_{0,0}^{\left( 0,0\right) }=\mathrm{U}_{0,0}^{\ast }\mathfrak{Z}_{%
\mathrm{at}}\mathrm{e}^{\alpha \left( \frac{\eta }{2}\mathfrak{L}_{\mathrm{p}%
}^{\ast }+\lambda ^{2}\mathfrak{L}_{\mathcal{R}}^{\ast }\right) }\mathfrak{Z}%
_{\mathrm{at}}^{-1}\mathrm{U}_{0,0}\mathcal{P}_{0,0}^{\left( 0,0\right) }\ .
\end{equation*}%
By (\ref{unitary transform}), it follows that, for any observable $A\in
\mathfrak{D}\subset \mathcal{B}(\mathbb{C}^{d})$ and initial density matrix $%
\rho _{\mathrm{at}}\in \mathfrak{D}$,%
\begin{eqnarray}
\left\langle \Omega ,\mathrm{e}^{\alpha (\Lambda ^{\left( \lambda ,\eta
\right) })^{\ast }}\left( A\otimes \mathbf{1}_{\mathfrak{H}_{\mathcal{R}%
}}\Omega \right) \right\rangle _{\mathcal{H}} &=&\left\langle \rho _{\mathrm{%
at}}^{1/2},\mathfrak{Z}_{\mathrm{at}}\mathrm{e}^{\alpha \left( \frac{\eta }{2%
}\mathfrak{L}_{\mathrm{p}}^{\ast }+\lambda ^{2}\mathfrak{L}_{\mathcal{R}%
}^{\ast }\right) }\mathfrak{Z}_{\mathrm{at}}^{-1}A\rho _{\mathrm{at}%
}^{1/2}\right\rangle _{\mathfrak{H}_{\mathrm{at}}}  \notag \\
&=&\left\langle \rho _{\mathrm{at}},\mathrm{e}^{\alpha \left( \frac{\eta }{2}%
\mathfrak{L}_{\mathrm{p}}^{\ast }+\lambda ^{2}\mathfrak{L}_{\mathcal{R}%
}^{\ast }\right) }A\right\rangle _{\mathfrak{H}_{\mathrm{at}}}  \notag \\
&=&\left\langle \mathrm{e}^{\alpha \left( \frac{\eta }{2}\mathfrak{L}_{%
\mathrm{p}}+\lambda ^{2}\mathfrak{L}_{\mathcal{R}}\right) }\rho _{\mathrm{at}%
},A\right\rangle _{\mathfrak{H}_{\mathrm{at}}}  \notag \\
&=&\left\langle P_{\mathfrak{D}}(\mathrm{e}^{\alpha \left( \frac{\eta }{2}%
\mathfrak{L}_{\mathrm{p}}+\lambda ^{2}\mathfrak{L}_{\mathcal{R}}\right)
}\rho _{\mathrm{at}}),A\right\rangle _{\mathfrak{H}_{\mathrm{at}}}  \notag \\
&=&\left\langle P_{\mathfrak{D}}(\mathrm{e}^{\alpha \tilde{\Lambda}^{\left(
\lambda ,\eta \right) }}\rho _{\mathrm{at}}),A\right\rangle _{\mathfrak{H}_{%
\mathrm{at}}}\ ,  \label{eq final4eq final4}
\end{eqnarray}%
where
\begin{equation*}
\tilde{\Lambda}^{\left( \lambda ,\eta \right) }=\frac{\eta }{2}\mathfrak{L}_{%
\mathrm{p}}+\lambda ^{2}\mathfrak{L}_{\mathcal{R}}
\end{equation*}%
is the operator of \cite[Theorem 4.6 (i)]{BruPedraWestrich2011a}. We used in
the second equality the identities
\begin{equation*}
\mathcal{P}_{0,0}^{\left( 0,0\right) }((A\rho _{\mathrm{at}}^{1/2})\otimes
\Omega _{\mathcal{R}})=(A\rho _{\mathrm{at}}^{1/2})\otimes \Omega _{\mathcal{%
R}}\quad \text{and}\quad \mathrm{U}_{0,0}((A\rho _{\mathrm{at}%
}^{1/2})\otimes \Omega _{\mathcal{R}})=A\rho _{\mathrm{at}}^{1/2}
\end{equation*}
when $A\in \mathfrak{D}$. [Cf. definition (\ref{unitary transform}%
).]\smallskip

\noindent \textbf{3.} Therefore, Theorem \ref{ninja thm cool} follows from
Equations (\ref{eq final1})--(\ref{eq final4eq final4}) and \cite[Corollary
4.5, Eqs. (4.37)--(4.38)]{BruPedraWestrich2011a}.

\section{Appendix\label{sectino appendix}}

\textbf{1.} For the reader's convenience, we review here one definition and
a result from the theory of analytic families of closed operators used in
Section \ref{section spectral deformation} to study the analytic deformation
of the Howland operator $\mathcal{G}$.

\begin{definition}[Families of type A]
\label{Families of type A}\mbox{ }\newline
For an open subset $\mathcal{S}$ of $\mathbb{C}$, a family $\{F_{z}\}_{z\in
\mathcal{S}}$ of closed operators defined on a Banach space $\mathcal{X}$
and with non--empty resolvent set is of type A when $\mathrm{Dom}(F_{z})=%
\mathcal{Y}\subset \mathcal{X}$ of $F_{z}$ is independent of $z\in \mathcal{S%
}$ and the map $z\mapsto F_{z}x$ is strongly analytic on $\mathcal{S}$ for
all $x\in \mathcal{Y}$.
\end{definition}

\noindent Families of type A are useful in the context of Kato's
perturbation theory as they form a special case of analytic family (in the
sense of Kato). In our proofs, we use the following well--known result on
type A families (see \cite[Sect. VII.1, Theorem 1.3]{Kato}):

\begin{lemma}[Type A families and analycity of resolvents]
\label{typeA}\mbox{ }\newline
Let $\{F_{z}\}_{z\in \mathcal{S}}$ be a closed operator family of type A and
let $\zeta \in \varrho (F_{z_{0}})$, where $\varrho (F_{z_{0}})$ is the
resolvent set of $F_{z_{0}}$ with $z_{0}\in \mathcal{S}$. Then the map $%
z\mapsto (\zeta -F_{z})^{-1}$ is analytic in some neighborhood of $z_{0}$.
\end{lemma}

\noindent \textbf{2.} We used a representation of strongly continuous
one--parameter semigroups w.r.t. resolvents. Indeed, resolvents and
semigroups are related to each other through the Laplace transform (see,
e.g., \cite[Chap. III, Corollary 5.15]{Nagel-engel}):

\begin{lemma}[Semigroups as Laplace transforms]
\label{typeAbis copy(1)}\mbox{ }\newline
Let $A$ be a (possibly unbounded) generator of a strongly continuous
semigroup $\left\{ \mathrm{e}^{\alpha A}\right\} _{\alpha \geq 0}$ on a
Banach space $\mathcal{X}$ with growth bound $\mathbb{\varsigma }\in \mathbb{%
R\cup }\left\{ -\infty \right\} $. Then, for all $x\in \mathrm{Dom}\left(
A\right) $, $w>\mathbb{\varsigma }$ and $\alpha \in \mathbb{R}^{+}$,%
\begin{equation*}
\mathrm{e}^{\alpha A}x=\underset{N\rightarrow \infty }{\lim }\left\{ \frac{1%
}{2\pi i}\int_{w-iN}^{w+iN}\mathrm{e}^{\alpha \zeta }(\zeta -A)^{-1}x\mathrm{%
d}\zeta \right\} \ .
\end{equation*}
\end{lemma}

\noindent \textbf{3.} In Kato's perturbation theory, a perturbed Riesz
projection $P\left( y\right) $ is related to the (non--perturbed Riesz)
projection $P\left( 0\right) $ through some bounded operator $U\left(
y\right) $ with bounded inverse $V\left( y\right) $ by the relation $P\left(
y\right) =U\left( y\right) P\left( 0\right) V\left( y\right) $, provided the
coupling constant $y\in \mathbb{R}$ measuring the strength of the
perturbation is small enough in absolute value. This fact is used to prove
Lemma \ref{typeAbis copy(3)}. The following relations between pairs of
projections (see \cite[I.4.6. p. 33]{Kato}) are important in that proof: Let
$P,Q$ be two projections acting on a Banach space $\mathcal{X}$. Then, the
bounded operator
\begin{equation*}
U^{\prime }:=QP+\left( 1-Q\right) \left( 1-P\right)
\end{equation*}%
maps $\mathrm{ran}\left( P\right) $ into $\mathrm{ran}\left( Q\right) $,
whereas
\begin{equation*}
V^{\prime }:=PQ+\left( 1-P\right) \left( 1-Q\right)
\end{equation*}%
maps $\mathrm{ran}\left( Q\right) $ into $\mathrm{ran}\left( P\right) $.
Moreover, $V^{\prime }U^{\prime }=U^{\prime }V^{\prime }=1-\mathcal{R}$,
where $\mathcal{R}:=\left( P-Q\right) ^{2}$. In Kato's perturbation theory,
the operator $(P\left( y\right) -P\left( 0\right) )$ has typically a small
norm for small $\left\vert y\right\vert $ and $\left( 1-\mathcal{R}\right)
^{-1/2}$ exists as an absolutely convergent power series in $\mathcal{R}$.
Under this assumption and since $\mathcal{R}$ always commutes with $P$ and $%
Q $, we can relate both projections to each other, as explained above:

\begin{lemma}[Pairs of near projections]
\label{typeAbis copy(5)}\mbox{ }\newline
Let $P,Q$ be two projections and $\mathcal{R}:=\left( P-Q\right) ^{2}$. If $%
\left( 1-\mathcal{R}\right) ^{-1/2}$ exists and is bounded then the bounded
operator%
\begin{equation*}
U:=U^{\prime }\left( 1-\mathcal{R}\right) ^{-1/2}=\left( 1-\mathcal{R}%
\right) ^{-1/2}U^{\prime }
\end{equation*}%
maps $\mathrm{ran}\left( P\right) $ into $\mathrm{ran}\left( Q\right) $,
whereas
\begin{equation*}
V:=V^{\prime }\left( 1-\mathcal{R}\right) ^{-1/2}=\left( 1-\mathcal{R}%
\right) ^{-1/2}V^{\prime }
\end{equation*}%
maps $\mathrm{ran}\left( Q\right) $ into $\mathrm{ran}\left( P\right) $.
Moreover, $VU=UV=1$, i.e., $V=U^{-1}$, and $P=VQU$, $Q=UPV$.
\end{lemma}

\noindent \textit{Acknowledgments:} We would like to thank Volker Bach for
his support as well as Matthias\ Westrich for discussions and his help in
Section \ref{Time--dependent C--Liouvilleans}. We thank the referees and the
editor for useful hints. This research is supported by a grant of the
\textquotedblleft Inneruniversit\"{a}re Forschungsf\"{o}rderung%
\textquotedblright\ of the Johannes Gutenberg University, by the agencies
MINECO and FAPESP under Grants MTM2010-16843 and 2013/13215-5 as well as by
the Basque Government through the grant IT641-13 and the BERC 2014-2017
program and by the Spanish Ministry of Economy and Competitiveness MINECO:
BCAM Severo Ochoa accreditation SEV-2013-0323.

\end{document}